\documentclass[a4paper,11pt]{article}
\usepackage{jheppub} 
\usepackage{lipsum}
\usepackage{appendix}

\newcommand{\be}{\begin{equation}}
\newcommand{\ee}{\end{equation}}
\newcommand{\ba}{\begin{array}}
\newcommand{\ea}{\end{array}}
\newcommand{\bea}{\begin{eqnarray}}
\newcommand{\eea}{\end{eqnarray}}

\newcommand{\beq}{\begin{equation}}
\newcommand{\eeq}{\end{equation}}
\newcommand{\beqa}{\begin{eqnarray}}
\newcommand{\eeqa}{\end{eqnarray}}

\newcommand{\muon}{{|\nu_{\mu}\rangle}}

\newcommand{\tx}{{\theta_{12}}}
\newcommand{\ty}{{\theta_{13}}}
\newcommand{\tz}{{\theta_{23}}}

\newcommand{\dl}{{\Delta_{31}}}
\newcommand{\ds}{{\Delta_{21}}}

\newcommand{\ahat}{\hat{A}}
\newcommand{\dlhat}{\hat{\Delta}_{31}}
\newcommand{\dshat}{\hat{\Delta}_{21}}

\newcommand{\dcp}{\delta_{\mathrm{CP}}}
\newcommand{\nova}{NO$\nu$A~}

\newcommand{\pme}{P_{\mu e}}

\newcommand{\pmebar}{P_{\bar{\mu} \bar{e}}}
\newcommand{\pmt}{P_{\mu \tau}}
\newcommand{\pmtbar}{P_{\bar{\mu} \bar{\tau}}}

\newcommand{\dchsq}{\Delta\chi^2}

\title{\boldmath Effects of tau-neutrino detection on non-standard interactions at DUNE with a short discussion on the nature of neutrino mixing}

\author{Xin Yue Yu,}
\emailAdd{xyz.yu@mail.utoronto.ca}

\author{Zishen Guan,}
\emailAdd{nick.guan@mail.utoronto.ca}

\author{William Dallaway,}
\emailAdd{william.dallaway@mail.utoronto.ca}

\author{Ushak Rahaman,}
\emailAdd{ushak.rahaman@cern.ch}

\author{Nikolina Ilic,}
\affiliation{Department of Physics, University of Toronto, Toronto, ON M5S 1A7, Canada}

\emailAdd{Nikolina.Ilic@cern.ch}

\abstract{In this paper, we investigate the effects of $\nu_\tau$ and $\bar{\nu}_\tau$ detection at the DUNE far detector on the experiment’s sensitivity to Non-Standard Interactions (NSI) in neutrino propagation.  We show that the strongest observable NSI effect in the $\nu_\tau$ and $\bar{\nu}_\tau$ appearance probabilities arises from $\epsilon_{\mu\tau}$. We have studied the hierarchy sensitivity, CP violation sensitivity and octant sensitivity of DUNE from $\nu_\tau$ and $\bar{\nu}_\tau$ appearance channels in presence of NSI. We have also studied the detection sensitivity of NSI phases and the future constaints on NSI parameters from the tau neutrino appearance channels in DUNE. Additionally, we examine the role of $\nu_\tau$ detection in constraining the unitary nature of the PMNS matrix. These studies emphasize the importance of incorporating $\nu_\tau$ detection in long-baseline neutrino experiments such as DUNE.}

\begin{document}

\maketitle
\flushbottom
\section{Introduction}
\label{intro}
The neutrino oscillation phenomenon is one of those few windows, in the otherwise complete standard model (SM), which gives the particle physics community an opportunity to look into the promising world of the beyond standard model physics. In the simplest extension of SM, in order to accommodate neutrino oscillation, the neutrino mass and flavour states mix through a unitary transformation, defined by the unitary PMNS mixing matrix
\begin{equation}
    U=\left(
\begin{array}{ccc}
c_{13}c_{12} & s_{12}c_{13} & s_{13}e^{-i\dcp}\\
-s_{12}c_{23}-c_{12}s_{23}s_{13}e^{i\dcp} & c_{12}c_{23}-s_{12}s_{23}s_{13}e^{i\dcp} & s_{23}c_{13}\\
s_{12}s_{23}-s_{13}c_{12}c_{23}e^{i\dcp} & -c_{12}s_{23}-s_{13}c_{23}s_{12}e^{i\dcp} & c_{23}c_{13}\\
\end{array}
\right),
\label{PMNS}
\end{equation}
where $c_{ij}=\cos \theta_{ij}$, and $s_{ij}=\sin \theta_{ij}$, with $i,\,j=1,\,2,\,3$. Neutrino oscillation probabilities are functions of three mixing angles $\tx$, $\ty$, and $\tz$, one CP violating phase $\dcp$, and two mass-squared differences $\ds=m_{2}^{2}-m_{1}^{2}$ and $\dl=m_{3}^{2}-m_{1}^{2}$. Decades of neutrino oscillation experiments have enabled the particle physics community to measure most of the neutrino oscillation parameters, except the sign of $\dl$, octant of $\tz$, and $\dcp$ \footnote{Readers can see this review article \cite{Rahaman:2022rfp} to have an idea about the determination of other neutrino oscillation parameters.} The current global best-fit values of the neutrino oscillation parameters have been listed in table \ref{bestfit}. Depending on the sign of $\dl$, there can be two possibilities: 1. Normal hierarchy (NH): $m_3>>m_2>m_1$, and 2. Inverted hierarchy (IH): $m_2>m_1>>m_3$, where $m_i$s are the mass of the neutrino mass eigenstates $\nu_i$s. Similarly, there can be two possibilities for the octants of $\tz$: 1. Higher octant (HO): $\sin^2 \tz>0.5$, and 2. Lower octant (LO): $\sin^2 \tz<0.5$. 

The presently running long-baseline accelerator neutrino experiments \nova \cite{NOvA:2007rmc} and T2K \cite{T2K:2001wmr} are expected to measure the unknown neutrino oscillation parameters by measuring the $\nu_\mu \to \nu_e$ appearance and $\nu_\mu \to \nu_\mu$ survival probabilities in both neutrino, and anti-neutrino modes. According to the present T2K result \cite{T2Kapp, T2Kdisapp}, $|\Delta_{32}|=(2.494^{+0.041}_{-0.058})\times 10^{-3}\, {\rm eV}^2$ ($|\Delta_{31}|=(2.463^{+0.042}_{-0.056})\times 10^{-3}\, {\rm eV}^2$), $\sin^2 \tz=0.561^{+0.019}_{-0.038}$ ($0.563^{+0.017}_{-0.032}$), $\dcp=-1.97^{+0.97}_{-0.62}$ ($-1.44^{+0.56}_{-0.59}$) for NH (IH). On the other hand, the present \nova data yields  $|\Delta_{32}|=(2.433^{+0.035}_{-0.036})\times 10^{-3}\, {\rm eV}^2$ ($2.473\pm0.035$), $\sin^2 \tz=0.546^{+0.032}_{-0.075}$ ($0.539^{+0.028}_{-0.075}$), $\dcp/\pi=0.88$ ($1.51$) for NH (IH) \cite{Wolcott:2024}. There exists tension between the \nova and T2K data regarding the best-fit value of $\dcp$ when the mass hierarchy is normal, and the two experiments rule out the $1\,\sigma$ allowed regions of each other on the $\sin^2\tz-\dcp$ planes. This tension arises mainly from the $\nu_\mu \to \nu_e$ appearance channel \cite{Rahaman:2021zzm}, and has become stronger with time \cite{Rahaman:2022rfp}. Several studies have been conducted to resolve this tension with the introduction of beyond standard model (BSM) physics \cite{Rahaman:2021leu, Miranda:2019ynh, Chatterjee:2020kkm, Denton:2020uda}. 

The data collected at future long-baseline accelerator neutrino experiments such as DUNE \cite{Abi:2018dnh} and T2HK \cite{Ishida:2013kba} will play a key role to resolving this tension as well as searching for possible BSM physics. Many studies have been conducted on BSM physics in future long-baseline experiments \cite{Friedland:2012tq, Ge:2016xya, Escrihuela:2016ube, Kaur:2021rau}. In ref.~\cite{Chatterjee:2020kkm, Denton:2020uda, Chatterjee:2024kbn}, the authors have explored non-standard interactions (NSI) during neutrino propagation as a possible solution to the tension between \nova and T2K. In these references, the effects of NSI in the $e-\mu$ and $e-\tau$ sector have been considered. Since vector NSI results in the addition of a matter-like effect term to the neutrino propagation Hamiltonian, the introduction of NSI does not drastically change the T2K result. This is because for a T2K baseline of $295\, {\rm km}$, and flux peaking at an energy of $0.6\, {\rm GeV}$, the matter effect in T2K is small and hence the NSI effect is small as well. However, the introduction of NSI in the $e-\mu$ and $e-\tau$ sectors can make significant changes to the oscillation probability $\nu_\mu \to \nu_e$ ($\pme$) and $\bar{\nu}_\mu \to \bar{\nu}_e$ ($\pmebar$) and affect the \nova result, bringing its best fit point closer to the best fit point of T2K. According to reference \cite{Chatterjee:2024kbn}, for $|\epsilon_{e\mu}|=0.125$, and $\phi_{e\mu}=1.35\pi$, \nova (T2K) has $\dcp$ best-fit point around $0.9\pi$ ($1.5\pi$) and both experiments have a large overlap between their allowed regions at $1\,\sigma$ C.L. For $|\epsilon_{e\tau}|=0.22$, and $\phi_{e\tau}=1.70\pi$, both experiments have a $\dcp$ best-fit point around $1.5\pi$. Here $\epsilon_{e\mu}=|\epsilon_{e\mu}|e^{i \phi_{e\mu}}$ and $\epsilon_{e \tau}=|\epsilon_{e \tau}|e^{i \phi_{e \tau}}$ are NSI parameters representing the $e-\mu$ and $e-\tau$ sectors respectively.

Looking for $\nu_\mu \to \nu_\tau$, and $\bar{\nu}_\mu \to \bar{\nu}_\tau$ oscillations in experiments can provide a useful additional handle in discovering NSI. However, direct measurement of $\nu_\tau$ events is difficult in both beam and atmospheric neutrino oscillation experiments. Neutrino experiments fall into the category of fixed target experiments. This characteristic coupled with a large $\tau$ mass yields large threshold energies for $\nu_\tau$ scattering of ordinary matter. The threshold energy, $E_{\rm Th}$, is greater than $3.35\, {\rm GeV}$ ($3.1\,{\rm GeV}$) for $\nu_\tau+N\to \tau+N$ ($\nu_\tau+e\to\tau+\nu_e$), where $N$ is a nucleon \cite{DeGouvea:2019kea}. Due the large beam energies of DUNE, motivated by the requirement of observing large $\dl$ driven oscillations at the far detector, DUNE will be able to observe $\tau$ events above this threshold energy. However, the phase suppression effects are large at DUNE, and in its lifetime, the experiment is expected to contain between $100-1000$ tau events. Moreover, identifying and reconstructing $\nu_{\tau}$ events in the liquid argon of the DUNE far detector is challenging, because the decay of a $\tau$ lepton involves a $\nu_{\tau}$ in the final state, which carries away significant undetectable energy. The beam mode of DUNE, due to its large detector fiducial volume, powerful neutrino beam and exquisite track reconstruction capability, is capable of detecting $\nu_\tau$ in spite of the challenges. In ref.~\cite{Machado:2020yxl} a novel approach to improving the $\nu_\tau$ signal (S) over background (B) has been proposed for DUNE. With a $\tau$ optimized beam, $S/\sqrt{B}=8.8$ ($11$) for hadronic (leptonic) decay of the $\tau$ lepton could be achieved in DUNE. In ref.~\cite{DeGouvea:2019kea, Masud:2017bcf}, the authors have discussed several physics possibilities with $\nu_\tau$ in DUNE, including NSI. However, they used an outdated flux prediction, and only addressed the future sensitivity of NSI parameters. In ref.~\cite{Ghoshal:2019pab}, the authors have investigated future sensitivities of the NSI parameters using $\nu_\tau$ and $\bar{\nu}_\tau$ appearances. However, they also used  older flux files, and their detector simulations and energy resolution functions are quite different from our paper. Moreover, in our work, we have investigated the implications of tau neutrino detection at DUNE from theoretical point of view in more details. In the appendix~\ref{flux+sim}, we have discussed the relative comparison between the flux files used in this paper and those used in ref.~\cite{DeGouvea:2019kea, Masud:2017bcf, Ghoshal:2019pab}. 

In this paper, we explore NSI effects with $\nu_\tau$ in DUNE. We calculate the hierarchy sensitivity, CP violation sensitivity and octant sensitivity of DUNE in presence of NSI. We have considered NSI due to $\epsilon_{e\mu}$ and $\epsilon_{e\tau}$. The presence of these particular values is motivated by the fact that their existence resolves the \nova and T2K tension as shown in ref.~\cite{Chatterjee:2020kkm, Denton:2020uda, Chatterjee:2024kbn}. We have also considered the effects due to $\epsilon_{\mu\mu}$, $\epsilon_{\mu \tau}$ and $\epsilon_{\tau \tau}$. The detailed physics effects of these parameters on the oscillation probabilities are discussed in the next section. We have also provide future expected constraints on different NSI parameters when the $\nu_\mu \to \nu_\tau$ oscillation is considered as part of the signal. The issue of determining NSI phases in the case of NSI being present due to either of $\epsilon_{e\mu}$, $\epsilon_{e\tau}$, and $\epsilon_{\mu\tau}$ is also addressed. In section \ref{theory}, theoretical background and motivations are discussed. The details of analysis, including $\nu_\tau$ simulation, are discussed in section \ref{analysis}. The final results are provided in section \ref{results}. An important physics motivation for the detection $\nu_\tau$ in DUNE is to constrain the unitary property of the $\nu_\tau$ mixing with the mass eigenstates. A short discussion on the role of the $\tau$ neutrino in determining the unitary property of the PMNS mixing matrix is presented in section~\ref{unitary}. The final conclusions are drawn in section \ref{conclusion}.

\begin{table}
  \begin{center}
\begin{tabular}{|c|c|c|}
  \hline
  Parameters & NH &
  IH\\
  
  \hline
  $\tx/^\circ$ & $33.68^{+0.73}_{-0.70}$ & $33.68^{+0.73}_{-0.70}$\\
  \hline
  $\tz/\circ$ & $43.3^{+1.0}_{-0.8}$ & $47.9^{+0.7}_{-0.9}$\\
  \hline
  $\ty/\circ$ & $8.56^{+0.11}_{-0.11}$ & $8.59^{+0.11}_{-0.11}$ \\
  \hline
 $ \dcp/\circ$ & $212^{+26}_{-41}$ & $274^{+22}_{-25}$ \\
 \hline
  $ \frac{\Delta_{21}}{10^{-5}\, {\rm eV}^2}$  & $7.49^{+0.19}_{-0.19}$ & $7.49^{+0.19}_{-0.19}$ \\
 \hline
 $ \frac{\Delta_{3l}}{10^{-3}\, {\rm eV}^2}$  & $2.513^{+0.021}_{-0.019}$ & $-2.484^{+0.020}_{-0.020}$ \\
 \hline
\end{tabular}
\end{center}
 \caption{Best-fit values of the neutrino oscillation parameters according to the present global fits \cite{Esteban:2024eli, nufit}. $\Delta_{3l}=\dl>0$ for NH, $\Delta_{3l}=\Delta_{32}<0$ for IH. }
  \label{bestfit}
\end{table}
\section{Theoretical background and motivations}
\label{theory}
In the 3-neutrino paradigm, the $\nu_\mu\to \nu_\tau$ oscillation probability in vacuum for DUNE with a baseline of $1300$ km, and energy above $\tau$ threshold ($E>3.4$ GeV) can be written as
\begin{eqnarray}
    P_{\mu \tau}^{\rm vac}&=&4 |U_{\mu 3}|^2 |U_{\tau 3}|^2 \sin^2\left(\frac{\dl L}{4E}\right)+{\rm subleading} \nonumber \\
    &=& \sin^2 2\theta_{23}\cos^4 \theta_{13} \sin^2\left(\frac{\dl L}{4E}\right)+{\rm subleading}
    \label{mu-tau}
\end{eqnarray}
Using the best-fit values from ref.~\cite{Esteban:2024eli, nufit}, we obtain $4 |U_{\mu 3}|^2 |U_{\tau 3}|^2 \approx 0.95$. The subleading term consists of the solar neutrino oscillation, and an interference term driven by the ratio of two mass-squared differences: $\ds/\dl \sim 0.03$. This is the major difference with respect the $\nu_\mu \to \nu_e$ oscillation probability. In the later case, the leading term is proportional to $|U_{e3}|^2\approx 0.022$, thus the interference term is of the same order as the leading term. 

Throughout our analysis, we have include the matter effect. The matter effect is parameterized by $A=\sqrt{2}G_Fn_eE$, where $G_F$ is the Fermi constant, and $n_e$ is the electron density in matter. For DUNE, $A\sim 5.8\times 10^{-4}E\, {\rm eV}^2$. Thus, both $\dlhat\equiv \dl/(2E)$, and $\hat{A}\equiv A/E$ are much larger than $\dshat\equiv \ds/(2E)$, for the neutrino energy above 3.4 GeV. Hence, $\hat{\ds}\to0$ is a good approximation, and under this approximation, 
\begin{equation}
    P_{\mu \tau}=\sin^2 2\tz \left|e^{\left(i\frac{\Delta_M L}{2}\right)}\cos^2\theta_M\sin \left(\frac{\Delta_{+}L}{2}\right)+\sin^2 \theta_{M}\sin\left(\frac{\Delta_-L}{2}\right)\right|^2,
    \label{mu-tau-mat}
\end{equation}
where
\begin{eqnarray}
    \Delta_{\pm}&&=\frac{\ahat+\dlhat}{2}\pm \frac{\Delta_M}{2},\\
    \Delta_M&&=\sqrt{(\ahat-\dlhat \cos 2\ty)^2+\dl\sin^2 2\ty},\\
    \tan 2\theta_M&&=\frac{\dlhat \sin 2\ty}{\dlhat\cos2\ty-A}.
\end{eqnarray}
At the $\tau$ threshold energy, $\dlhat \approx 3.6/10^{-4}\, {\rm eV}^2/{\rm GeV}$. Thus, $\ahat>\dlhat$ above the $\tau$ threshold energy. In this limit, and when $\dlhat L<<1$, $\ahat>>\dlhat$, eq.~\ref{mu-tau-mat} can be well approximated by eq.~\ref{mu-tau}. In ref.~\cite{Denton:2016wmg}, the non-trivial effect of non-zero $\ds$ have been considered, and it can be seen numerically that the inclusion of matter effect does not change the physics to any significant level.

\begin{figure}[htbp]
\centering
\includegraphics[width=1.0\textwidth]{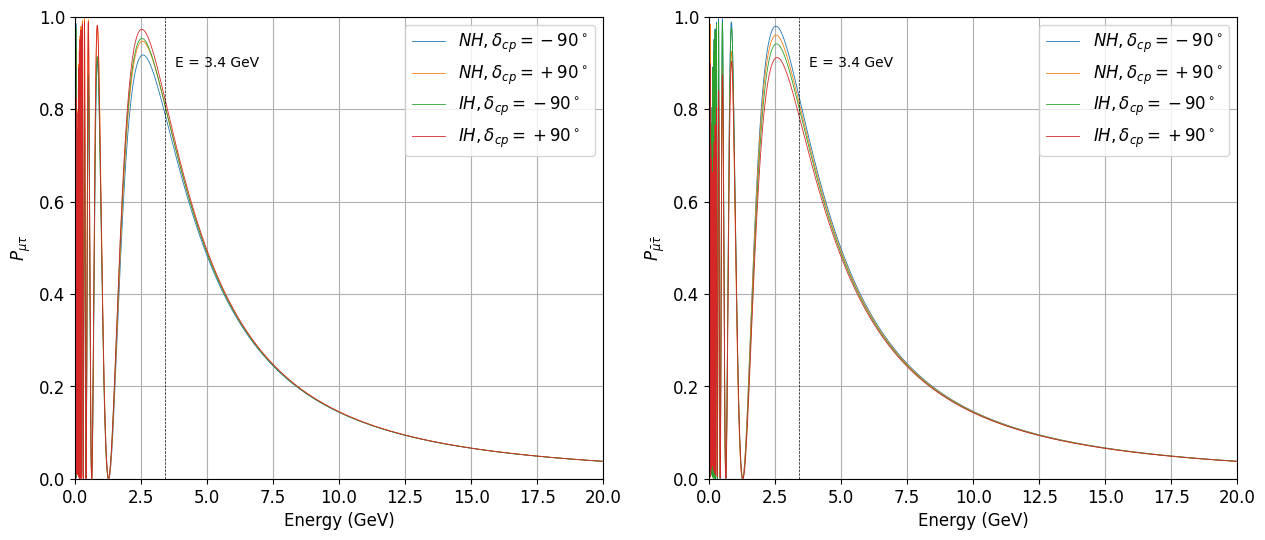}
\caption{\footnotesize{$\nu_\mu \to \nu_\tau$ oscillation probability as a function of energy with different hierarchy and $\dcp$ values for (anti-) neutrino in the (right) left panel. The hierarchy-$\dcp$ combinations for different coloured lines have been mentioned on the plots. The other oscillation parameters have been fixed to the best-fit values taken from ref.~\cite{Esteban:2024eli, nufit}.}}
\label{prob-mu-tau-hierarchy}
\end{figure}

\begin{figure}[htbp]
\centering
\includegraphics[width=1.0\textwidth] {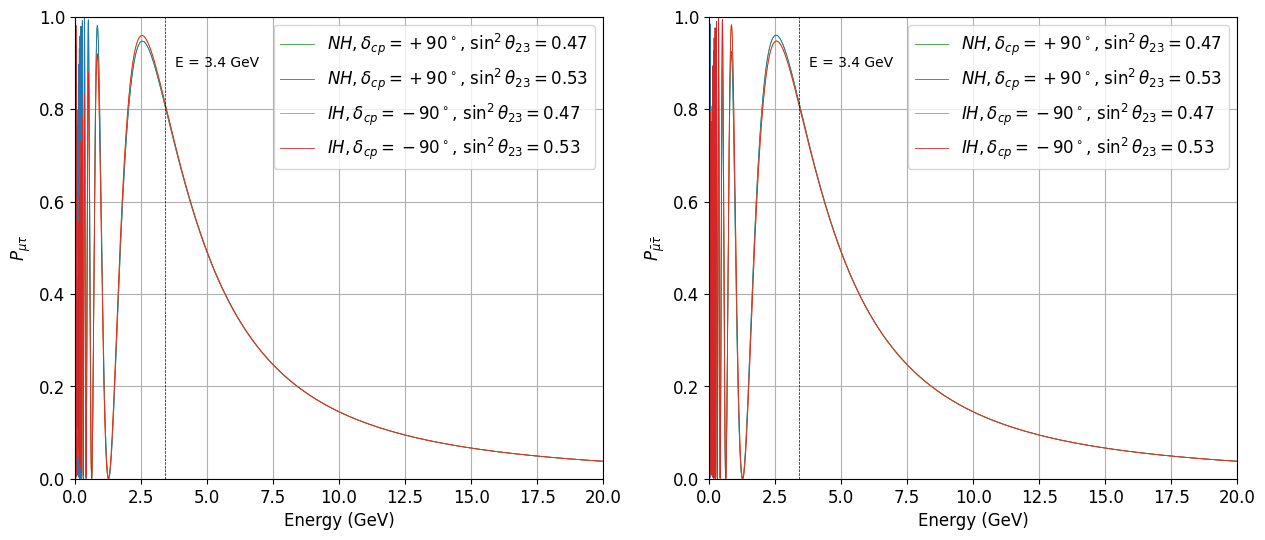}
\caption{\footnotesize{$\nu_\mu \to \nu_\tau$ oscillation probability as a function of energy with different  NH-$\dcp=+90^\circ$ and IH-$\dcp=-90^\circ$ and $\tz$ in LO and HO for (anti-) neutrino in the (right) left panel. The hierarchy-$\dcp$ combinations, and $\sin^2\tz$ values for different coloured lines have been mentioned on the plots. The other oscillation parameters have been fixed to the best-fit values taken from ref.~\cite{Esteban:2024eli, nufit}.}}
\label{prob-mu-tau-octant-1}
\end{figure}

\begin{figure}[htbp]
\centering
\includegraphics[width=1.0\textwidth] {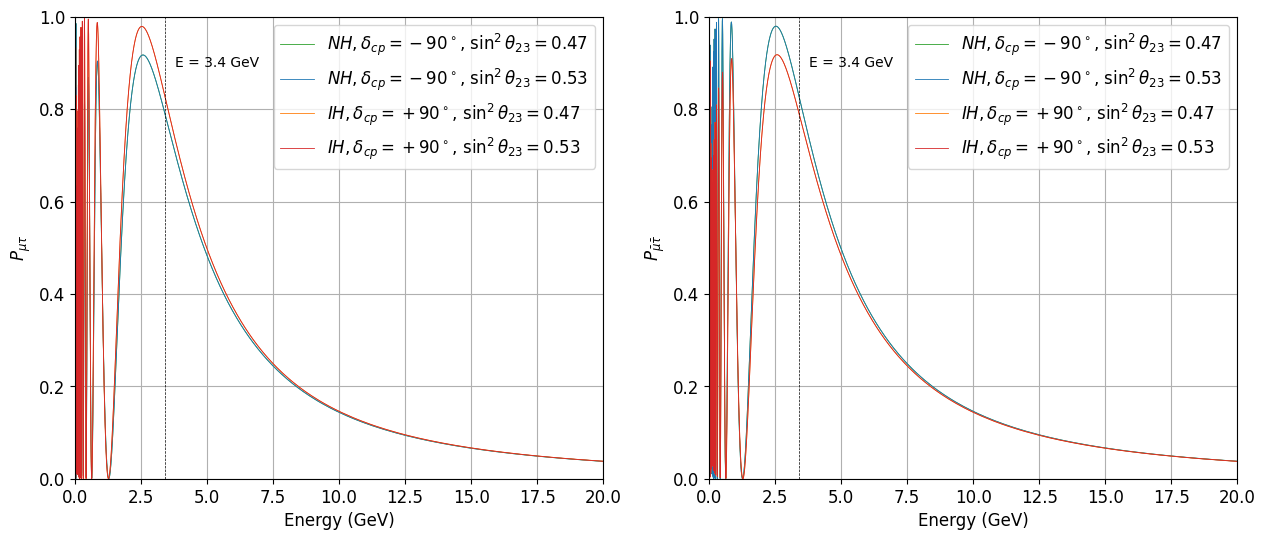}
\caption{\footnotesize{$\nu_\mu \to \nu_\tau$ oscillation probability as a function of energy with different  NH-$\dcp=-90^\circ$ and IH-$\dcp=90^\circ$ and $\tz$ in LO and HO for (anti-) neutrino in the (right) left panel. The hierarchy-$\dcp$ combinations, and $\sin^2\tz$ values for different coloured lines have been mentioned on the plots. The other oscillation parameters have been fixed to the best-fit values taken from ref.~\cite{Esteban:2024eli, nufit}.}}
\label{prob-mu-tau-octant-2}
\end{figure}

From eq.~\ref{mu-tau}, we can comment that the $\nu_\mu \to \nu_\tau$ oscillation probability depends on $|\dl|$ and $\sin^2 2\theta_{\mu \tau}\equiv |U_{\mu 3}|^2 |U_{\tau 3}|^2=1/4 \sin^2 2\tz\cos^4 \ty$. Therefore, at the DUNE baseline length, for energies higher than the $\tau$ threshold energy, the $\nu_\tau$ appearance channel does not have hierarchy, $\dcp$ or octant sensitivity.  This point has been illustrated in figs.~\ref{prob-mu-tau-hierarchy}-\ref{prob-mu-tau-octant-2}. From fig.~\ref{prob-mu-tau-hierarchy}, we can see that $\pmt$ is maximum (minimum) for the IH-$\dcp=90^\circ$ (NH-$\dcp=-90^\circ$) hierarchy-$\dcp$ combination. For the anti-neutrino probability $\pmtbar$, the opposite holds. Other hierarchy-$\dcp$ combinations fall in between these two extreme cases. However, we can see that above the $\tau$ threshold energy of 3.4 GeV, it is difficult to make any significant differentiation between $\pmt$ ($\pmtbar$) for the different hierarchy and $\dcp$ combinations. 

In fig.~\ref{prob-mu-tau-octant-1}, we show the oscillation probabilities $\pmt$ and $\pmtbar$ for NH-$\dcp=+90^\circ$ and IH-$\dcp=-90^\circ$. We have used $\sin^2\tx=0.47$ ($0.53$) for $\tz$ in LO (HO). At the oscillation probability level, the $\mu-\tau$ oscillation channels do not have any octant sensitivity above the $\tau$ threshold energy. Similar conclusions can be drawn from fig.~\ref{prob-mu-tau-octant-2} for NH-$\dcp=-90^\circ$ and IH-$\dcp=90^\circ$.

Non-standard interactions can arise as a low-energy manifestation of new heavy states of a more complete model at high energy \cite{Farzan:2017xzy, Biggio:2009nt, Ohlsson:2012kf, Miranda:2015dra, Proceedings:2019qno} or it can arise due to light mediators \cite{Farzan:2015doa, Farzan:2015hkd}. NSI can modify the neutrino and antineutrino flavour conversion in matter \cite{msw1, Mikheev:1986gs, Mikheev:1986wj}. Neutral current NSI during neutrino propagation can be represented by a dimension 6 operator \cite{msw1}: 
\begin{equation}
    \mathcal{L}_{\rm NC-NSI}= -2\sqrt{2} G_F \epsilon_{\alpha \beta}^{fC}\left( \bar{\nu}_{\alpha}\gamma^\mu P_L \nu_\beta  \right)\left( \bar{f}\gamma_\mu P_C f \right)
    \label{lag-nsi},
\end{equation}
where $\alpha,\, \beta=e,\, \mu,\, \tau$ denote the neutrino flavour, $f=e,\, u,\, d$ denotes the fermions inside matter, $P$ is the projection operator with the superscript $C$ referring to the $L$ or $R$ chirality of the $ff$ current, and $\epsilon_{\alpha \beta}^{fC}$ denotes the strength of the NSI. From the hermiticity of the interaction,
\begin{equation}
    \epsilon_{\beta \alpha}^{fC}=\left( \epsilon_{\alpha \beta}^{fC} \right)^* \,.
\end{equation}
For neutrino propagation through earth matter, the relevant expression is
\begin{equation}
    \epsilon_{\alpha \beta} \equiv \sum_{f=e,u,d} \epsilon_{\alpha \beta}^{f}\frac{N_f}{N_e} \equiv \sum_{f=e,u,d} \left(\epsilon_{\alpha \beta}^{fL}+\epsilon_{\alpha \beta}^{fR}\right)\frac{N_f}{N_e},
\end{equation}
where $N_f$ is the density of $f$ fermion. If we consider earth matter to be neutral and isoscalar, then $N_n \simeq N_p=N_e$. Thus, 
\begin{equation}
    \epsilon_{\alpha \beta}\simeq \epsilon_{\alpha \beta}^{e}+3 \epsilon_{\alpha \beta}^{u}+ 3 \epsilon_{\alpha \beta}^{d} \,.
\end{equation}
The effective Hamiltonian for neutrino propagation in matter in presence of NSI can be written in the flavour basis as 
\begin{equation}
    H=H_{\rm vac}+H_{\rm mat}+H_{\rm NSI},
    \label{Ham-NSI}
\end{equation}
where 
\begin{equation}
    H_{\rm vac}=\frac{1}{2E}U\left[
\begin{array}{ccc}
m_{1}^{2} & 0 & 0\\
0 & m_{2}^{2} & 0\\
0 & 0 & m_{3}^{2}\\
\end{array}
\right]U^\dagger; H_{\rm mat}=\sqrt{2}G_FN_e \left[
\begin{array}{ccc}
1 & 0 & 0\\
0 & 0 & 0\\
0 & 0 & 0\\
\end{array}
\right];
\end{equation}
\begin{equation}
    H_{\rm NSI}=\sqrt{2}G_FN_e \left[
\begin{array}{ccc}
\epsilon_{ee} & \epsilon_{e\mu} & \epsilon_{e\tau}\\
\epsilon_{e\mu}^{*} & \epsilon_{\mu \mu} & \epsilon_{\mu \tau}\\
\epsilon_{e\tau}^{*} & \epsilon_{\mu\tau}^{*} & \epsilon_{\tau \tau}\\
\end{array}
\right].
\label{Ham-NSI3}
\end{equation}
Calculating the oscillation probabilities with NSI in a 3-flavour oscillation scheme is a no-trivial and cumbersome job. Different perturbative approaches have been taken to calculate the oscillation probabilities with NSI in references \cite{Kikuchi:2008vq, Martinez-Soler:2018lcy, Chaves:2018sih}. Following ref. \cite{Kikuchi:2008vq}, we can write $\pmt$ with NSI up to second order pertubative term as
\begin{equation}
    \pmt^{\rm NSI}= \pmt^{2 {\rm vac}}+\pmt^{\epsilon_{e\mu},\epsilon_{e\tau}}+\pmt^{\epsilon_{\mu\mu},\epsilon_{\mu\tau},\epsilon_{\tau \tau}}
    \label{mu-tau-nsi}
\end{equation},
where the leading order term is the $\nu_\mu \to \nu_\tau$ oscillation probability in vacuum with a 2-flavour scheme, and given by
\begin{equation}
    \pmt^{2 {\rm vac}}= 4 \cos^2 \tz \sin^2 \tz \sin^2 \frac{\dl L}{4E}.
\end{equation}
As expected, the leading order term does not have dependency on hierarchy, octant or $\dcp$. After the first term, the second term $\pmt^{\epsilon_{e\mu},\epsilon_{e\tau}}$, which is a function of $\epsilon_{e\mu}$ and $\epsilon_{e\tau}$, the two NSI terms responsible for resolving the tension between \nova and T2K. The third term $\pmt^{\epsilon_{\mu\mu},\epsilon_{\mu\tau},\epsilon_{\tau \tau}}$ is a function of $\epsilon_{\mu\mu}$, $\epsilon_{\mu\tau}$, and $\epsilon_{\tau \tau}$. From equations B6 and B8 of ref.~\cite{Kikuchi:2008vq}, we can see that the maximum contribution comes from $\epsilon_{\mu \tau}$, and in that case the oscillation probability becomes sensitive to hierarchy. It is because in the oscillation probability, effect of $\epsilon_{\mu \tau}$ arises in the form $\sin^22\tz|\epsilon_{\mu\tau}|\cos\phi_{\mu\tau}\sin(\dl L/2E)$. For $|\epsilon_{\mu\tau}|=0.2$, and $\phi_{\mu\tau}=0$, we have a hierarchy sensitive term numerically proportional to $|\epsilon_{\mu\tau}|$, making its contribution $\sim10$ times larger, compared to the subleading term in eq.~\ref{mu-tau}. However, there is no octant sensitivity for the $\epsilon_{\mu \tau}$ terms. In case of other NSI parameters, $\epsilon_{\mu \mu}$ and $\epsilon_{\tau \tau}$ terms are proportional to $\cos^22\tz$ making their contribution negligible, whereas $\epsilon_{e\mu}$ and $\epsilon_{e\tau}$ terms always come in second order: $|\epsilon_{e\mu}|^2$ or $|\epsilon_{e\tau}|^2$-- making their contribution negligible compared to the leading order term. 

The $|\epsilon_{\mu\tau}|$ value of $0.2$ is larger than the $90\%$ limit on $|\epsilon_{\mu\tau}|$ from the IceCube data \cite{IceCubeCollaboration:2021euf}. However, throughout our analysis we have chosen the true values of all the NSI parameters to be $0.2$ to compare them with the best-fit values of $|\epsilon_{e\mu}|$ and $|\epsilon_{e\tau}|$ measured by the combined analysis of \nova and T2K. Moreover, the latest global fit results from \cite{Coloma:2019mbs} allow NSI parameter absolute values $\sim 10^{-1}$ at a $3\,\sigma$ limit. 

\begin{figure}[htbp]
\centering
\includegraphics[width=0.6\textwidth] {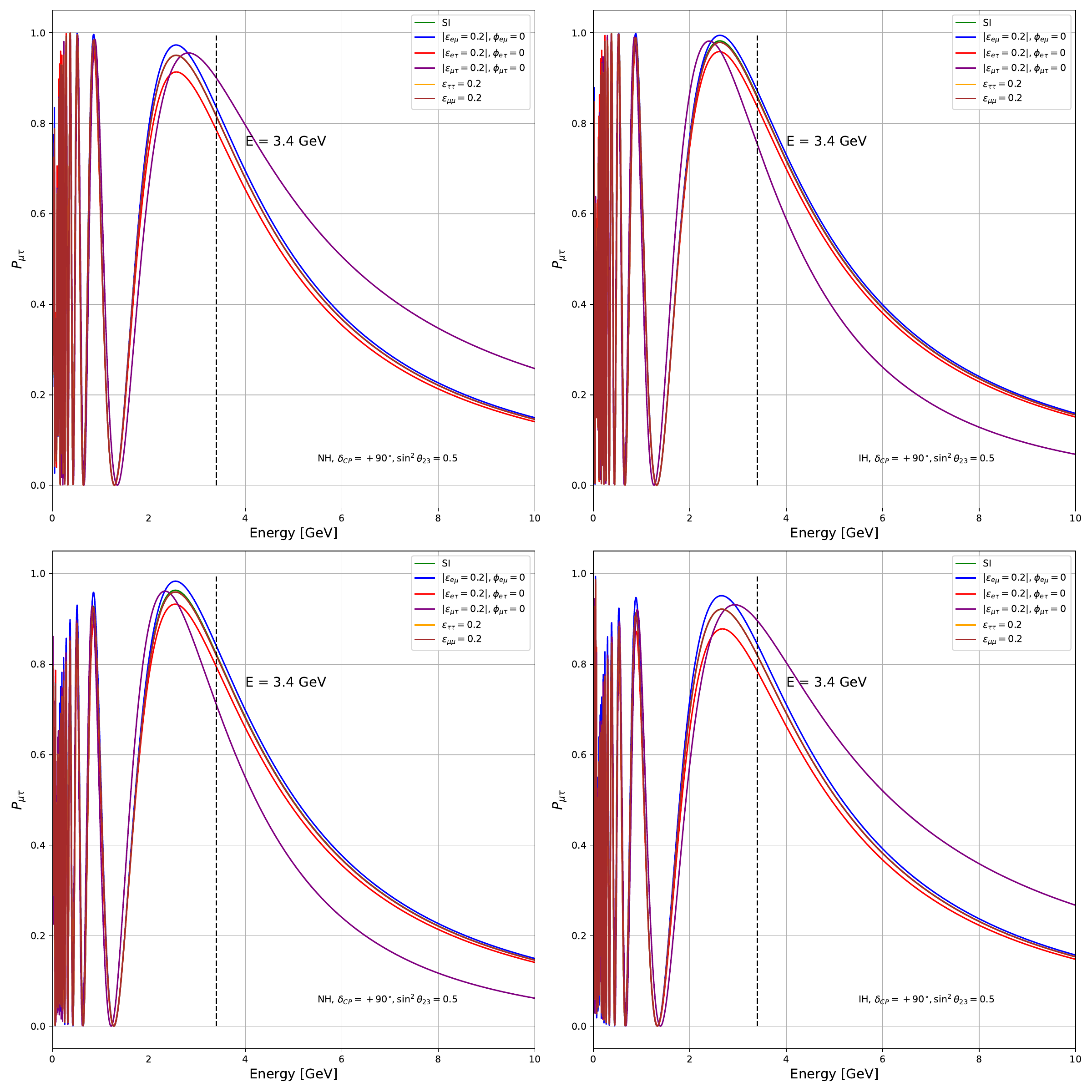}
\includegraphics[width=0.6\textwidth] {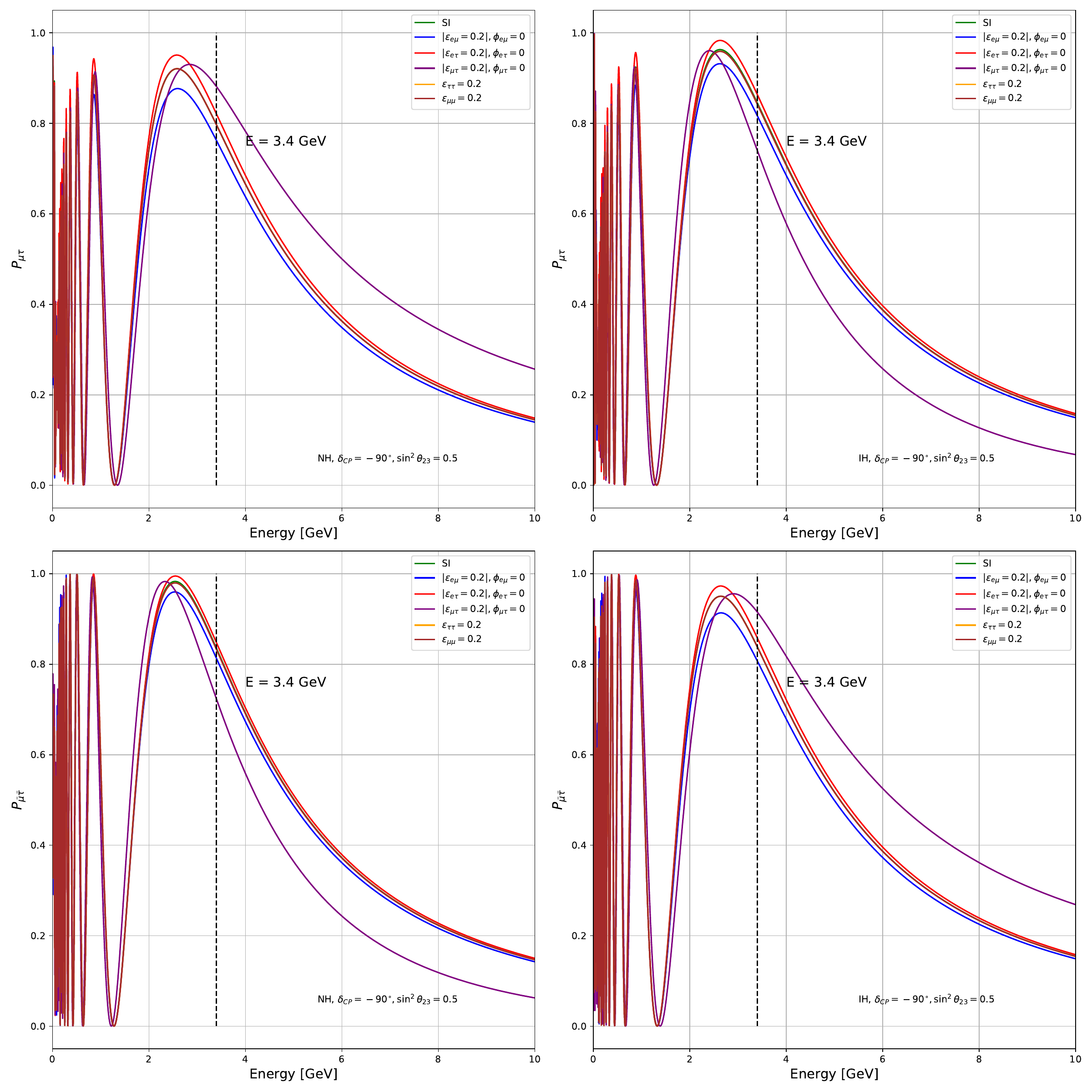}
\caption{\footnotesize{$\nu_\mu \to \nu_\tau$ oscillation probability for different NSI parameters as a function of energy with different  NH and IH with $\dcp=90^\circ$ and $-90^\circ$, and $\sin^2\tz=0.5$. The $\dcp$ and $\sin^2\tz$ values have been mentioned in the plots. The NSI parameter values for different coloured lines have been mentioned on the labels on the plots. While considering one NSI parameters, other NSI parameters have been fixed to 0. Other oscillation parameter values have been taken from ref.~\cite{Esteban:2024eli, nufit}.}}
\label{nsi-diff}
\end{figure}

\begin{figure}[htbp]
\centering
\includegraphics[width=0.55\textwidth] {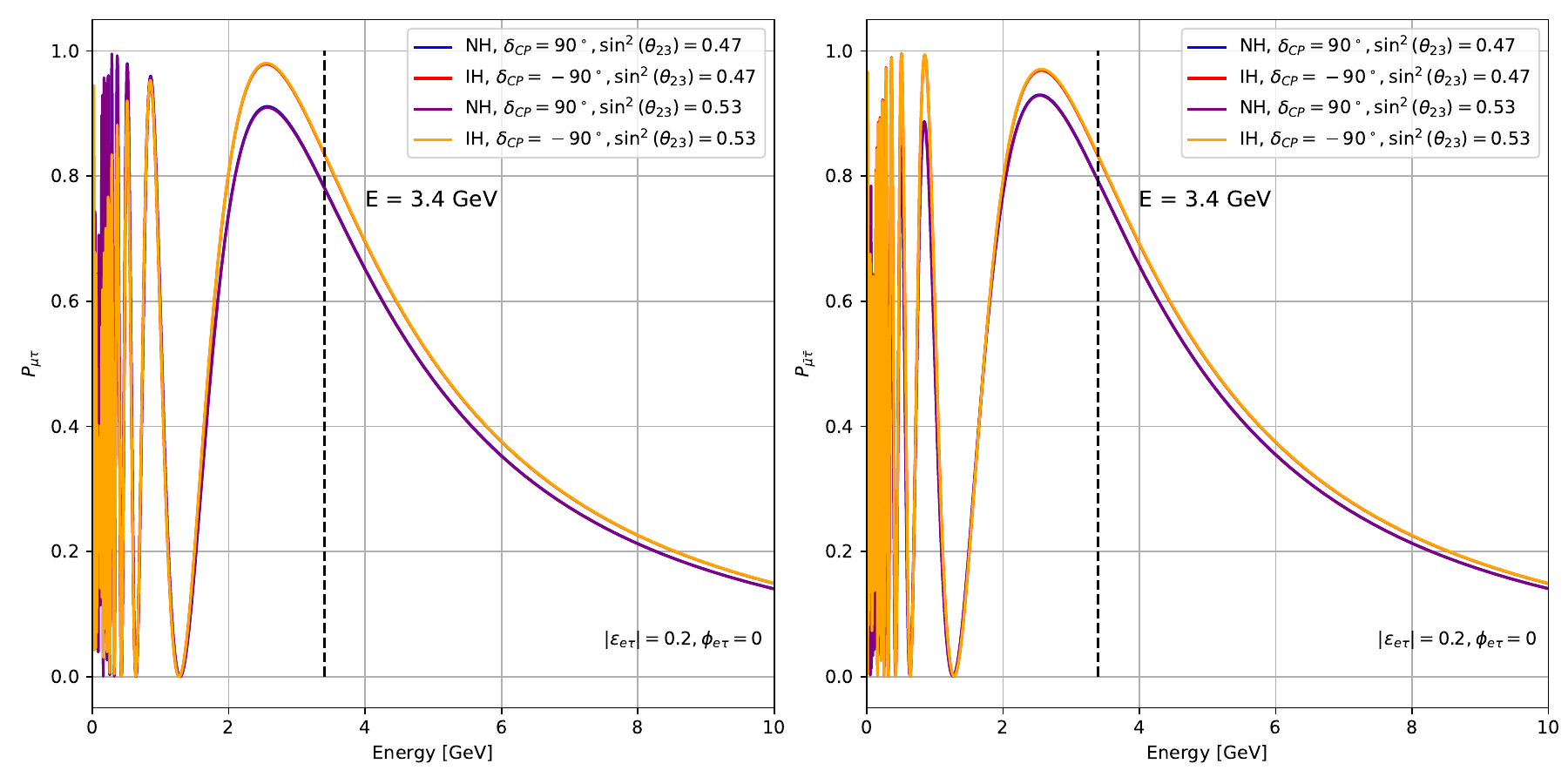}
\includegraphics[width=0.55\textwidth] {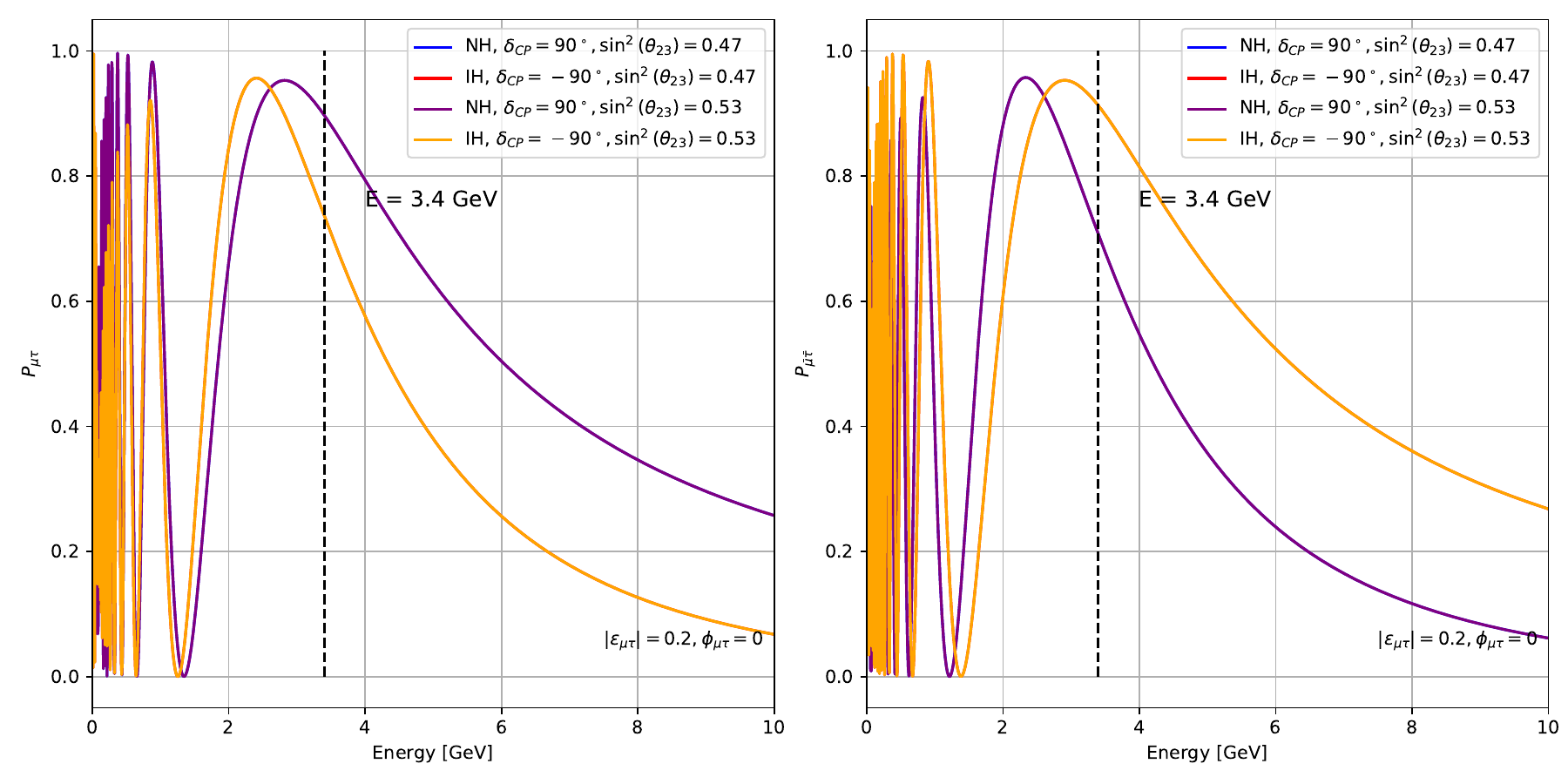}
\includegraphics[width=0.55\textwidth]{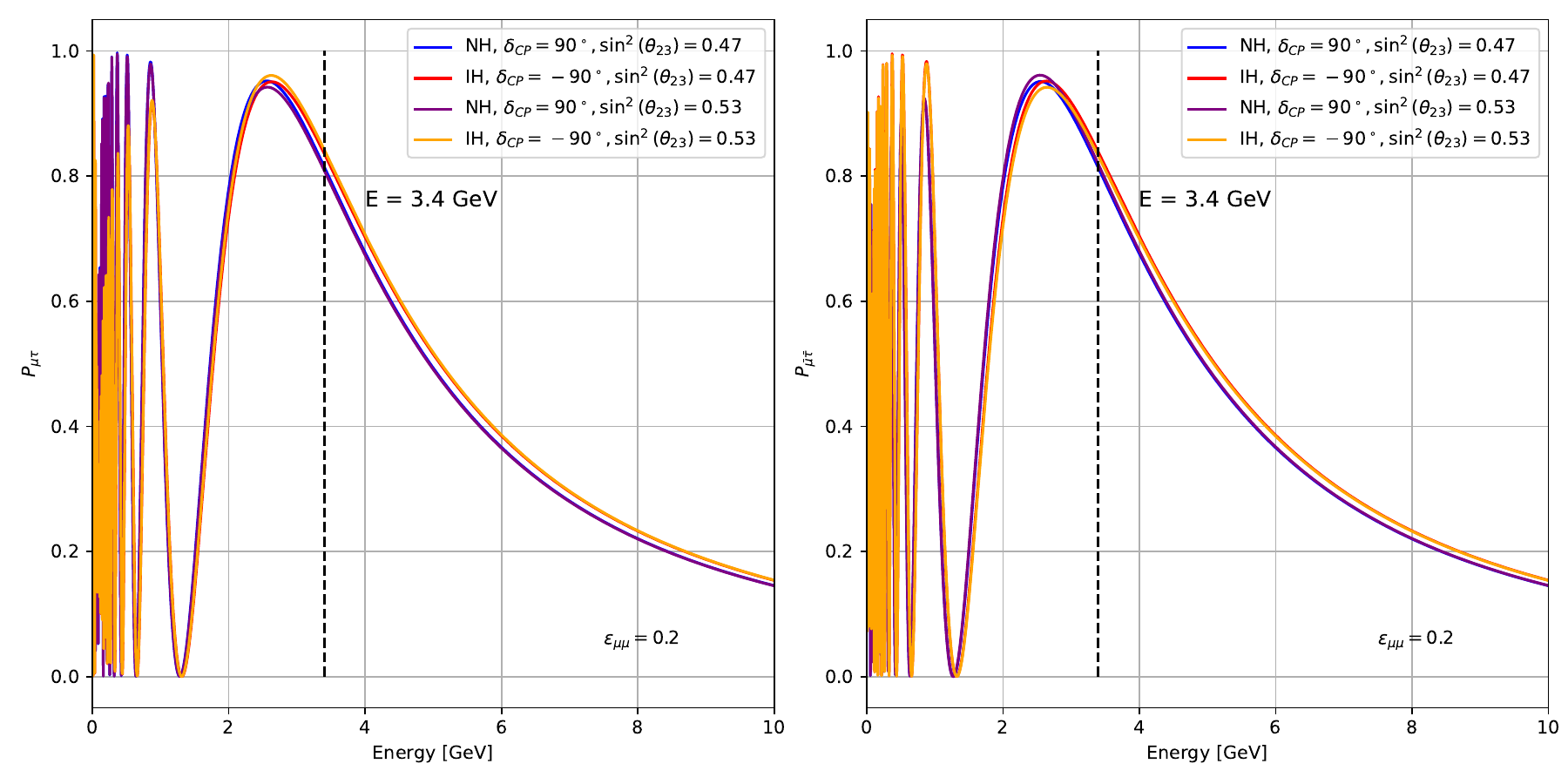}
\includegraphics[width=0.55\textwidth]{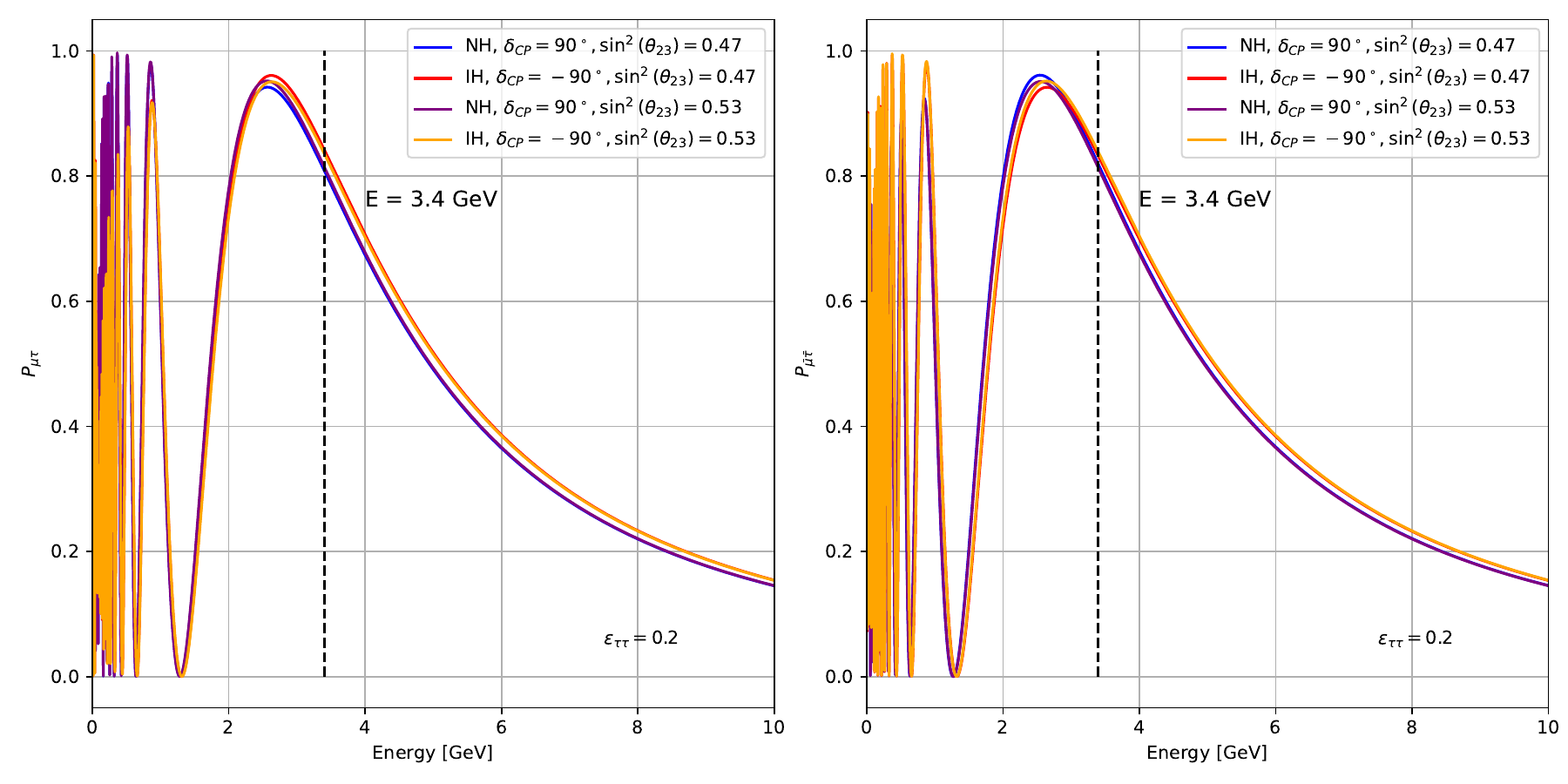}
\includegraphics[width=0.55\textwidth]{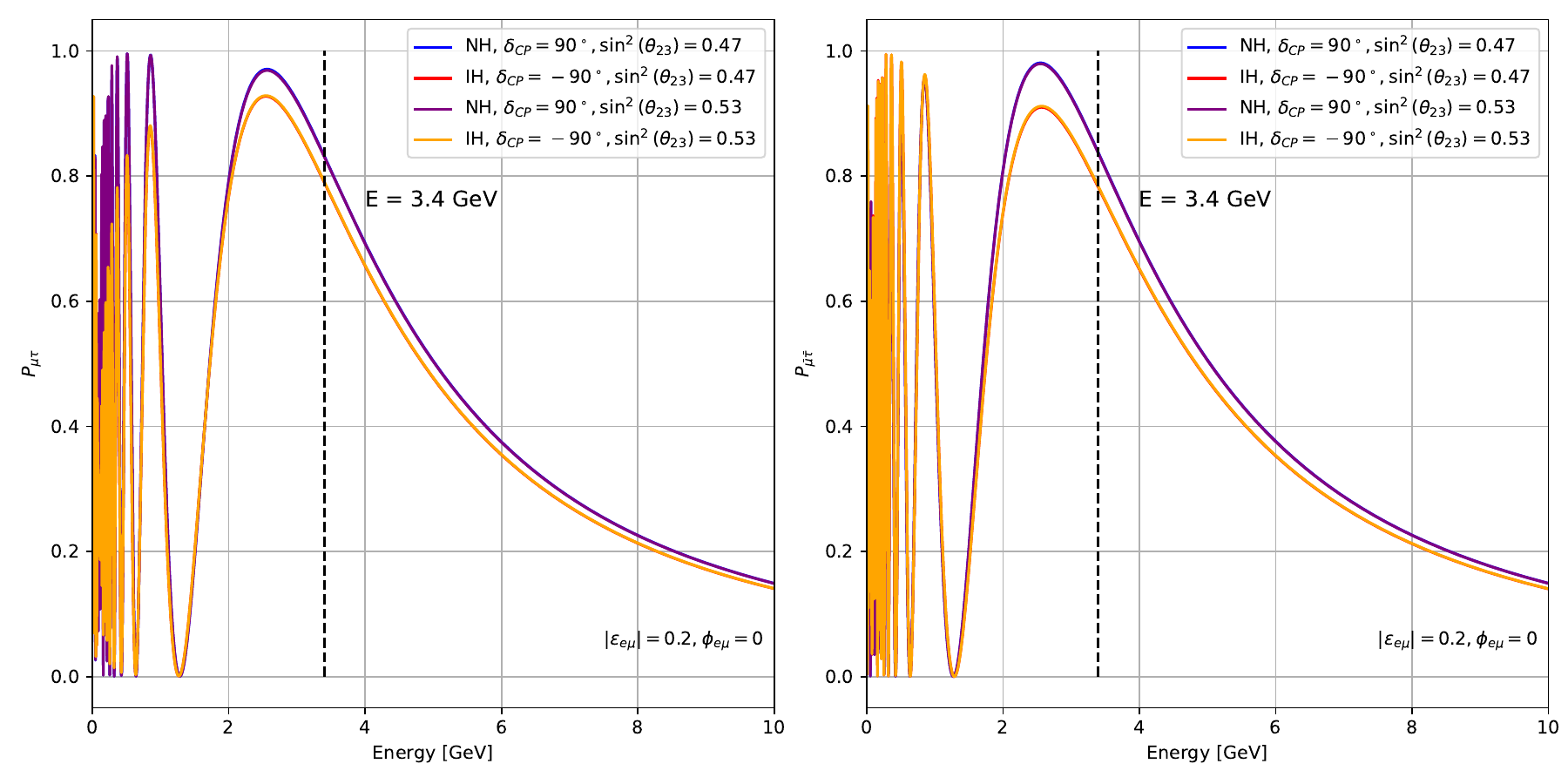}
\caption{\footnotesize{$\nu_\mu \to \nu_\tau$ oscillation probability for different NSI parameters as a function of energy with different  hierarchies, $\dcp$ values and octant of $\tz$. The NSI parameter values have been mentioned in the plots. The hierarchy, $\dcp$ and $\sin^2\tz$ values for different coloured lines have been mentioned on the labels on the plots. While considering one NSI parameters, other NSI parameters have been fixed to 0. Other oscillation parameter values have been taken from ref.~\cite{Esteban:2024eli, nufit}.}}
\label{nsi-hie-oct}
\end{figure}

These characteristics, discussed above, have been depicted in figs.~\ref{nsi-diff} and \ref{nsi-hie-oct}. In fig.~\ref{nsi-diff}, the first and third panels from the top represent $\pmt$ as a function of energy, whereas the second and fourth panels from the top represent $\pmtbar$ as a function of energy. We can see from this figure that the change in the $\pmt$ and $\pmtbar$ due to NSI is negligible for all NSI parameters, except $\epsilon_{\mu\tau}$. From fig.~\ref{nsi-hie-oct}, we conclude that when the NSI effect is due to $\epsilon_{\mu\tau}$, it is possible to discriminate between different hierarchies at the oscillation probability level. However, it remains true that even for this case, there is no octant or $\dcp$ sensitivity. The differences between probabilities due to the change in hierarchy is very small for $\epsilon_{e\mu}$ and $\epsilon_{e\tau}$, and is even smaller for NSI due to $\epsilon_{\mu\mu}$ and $\epsilon_{\tau \tau}$. There are no differences in the probabilities due to the change in the octant of $\tz$ or $\dcp$ for these parameters.

\section{Simulation details}
\label{analysis}
The future long baseline accelerator neutrino oscillation experiment DUNE \cite{Abe:2017vif} consists of a baseline of 1300 km. It has been designed to disentangle the changes due to CP violation from the changes due to the matter effect. Its baseline and correspondingly its energy are much longer compared to \nova and T2K and hence the matter effect is much larger. Therefore, it is expected to measure the unknown quantities with much better precision than the ongoing accelerator neutrino long baseline experiments. For our analysis, we have used the energy resolution, energy-dependent detector efficiencies for the signal and background, and systematic uncertainties in the case of $\nu_\mu$ and $\bar{\nu}_\mu$ disappearance, and $\nu_e$ and $\bar{\nu}_e$ appearance from ref.\cite{DUNE:2021cuw}. 

For the $\nu_\tau$ and $\bar{\nu}_\tau$ appearance simulations, we have followed ref.~\cite{DeGouvea:2019kea}. For the energy resolution function, we have used a bin based energy smearing function
\begin{equation}
R^c (E,E^\prime)=\frac{1}{\sqrt{2\pi}}e^{-\frac{(E-E^\prime)^2}{2\sigma^2(E)}},
\end{equation}
where $E^\prime$ is the reconstructed energy. The energy resolution function is given by 
\begin{equation}
\sigma(E)=\alpha E+\beta \sqrt{E}+\gamma.
\end{equation}
As suggested in ref.~\cite{DeGouvea:2019kea}, we have used $\alpha=0.25$, and $\beta=\gamma=0$. We have used the regular DUNE fluxes from ref.~\cite{DUNE:2021cuw}, and tau-optimized high energy flux in order to obtain a considerable number of events at energies above the$\tau$-threshold energy, from ref.~\cite{DUNE-tau-opt-flux}. 

For $\nu_\tau$ ($\bar{\nu}_\tau$) appearance, the signal comes from $\nu_\mu \to \nu_\tau$ ($\bar{\nu}_\mu \to \bar{\nu}_\tau$)oscillation from the $\nu_\mu$ ($\bar{\nu}_\mu$)beam, and the background arises from $\bar{\nu}_\mu \to \bar{\nu}_\tau$ ($\nu_\mu \to \nu_\tau$) oscillation from the $\bar{\nu}_\mu$ ($\nu_\mu$) impurities present in the neutrino (anti-neutrino) beam as well as the intrinsic $\bar{\nu}_\tau$ ($\nu_\tau$) present in the beam, and the neutral current backgrounds. The energy dependent detector efficiencies have been calculated by matching the event numbers with the event simulations provided in ref.~\cite{DeGouvea:2019kea}. In ref.~\cite{DeGouvea:2019kea} simulations for $\bar{\nu}_\tau$ appearance, for the high energy beam, were not provided. In our analysis, we have included $\bar{\nu}_\tau$ appearance for these high energy fluxes as well. In this case, the energy dependent detector efficiencies have been calculated assuming that the ratio of signals between neutrino and anti-neutrino beams, and that of background between the neutrino and anti-neutrino beams remain the same for regular DUNE fluxes as and high energy fluxes. 

For the i-th energy bin, by comparing the \textbf{true} experimental event rates, $N_{i}^{\rm true}$, and the \textbf{test} theoretical event rates, $N_{i}^{\rm test}$, the $\chi^2$ has been calculated using the GLoBES software \cite{Huber:2004ka, Huber:2007ji}. The cross-section files for tau neutrino and anti-neutrino have been taken from the \nova experiment cross-section files provided in the GLoBES website as well. The Poisionian $\chi^2$ between the true and test events is defined as
\begin{eqnarray}
\chi^2 &=& 2\sum_i \left\{
(1+z) N_i^{\rm th} - N_i^{\rm exp} + N_i^{\rm exp} 
\ln\left[ \frac{N_i^{\rm exp}}{(1+z) N_i^{\rm th}} \right]
\right\} + 2 \sum_j (1+z) N_j^{\rm th} + z^2 \,\, \nonumber \\
\label{poisionian}
\end{eqnarray}
where the index $i$ represents the bins for which $N_i^{\rm exp}\neq 0$, and $j$ represents the bins for which $N_j^{\rm exp} = 0$. The parameter $z$ defines the additional systematic uncertainties. Similar systematic uncertainties as in ref.~\cite{DeGouvea:2019kea} have been used in this analysis. The minimum of these $\chi^2$ values are subtracted from them to calculate $\dchsq$. For simulations, the minimum $\chi^2$ is $0$ when the true parameter values and test parameter values are equal, and hence $\chi^2$ and $\dchsq$ are the same for simulations. In this paper, we have investigated the hierarchy sensitivity and octant sensitivity at DUNE in the presence of NSI, determination of NSI phases, and the future sensitivity to NSI parameters without and with tau-neutrino determination. The details regarding the true parameter values considered, test parameter values, and priors used to accommodate results from previous experiments are discussed in the next chapter for each of the results presented. In our analysis, we have considered five different beam running schemes, which include:
\begin{enumerate}
    \item $5+5(\mu+e)$: $\nu_\mu+\bar{\nu}_\mu$ disappearance and $\nu_e+\bar{\nu}_e$ appearance after 5 years each of neutrino and anti-neutrino runs with regular DUNE fluxes.
    \item $5+5(\mu+\tau)$: $\nu_\mu+\bar{\nu}_\mu$ disappearance and $\nu_\tau+\bar{\nu}_\tau$ appearance after 5 years each of neutrino and anti-neutrino runs with regular DUNE fluxes.
    \item $5+5(\mu+e+\tau)$: $\nu_\mu+\bar{\nu}_\mu$ 
 disappearance, $\nu_e+\bar{\nu}_e$ appearance and $\nu_\tau+\bar{\nu}_\tau$ appearance after 5 years each of neutrino and anti-neutrino runs with regular DUNE fluxes.
    \item $5+5+1+1(\mu+\tau)$: $\nu_\mu+\bar{\nu}_\mu$ disappearance and $\nu_\tau+\bar{\nu}_\tau$ appearance after 5 years each of neutrino and anti-neutrino runs with regular DUNE fluxes, along with $\nu_\mu+\bar{\nu}_\mu$ disappearance and $\nu_\tau+\bar{\nu}_\tau$ appearance from 1 additional year each of neutrino and anti-neutrino running with the high energy fluxes.
    \item $5+5+1+1(\mu+e+\tau)$: $\nu_\mu+\bar{\nu}_\mu$ disappearance, $\nu_e+\bar{\nu}_e$ appearance and $\nu_\tau+\bar{\nu}_\tau$ appearance after 5 years each of neutrino and anti-neutrino runs with regular DUNE fluxes, along with $\nu_\mu+\bar{\nu}_\mu$ disappearance and $\nu_\tau+\bar{\nu}_\tau$ appearance from 1 additional year each of neutrino and anti-neutrino running with the high energy fluxes.
\end{enumerate}

For all of the cases considered, the number of protons on target is $1.1\times 10^{21}$ per year for both the regular DUNE flux and high energy $\tau$ optimized flux.
\section{Results and discussions}
\label{results}
\begin{figure}[htbp]
\centering
\includegraphics[width=0.8\textwidth] {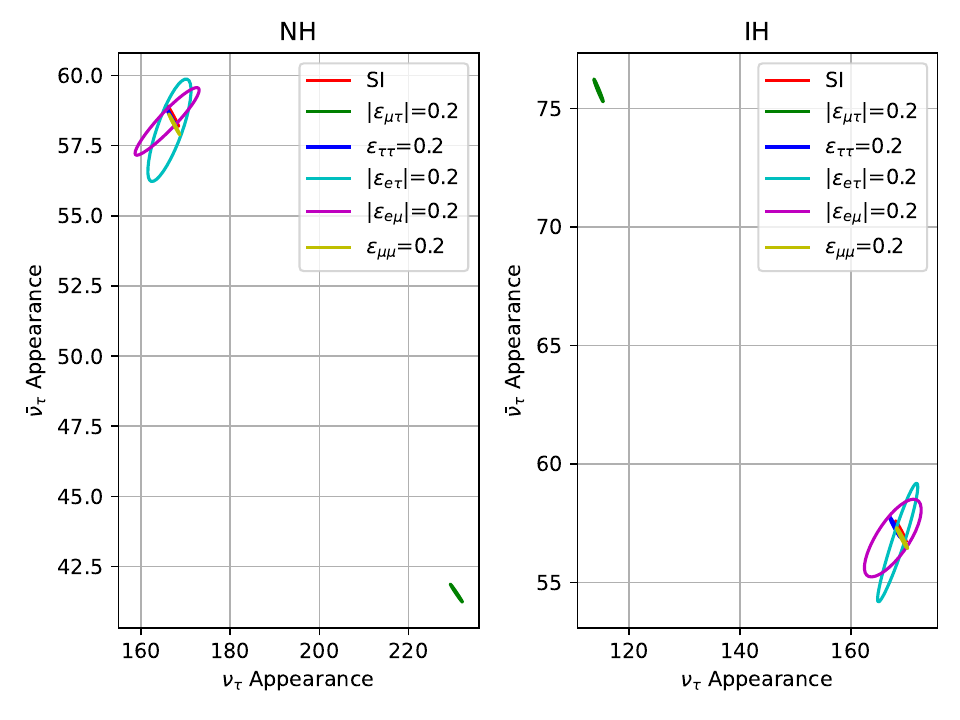}
\includegraphics[width=0.8\textwidth] {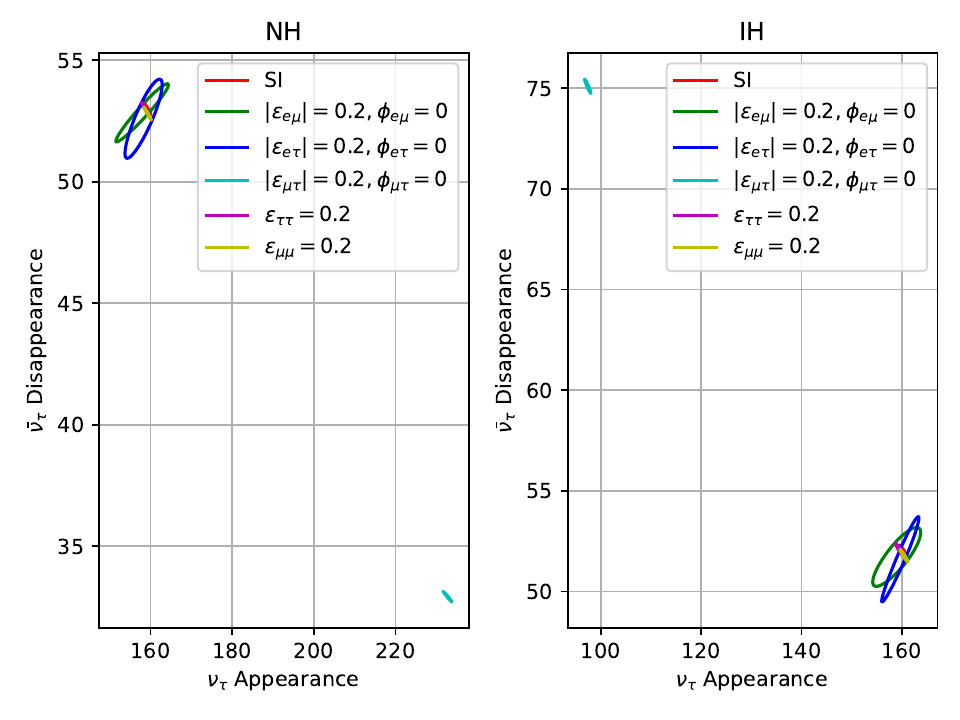}
\caption{\footnotesize{$\bar{\nu_\tau}$ vs $\nu_\tau$ event numbers after varying $\dcp$ in the range $[-180^\circ:180^\circ]$ for NH (IH) in the left (right) panel after 5 years (1 year) each of neutrino and anti-neutrino run with regular ($\tau$ optimized high energy) DUNE fluxes in the top (bottom) panel. All the other standard oscillation parameter values have been taken from ref.~\cite{Esteban:2024eli, nufit}. Different NSI parameter values have been mentioned on the figure.}}
\label{bievents1}
\end{figure}

\begin{figure}[htbp]
\centering
\includegraphics[width=0.8\textwidth] {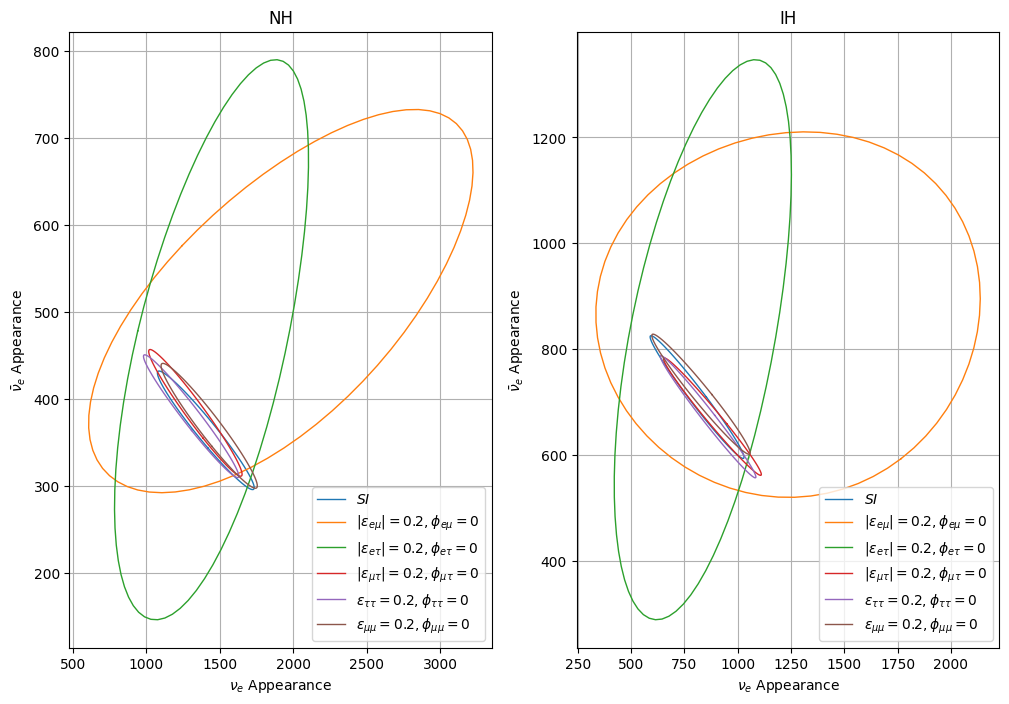}
\caption{\footnotesize{$\bar{\nu_e}$ vs $\nu_e$ event numbers after varying $\dcp$ in the range $[-180^\circ:180^\circ]$ for NH (IH) in the left (right) panel after 5 years each of neutrino and anti-neutrino run with regular DUNE fluxes. All the other standard oscillation parameter values have been taken from ref.~\cite{Esteban:2024eli, nufit}. Different NSI parameter values have been mentioned on the figure.}}
\label{bievents2}
\end{figure}

\begin{figure}[htbp]
\centering
\includegraphics[width=0.3\textwidth] {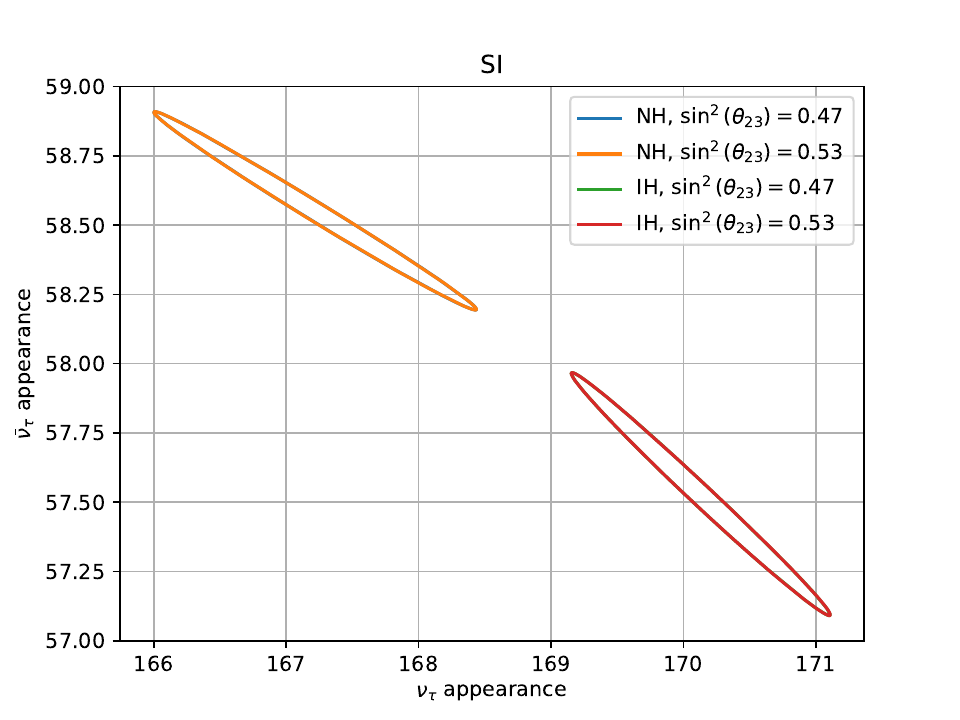}
\includegraphics[width=0.3\textwidth] {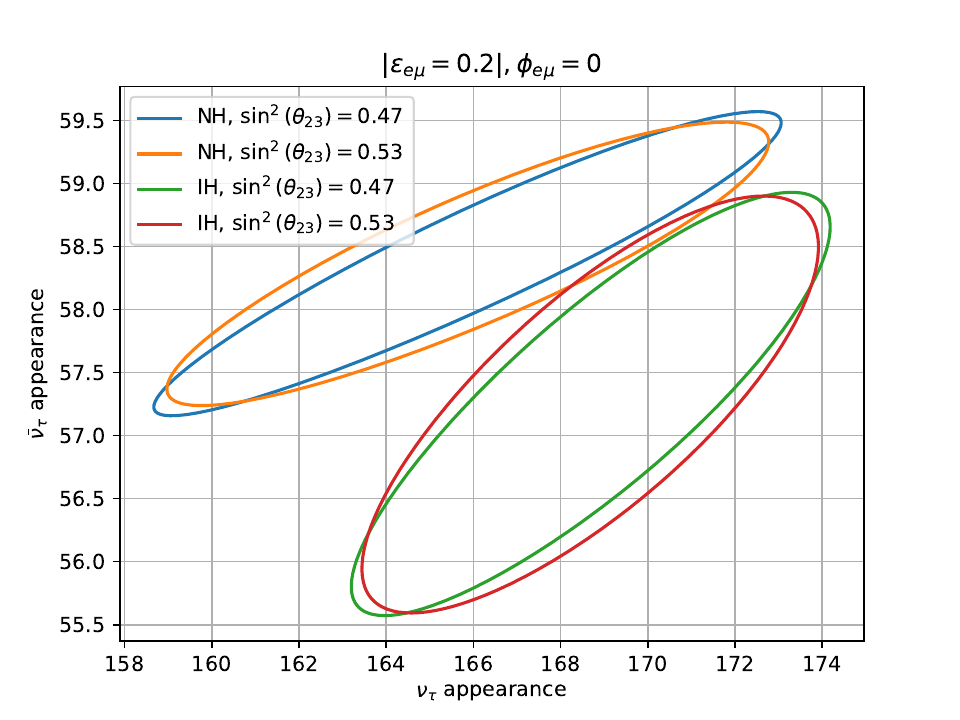}
\includegraphics[width=0.3\textwidth] {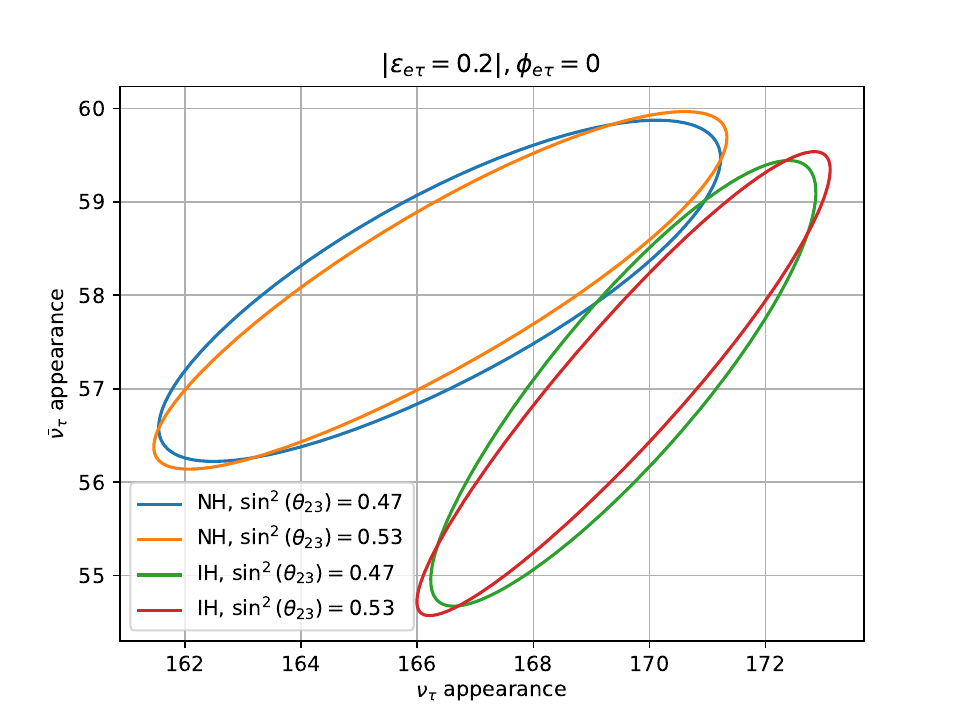}
\includegraphics[width=0.3\textwidth] {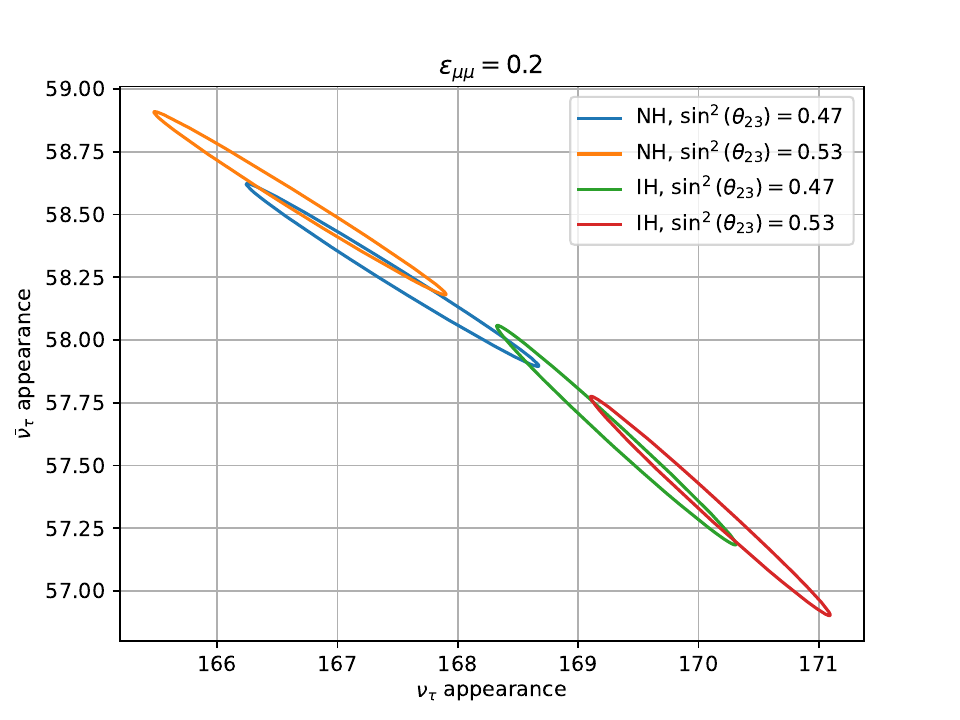}
\includegraphics[width=0.3\textwidth] {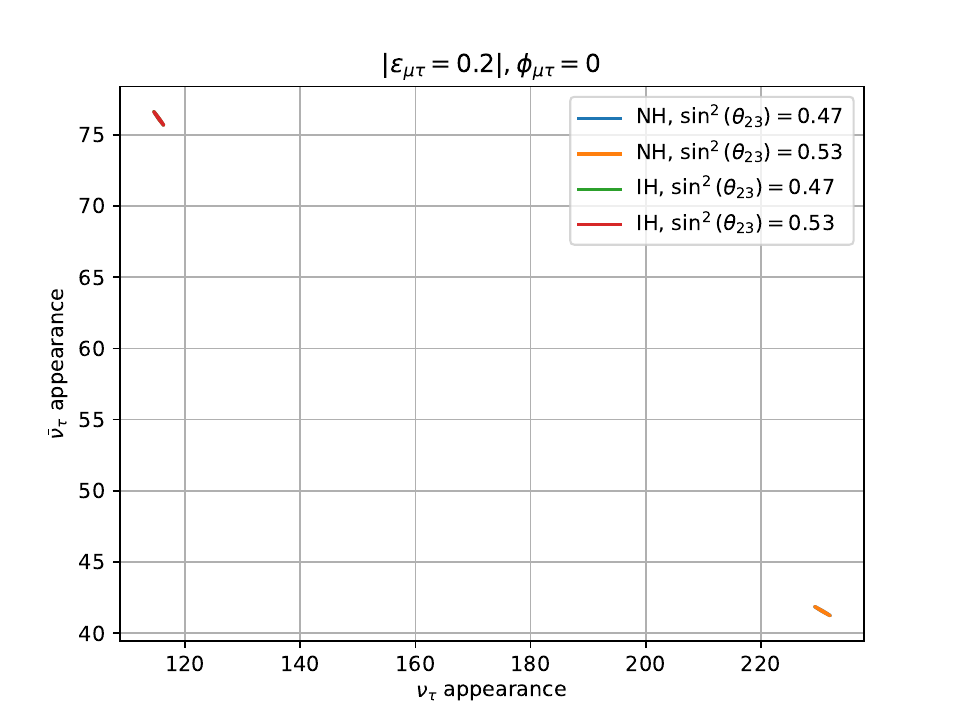}
\includegraphics[width=0.3\textwidth] {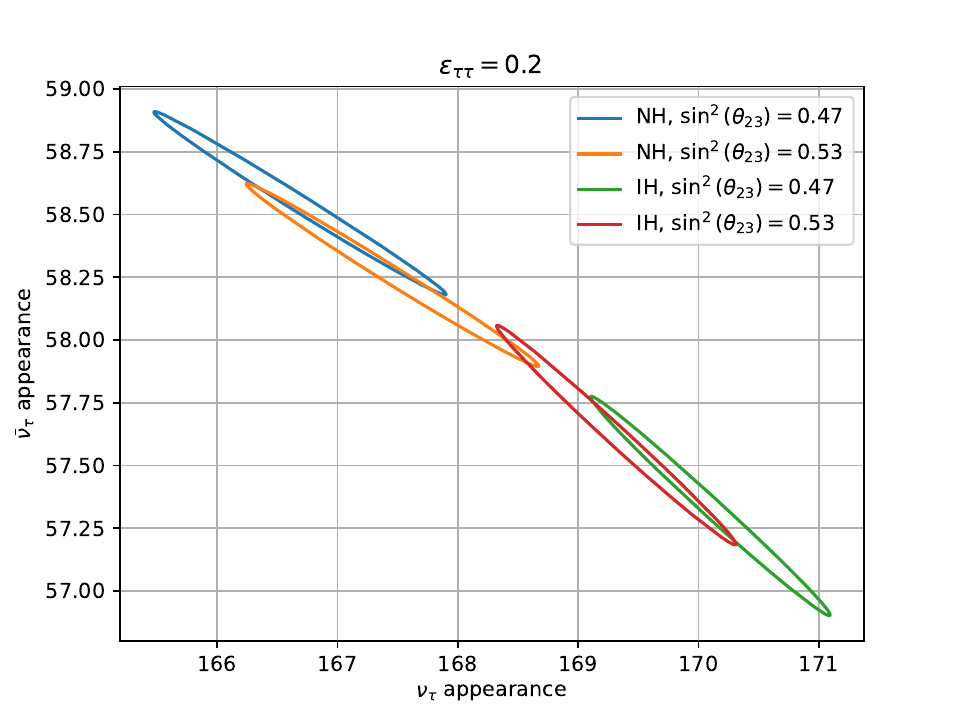}
\caption{\footnotesize{$\bar{\nu_\tau}$ vs $\nu_\tau$ event numbers after varying $\dcp$ in the range $[-180^\circ:180^\circ]$ for different NSI parameters with different hierarchy-octant combinations as mentioned in the panels, after 5 years each of neutrino and anti-neutrino run with regular DUNE fluxes. All the other standard oscillation parameter values have been taken from ref.~\cite{Esteban:2024eli, nufit}. Different NSI parameter values have been mentioned on the figure.}}
\label{bievents3}
\end{figure}

The bi-event plots for different NSI parameters are presented in figs.~\ref{bievents1}-\ref{bievents3}. In each case, by varying $\dcp$, the plot takes an elliptical shape. We have considered $\nu$ and $\bar{\nu}$ runs for 5 years each with regular DUNE fluxes as well as 1 year each with high energy DUNE fluxes. Fig.~\ref{bievents1} shows that the ellipses for different NSI parameters overlap with each other and with the ellipse due to the standard oscillation case. The only exception is when the NSI due to the $\epsilon_{\mu\tau}$ paramater. There are no overlaps between the ellipse due to $\epsilon_{\mu\tau}$ and any of the other ellipses. Hence, it can be concluded that there will be no large discrimination sensitivity to different NSI parameters, except $\epsilon_{\mu\tau}$ from the $\nu_{\tau}$ and $\bar{\nu}_\tau$ appearance channels. 

In fig.~\ref{bievents2}, we show the bi-event plots for $\nu_e$ and $\bar{\nu_e}$ appearance after neutrino and anti-neutrino runs of 5 years each with regular DUNE fluxes. It can be seen that for NSI with $\epsilon_{e\mu}$ and $\epsilon_{e\tau}$, there are very small overlaps with the standard oscillation case, leading to good sensitivity to these parameters from the $\nu_e$ and $\bar{\nu}_e$ appearance channels. However, the ellipses due to these two NSI parameters have large overlaps with each other, making it difficult to differentiate between them. The separations between different $\dcp$ values on these ellipses are large, and hence we can expect good CP violation sensitivity from the $\nu_e$ and $\bar{\nu}_e$ appearance channels for NSI due to $\epsilon_{e\mu}$ and $\epsilon_{e\tau}$. For all the NSI parameters, the ellipses overlap with the ellipse due to standard oscillation. Hence, we should not expect any sensitivity to these parameters from the $\nu_e$ and $\bar{\nu}_e$ appearance channels.
From fig.~\ref{bievents3}, we can see that for the standard oscillation scenario, after $\nu$ and $\bar{\nu}$ runs of 5 years each with regular DUNE fluxes, the ellipses due to different hierarchies are far apart from each other. The small differences between the two hierarchies at the probability level, manifests into a considerable difference in event numbers due to the large DUNE fluxes. However, for a particular hierarchy, the ellipses for different $\tz$ octants overlap on each other, leading to no octant sensitivity from the $\nu_{\tau}$ and $\bar{\nu}_\tau$ appearance channels for the standard oscillation. These characteristics remain unchanged for NSI due to different parameters as well. It is to be noted that the differences between event numbers for NH and IH is not more than $1\,\sigma$, except for $\epsilon_{\mu\tau}$. Therefore, apart from NSI due to $\epsilon_{\mu\tau}$, we should not expect more than $1\,\sigma$ hierarchy sensitivity from the $\nu_{\tau}$ and $\bar{\nu}_\tau$ appearance channels in DUNE.

\begin{figure}[htbp]
\centering
\includegraphics[width=0.6\textwidth] {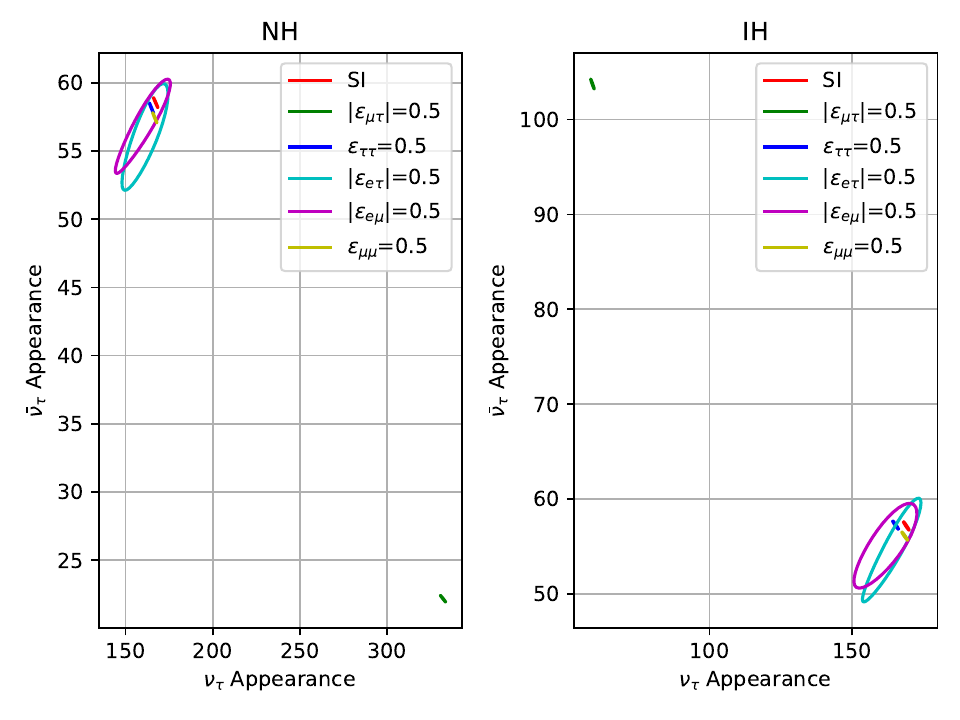}
\includegraphics[width=0.6\textwidth] {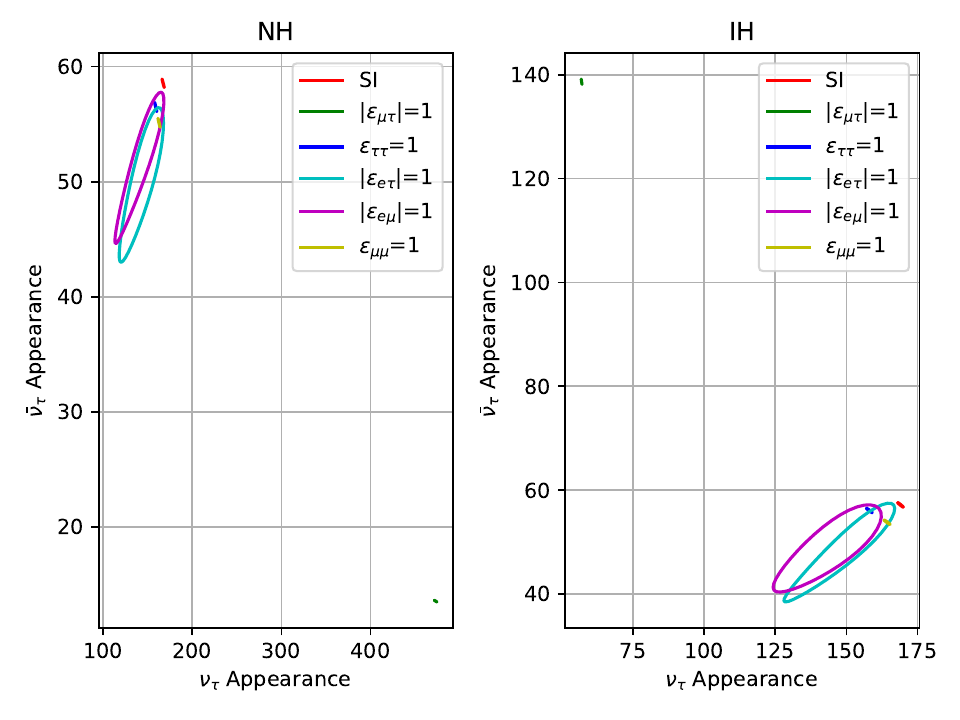}
\includegraphics[width=0.6\textwidth] {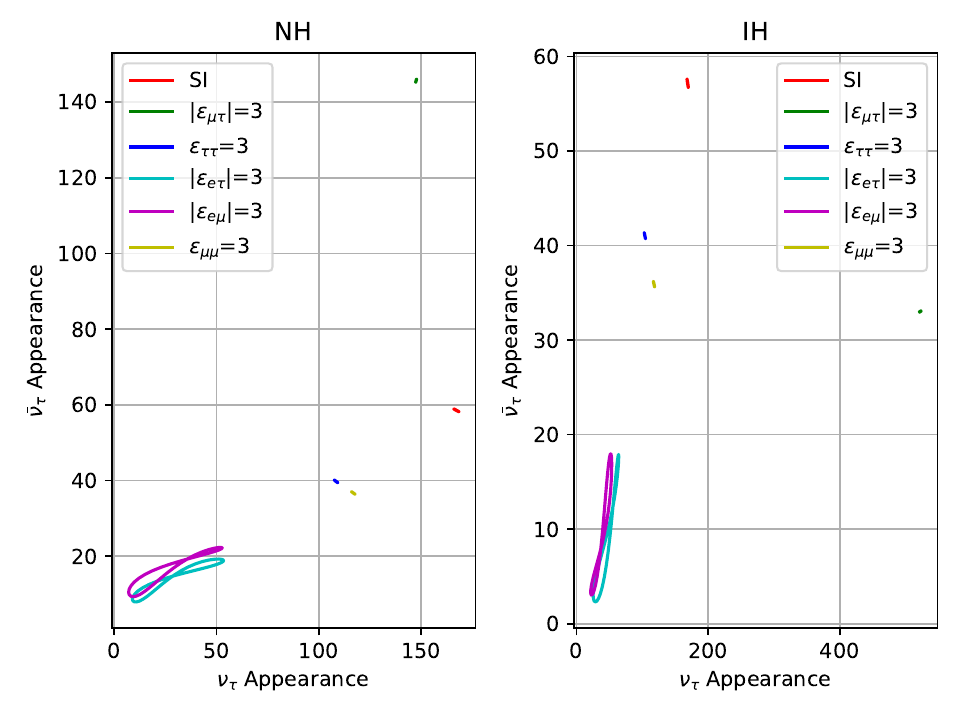}
\caption{\footnotesize{$\bar{\nu_\tau}$ vs $\nu_\tau$ event numbers after varying $\dcp$ in the range $[-180^\circ:180^\circ]$ for NH (IH) in the left (right) panel after 5 years each of neutrino and anti-neutrino run with regular DUNE fluxes, and for different values of NSI parameters. All the other standard oscillation parameter values have been taken from ref.~\cite{Esteban:2024eli, nufit}. Different NSI parameter values have been mentioned on the figure.}}
\label{bievents4}
\end{figure}

In fig.~\ref{bievents4}, we show the bi-event plots, similar to fig.~\ref{bievents1} with regular DUNE fluxes, but for higher order values of NSI parameters. It is evident that the effect of NSI parameters $\epsilon_{e\mu}$, $\epsilon_{e\tau}$, $\epsilon_{\mu\mu}$ and $\epsilon_{\tau \tau}$ on $\tau$ neutrino and anti-neutrino appearances can be separated from that of standard interaction only when the absolute values of the individual parameters are larger than 1. These large values of NSI parameters are ruled out at $3\,\sigma$ level by the present global fit \cite{Coloma:2019mbs}. Therefore, it is evident that the visible effects from NSI parameters $\epsilon_{e\mu}$, $\epsilon_{e\tau}$, $\epsilon_{\mu\mu}$ and $\epsilon_{\tau \tau}$ on $\nu_\tau$ and $\bar{\nu}_\tau$ appearances can only occur from unphysical values of these NSI parameters.
\subsection{Mass hierarchy sensitivity}
\label{hierarchy}
\begin{figure}[htbp]
\centering
\includegraphics[width=0.8\textwidth] {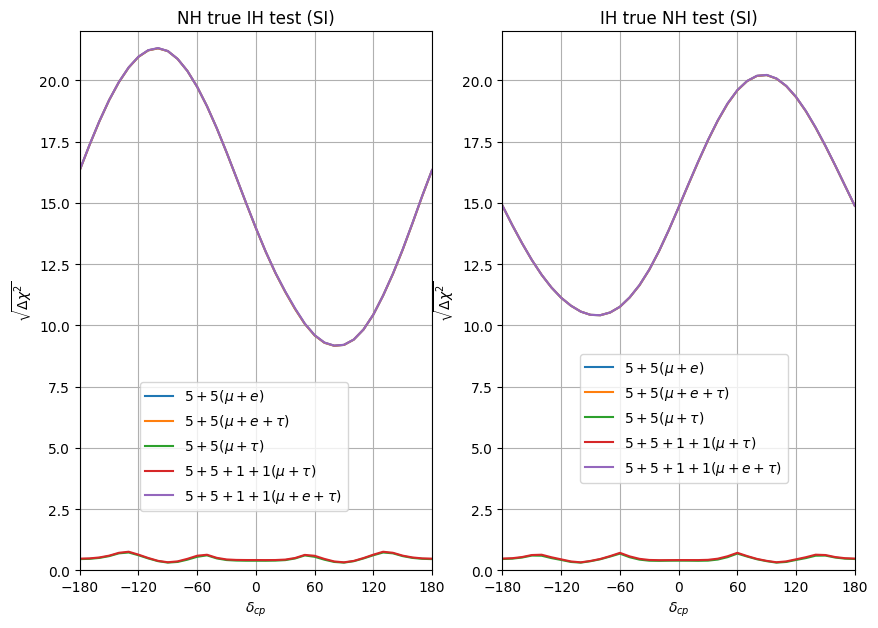}
\caption{\footnotesize{Hierarchy sensitivity of DUNE without the presence of any NSI and for different running schemes. The left (right) panel is for the true hierarchy being NH (IH).}}
\label{hierarchy-SM}
\end{figure}

\begin{figure}[htbp]
\centering
\includegraphics[width=0.45\textwidth] {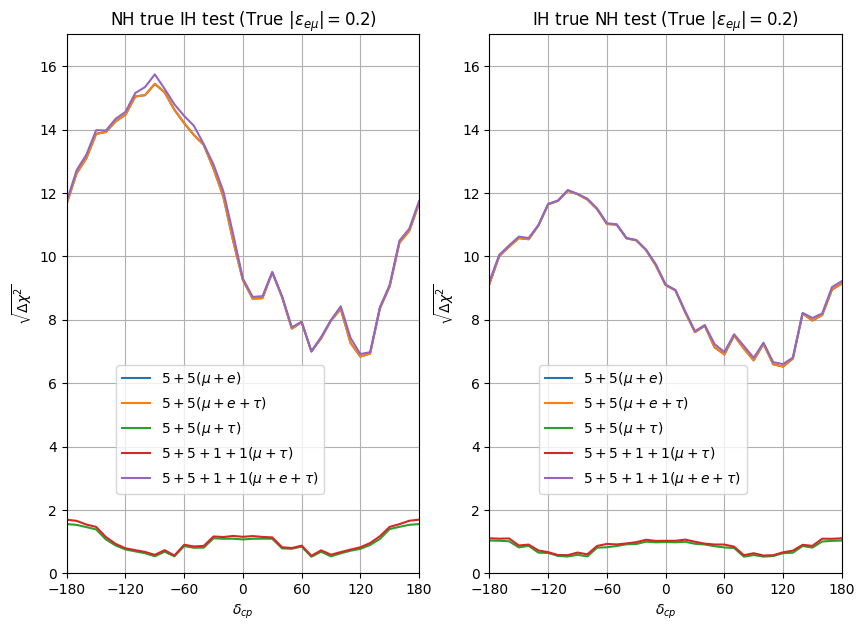}
\includegraphics[width=0.45\textwidth] {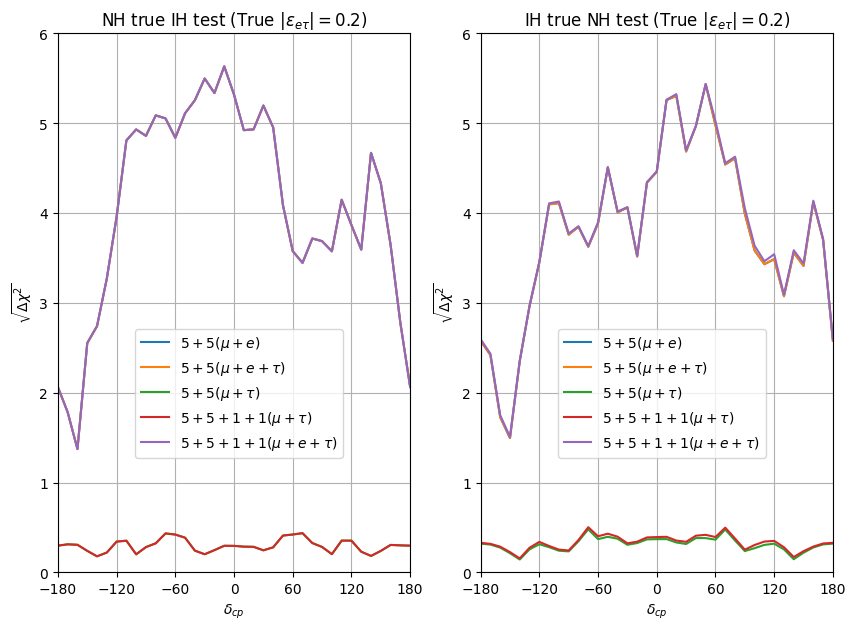}
\includegraphics[width=0.45\textwidth] {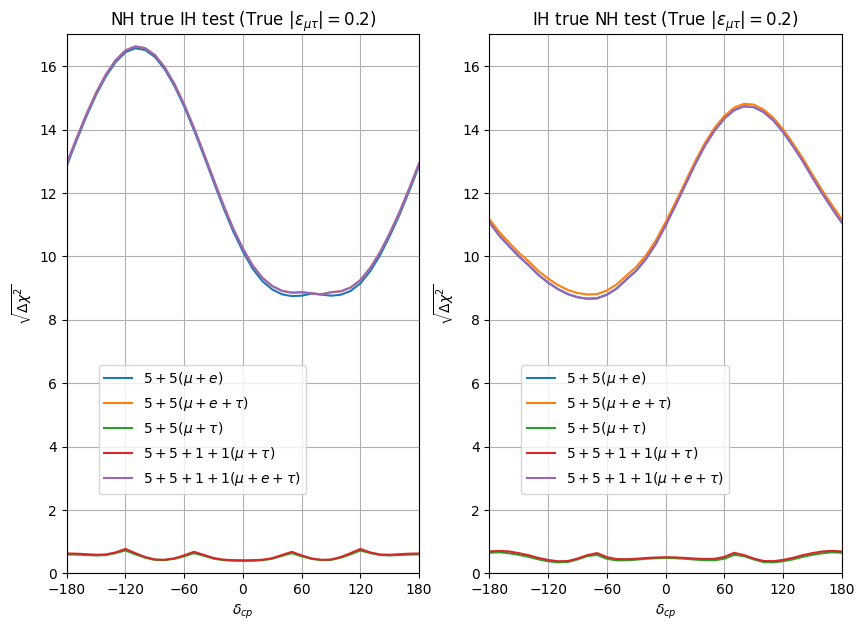}
\includegraphics[width=0.45\textwidth] {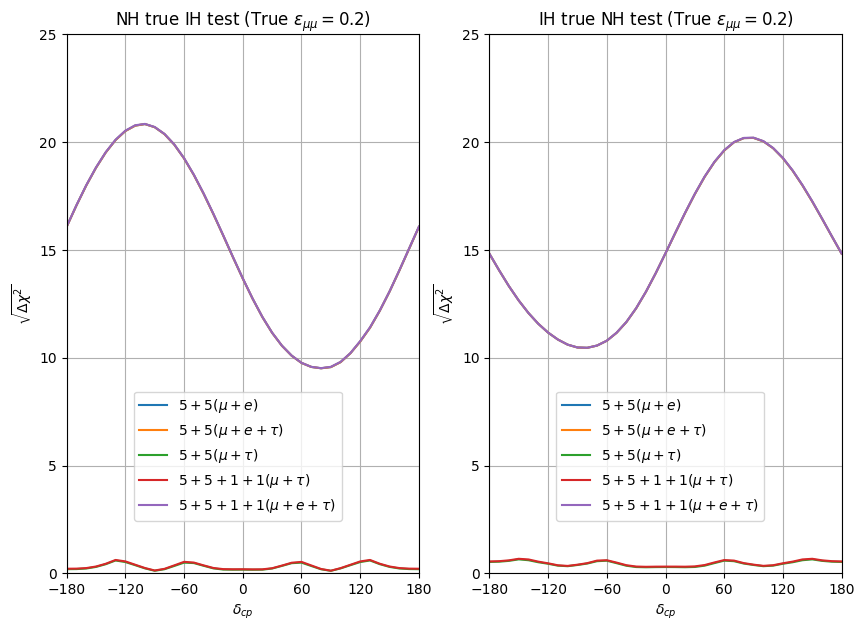}
\includegraphics[width=0.45\textwidth] {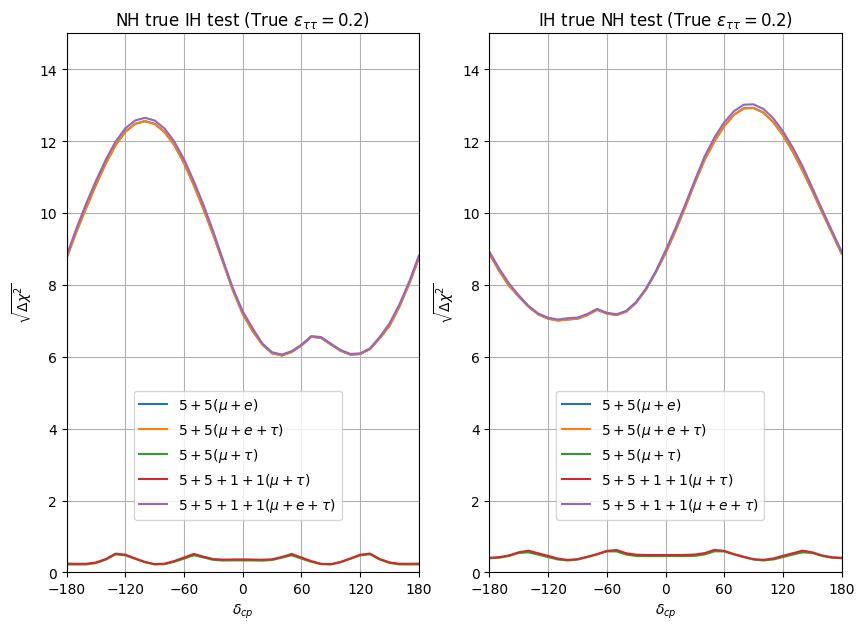}
\caption{\footnotesize{Hierarchy sensitivity of DUNE with the presence of NSI and for different running schemes. The left (right) panel is for the true hierarchy being NH (IH). The true values of NSI parameters have been mentioned on top of the each plot.} }
\label{hierarchy-NSI}
\end{figure}

\begin{figure}[htbp]
\centering
\includegraphics[width=0.8\textwidth] {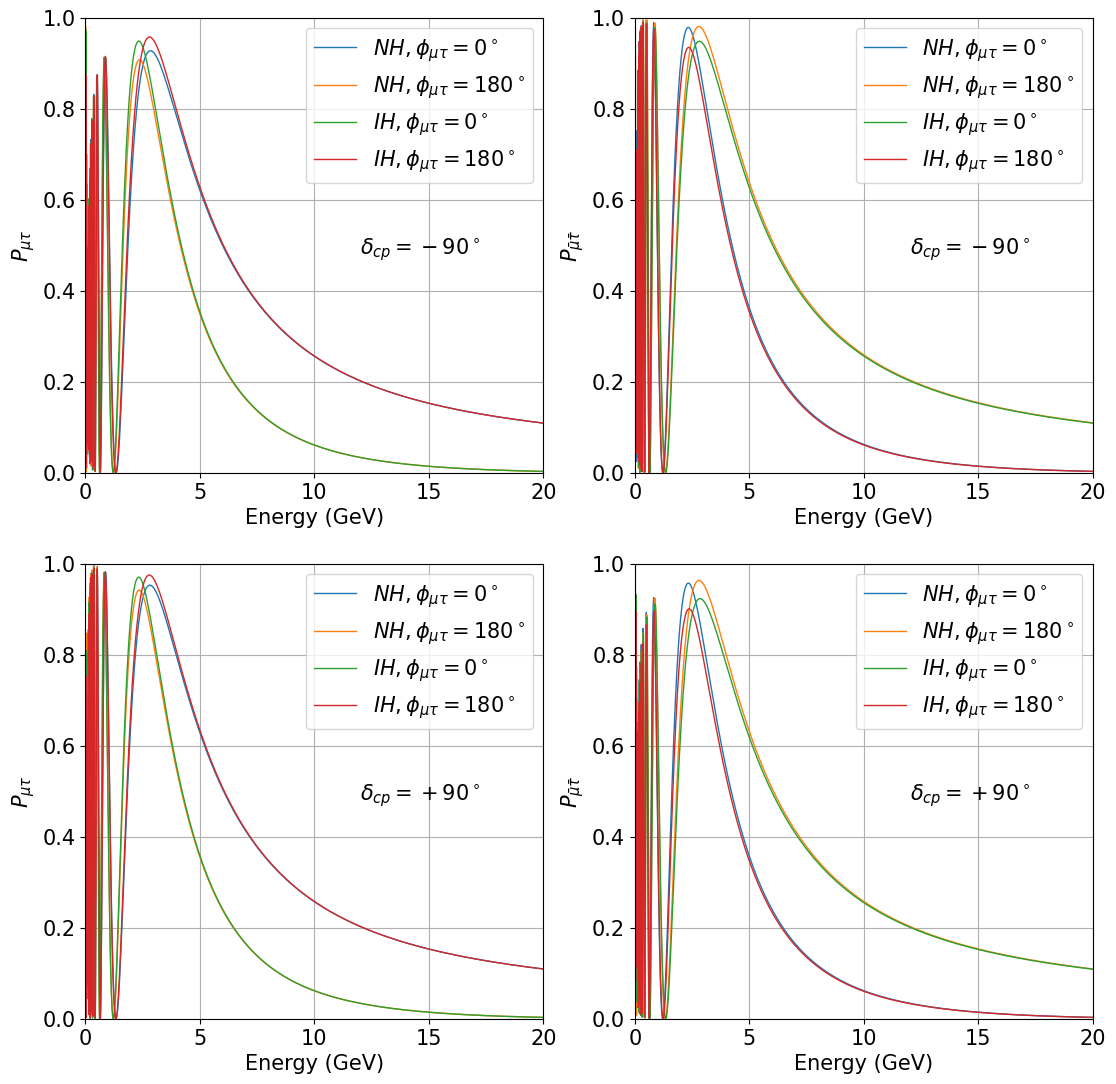}
\caption{\footnotesize{$\pmt$ ($\pmtbar$) in the left (right) panel as a function of neutrino energy for both the hierarchies and $\phi_{\mu\tau}$ being $0$ and $180^\circ$. The $\dcp$ value is $-90^\circ$ ($90^\circ$) in the top (bottom) panel.}}
\label{prob-mu-tau-nh-ih-phi-0-180}
\end{figure}

The first test involves the investigation of DUNE's sensitivity to the mass hierarchy with and without $\tau$ neutrino detection in the absence of NSI. To perform the test, we kept both the true and test values of the NSI parameters fixed to $0$. The true values of $\dcp$ have been varied in the range $[-180^\circ:180^\circ]$. The true values of other standard oscillation parameters have been fixed to their best-fit values taken from ref.~\cite{Esteban:2024eli}. For the test values, we have varied $\sin^2\tz$ and $|\dl|$ in their $3\,\sigma$ range taken from ref.~\cite{Esteban:2024eli}. The test values of $\dcp$ have been varied in their complete range. After calculating $\dchsq$ between the true and test event numbers, marginalisation has been performed over all of the test parameters and the result is shown in fig.~\ref{hierarchy-SM}. This figure shows that for standard oscillation, the detection of $\tau$ neutrinos does not have any effect on the determination of the mass hierarchy at DUNE. The $5+5(\mu+e)$ run alone can determine the mass hierarchy at more than $20\,\sigma$ when NH (IH) and $\dcp=-90^\circ$ ($+90^\circ$) are the true hierarchy-$\dcp$ combination, and close to (more than) $10\,\sigma$ when NH (IH) and $\dcp=+90^\circ$ ($-90^\circ$) are the true hierarchy-$\dcp$ combination. This sensitivity does not change with the addition of $\nu_\tau$ and $\bar{\nu}_\tau$ channels. 
In the next step, we have considered NSI with one parameter at a time. The true value of the NSI parameter has been fixed to $0.2$ with the true value of the corresponding phase, when necessary, fixed at $0$. We have already stated in section~\ref{theory}, that for $\epsilon_{\mu\mu}$ and $\epsilon_{\mu\tau}$, these values are outside the $90\%$ limit given by IceCube \cite{IceCubeCollaboration:2021euf}. Still we choose these values to be consistent with the best-fit values of $\epsilon_{e\mu}$ and $\epsilon_{e\tau}$ from the combined analysis of \nova and T2K, and because $\sim 10^{-1}$ values for $\epsilon_{\mu\mu}$ and $|\epsilon_{\mu\tau}|$ are allowed at $3\,\sigma$ level by the global fits of ref.~\cite{Coloma:2019mbs}. The test values of the NSI parameters have been varied in the range $[0:3]$ for the absolute values of the non-diagonal parameters $\epsilon_{e\mu}$, $\epsilon_{e\tau}$ and $\epsilon_{\mu\tau}$, with their phases varied in the range $[-180^\circ:180^\circ]$. In case of diagonal parameters $\epsilon_{\mu \mu}$ and $\epsilon_{\tau \tau}$, the test values have been varied in the range $[-3:3]$. The marginalisation of $\dchsq$ has been performed over both the standard and NSI test parameters. The results are shown in fig.~\ref{hierarchy-NSI}. It can be seen that the $\nu_\tau$ and $\bar{\nu}_\tau$ appearance channels do not improve the hierarchy sensitivity when compared to the $\nu_e$ and $\bar{\nu}_e$ appearance channels. The inclusion of NSI reduces the the hierarchy sensitivity for the $5+5(\mu+e)$ run for the case of $\epsilon_{e\mu}$, and more significantly for the case of $\epsilon_{e\tau}$. The fluctuations we see in these two cases, are due to the interference between $\dcp$ and the phases $\phi_{e\mu}$ and $\phi_{e\tau}$. In case of NSI due to $\epsilon_{\mu \tau}$, we do not see these fluctuations because the interference term between $\dcp$ and $\phi_{\mu \tau}$ do not exist in the expression of $\pme$ and $\pmebar$. Similarly, the fluctuations in the case of $\epsilon_{\mu\mu}$ and $\epsilon_{\tau\tau}$ are missing because of the absence of phases in these two real terms. The low hierarchy sensitivity in the $\nu_\tau$ and $\bar{\nu}_\tau$ appearance channels in case of NSI due to $\epsilon_{\mu\tau}$ is counter intuitive because we can see large separation in the bi-event plots for NH and IH from fig.~\ref{bievents3} and also from the probability plots in fig.~\ref{nsi-hie-oct}. However, the hierarchy sensitivity for $\epsilon_{\mu\tau}$ comes from the $\sin^22\tz|\epsilon_{\mu\tau}|\cos\phi_{\mu\tau}\sin(\dl L/2E)$ term in $\pmt$ and $\pmtbar$, as discussed in section~\ref{theory}. Therefore, the $\pmt$ and $\pmtbar$ for NH at $|\epsilon_{\mu\tau}|=0.2$, and $\phi_{\mu\tau}=0$, can be mimicked by choosing IH, $|\epsilon_{\mu\tau}|=0.2$, and $\phi_{\mu\tau}=180^\circ$. Similarly, the $\pmt$ and $\pmtbar$ for IH at $|\epsilon_{\mu\tau}|=0.2$ and $\phi_{\mu\tau}=0$, can be mimicked by choosing NH, $|\epsilon_{\mu\tau}|=0.2$, and $\phi_{\mu\tau}=180^\circ$. Similarly, for any true hierarchy and true value of $\phi_{\mu\tau}$, the hierarchy sensitivity can be cancelled by choosing a test $\phi_{\mu\tau}$ as $\phi_{\mu\tau}({\rm test})=180^\circ-\phi_{\mu\tau}({\rm true})$. Therefore, a new degeneracy in form of $\phi_{\mu\tau}$-hierarchy arises due to the $\cos\phi_{\mu\tau}$ dependency of the hierarchy sensitive part of the oscillation probability expression. Since we are marginalizing over $\phi_{\mu\tau}$, this degeneracy effectively reduces the hierarchy sensitivity in the $\tau$ appearance channels even in the case of $\epsilon_{\mu\tau}$. We have emphasized this argument in fig.~\ref{prob-mu-tau-nh-ih-phi-0-180}. The figure shows that the hierarchy sensitivity for NSI due to $\epsilon_{\mu\tau}$ is lost due to the $\phi_{\mu\tau}$-hierarchy degeneracy.

\subsection{CP violation sensitivity}
\label{CP}
In this section, we present the potential of DUNE to reject CP conserving $\dcp$ values, and how the detection of $\nu_\tau$ can affect the sensitivity for different running schemes. 

\begin{figure}[htbp]
\centering
\includegraphics[width=0.8\textwidth] {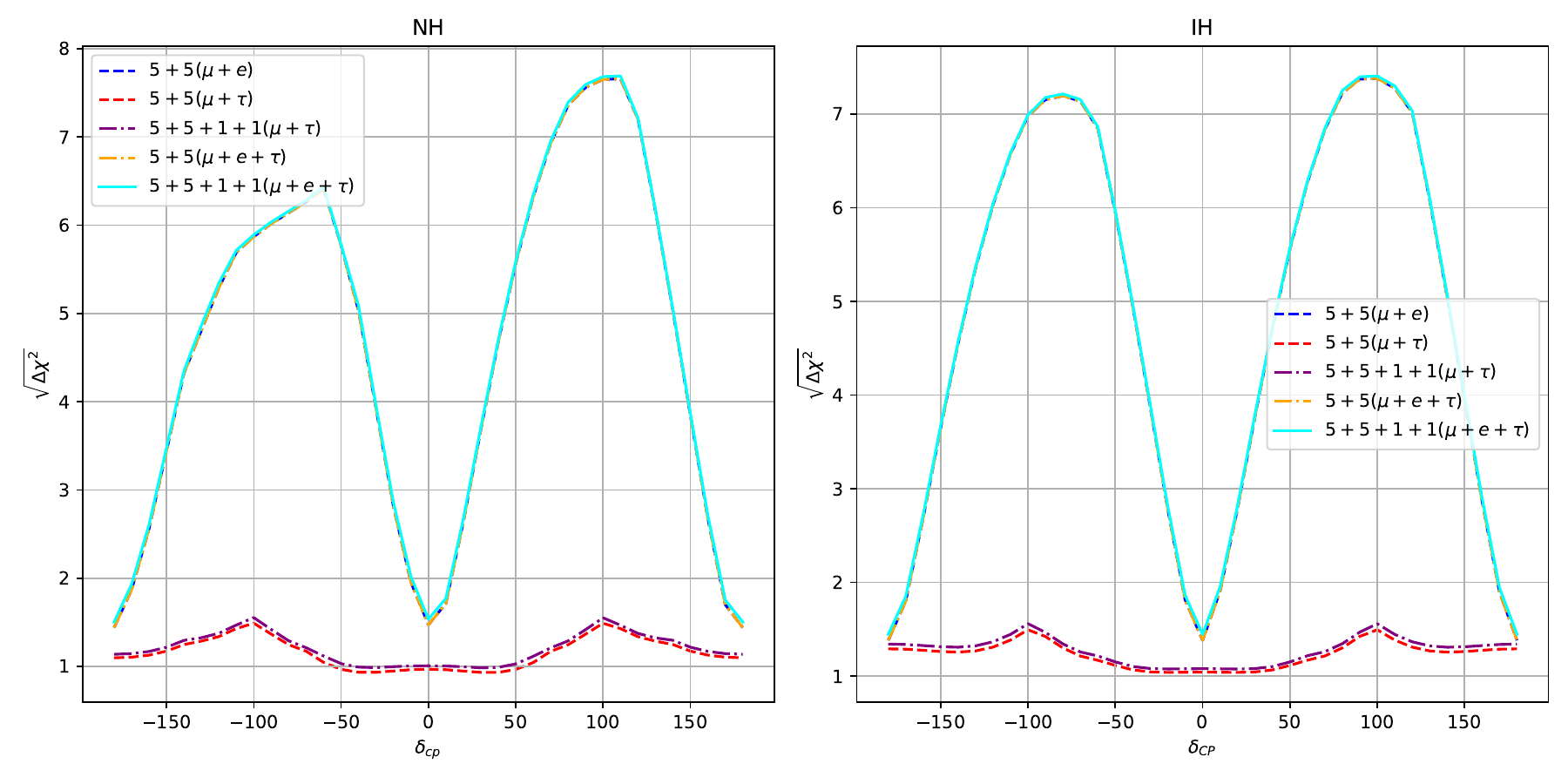}
\caption{\footnotesize{CP sensitivity of DUNE without the presence of any NSI and for different running schemes. The left (right) panel is for the true hierarchy being NH (IH).}}
\label{CP-SM}
\end{figure}

Firstly, we consider the standard 3-flavour oscillation without the presence of any NSI. For this case, we have fixed the true values of the standard oscillation parameters, except $\dcp$, at the global best-fit values of ref.~\cite{Esteban:2024eli}. The true values of $\dcp$ have been varied in the range $[-180^\circ:180^\circ]$. For the test values of oscillation parameters, we have varied $\sin^2\tz$ and $|\dl|$ in their $3\,\sigma$ range given in ref.~\cite{Esteban:2024eli}. The test values of $\dcp$ are the three CP concerving values, namely $0$, $-180^\circ$ and $180^\circ$. Both the hierarchies have been considered as the test hierarchy. The $\chi^2$ between true and test events numbers have been calculated. In this case also, $\chi^2$ and $\dchsq$ are basically same. The marginalisation of $\dchsq$ has been performed over the test parameter values, and the final result is presented using $\dchsq$ as a function of the true $\dcp$, as shown in fig.~\ref{CP-SM}. The figure shows that when NH is the true hierarchy, the maximum CP violation sensitivity can be established at $>6\,\sigma$ and $>7.5\,\sigma$ for $\dcp\sim -50^\circ$ and $\dcp\sim 150^\circ$ respectively in the case of $5+5(\mu+e)$ run. For the same run, when IH is the true hierarchy, CP sensitivity of maximum $7\,\sigma$ can be established at $\dcp=\pm 100^\circ$. Addition of $\nu_\tau$ and $\bar{\nu}_\tau$ will not have any significant changes to the CP violation sensitivity of DUNE. Only $1\,\sigma$ discovery of CP violation is possible with $5+5(\mu+\tau)$ and $5+5+1+1(\mu+\tau)$ run in case of $\dcp=\pm 100^\circ$ for both the hierarchies.

\begin{figure}[htbp]
\centering
\includegraphics[width=0.6\textwidth] {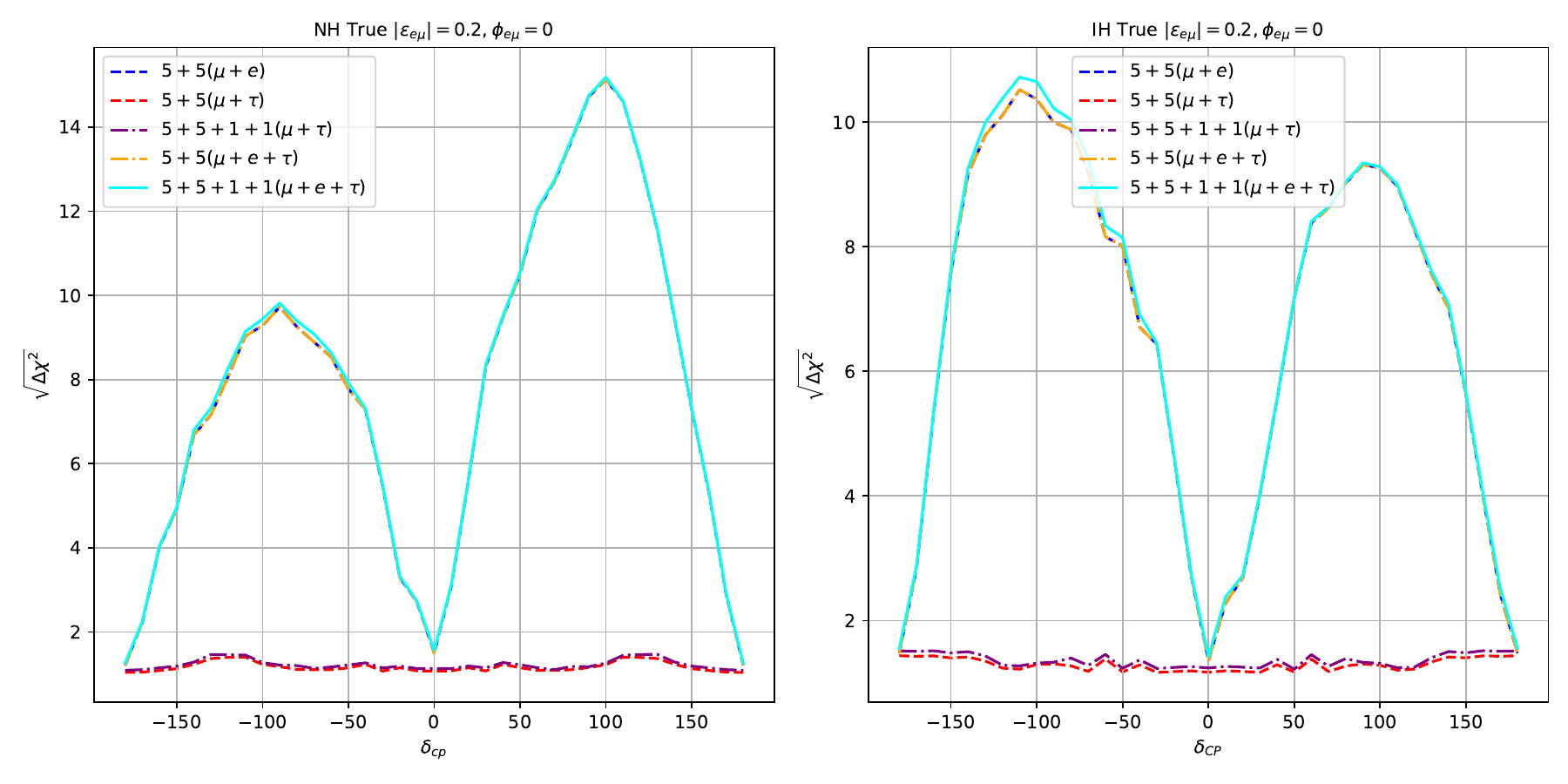}
\includegraphics[width=0.6\textwidth] {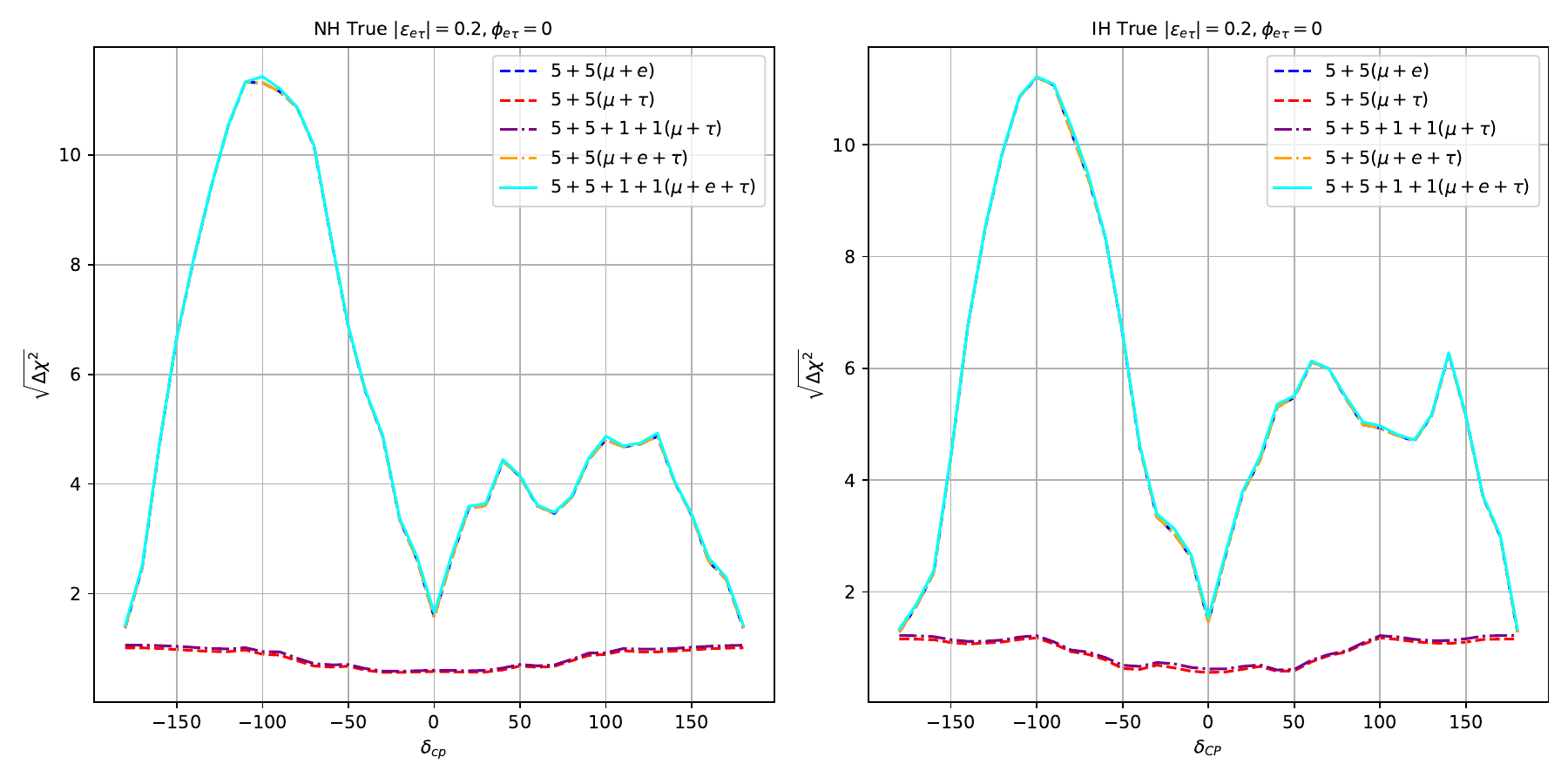}
\includegraphics[width=0.6\textwidth] {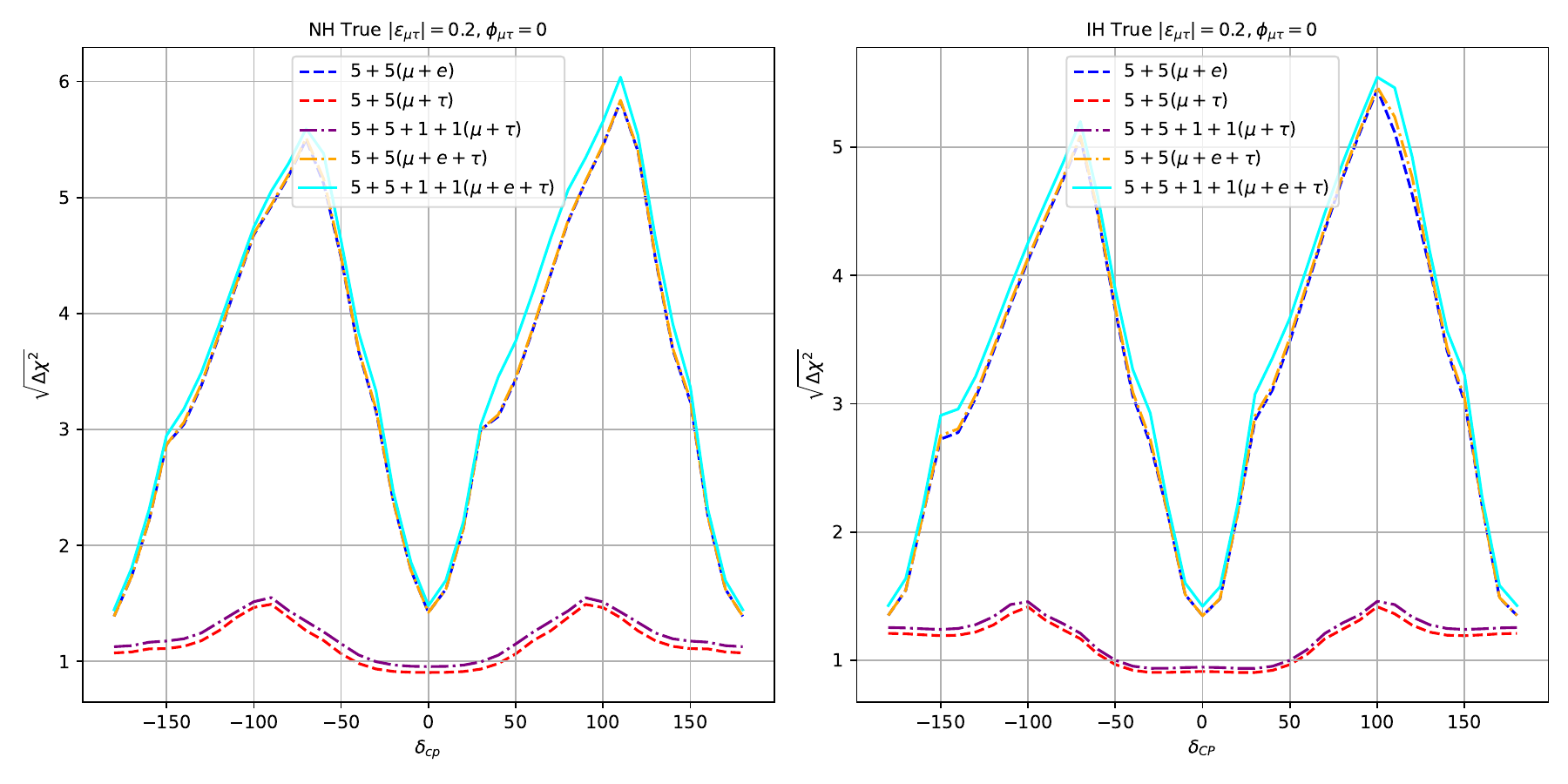}
\includegraphics[width=0.6\textwidth] {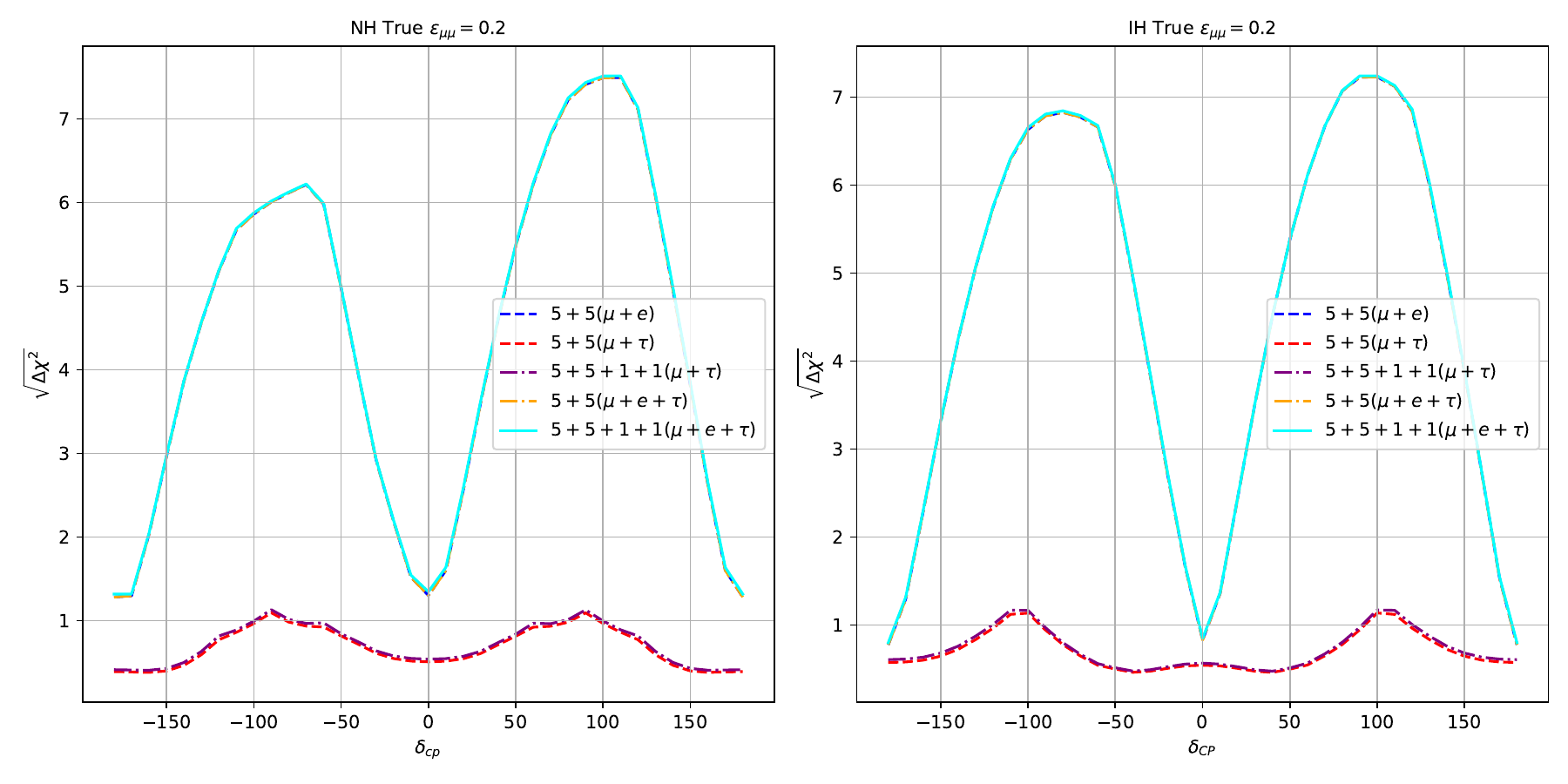}
\includegraphics[width=0.6\textwidth] {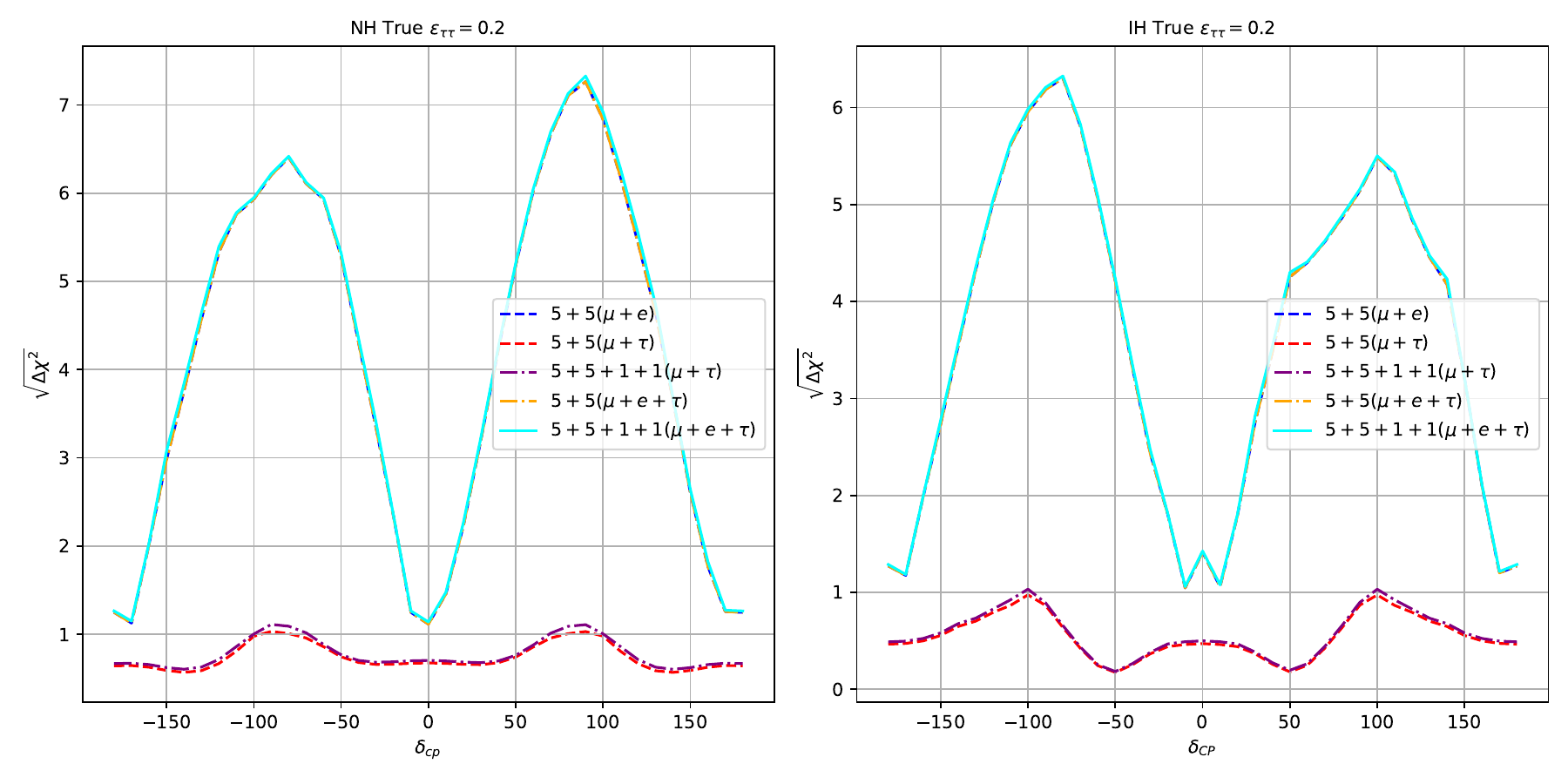}
\caption{\footnotesize{CP sensitivity of DUNE with the presence of NSI and for different running schemes. The left (right) panel is for the true hierarchy being NH (IH). The true values of NSI parameters have been mentioned on top of the each plot.} }
\label{CP-NSI}
\end{figure}

In the next step, we have considered NSI due to $\epsilon_{e\mu}$. The true values of $|\epsilon_{e\mu}|$ and $\phi_{e\mu}$ have been taken to be $0.2$ and $0$ respectively. The test values of $|\epsilon_{e\mu|}$ have been varied in the range $[0:3]$, whereas the test values of $\phi_{e\mu}$ have been varied in the range $[-180^\circ:180^\circ]$. The true and test values of standard parameters are the same as in the previous section. As in the previous case, the marginalisation of the $\dchsq$ has been performed over the test parameter values. A similar procedure has been repeated for $\epsilon_{e\tau}$ and $\epsilon_{\mu\tau}$ as well. For the case of $\epsilon_{\mu \mu}$ and $\epsilon_{\tau \tau}$, no phase has been considered and the test values of these two parameters have been varied in the range $[-3:3]$. The results have been shown in fig.~\ref{CP-NSI}. It is clear that just like in the case without NSI, with NSI the $5+5(\mu+\tau)$ and $5+5+1+1(\mu+\tau)$ running schemes have negligible sensitivity to CP violation. 

\subsection{Octant sensitivity}
\label{octant}
\begin{figure}[htbp]
\centering
\includegraphics[width=0.8\textwidth] {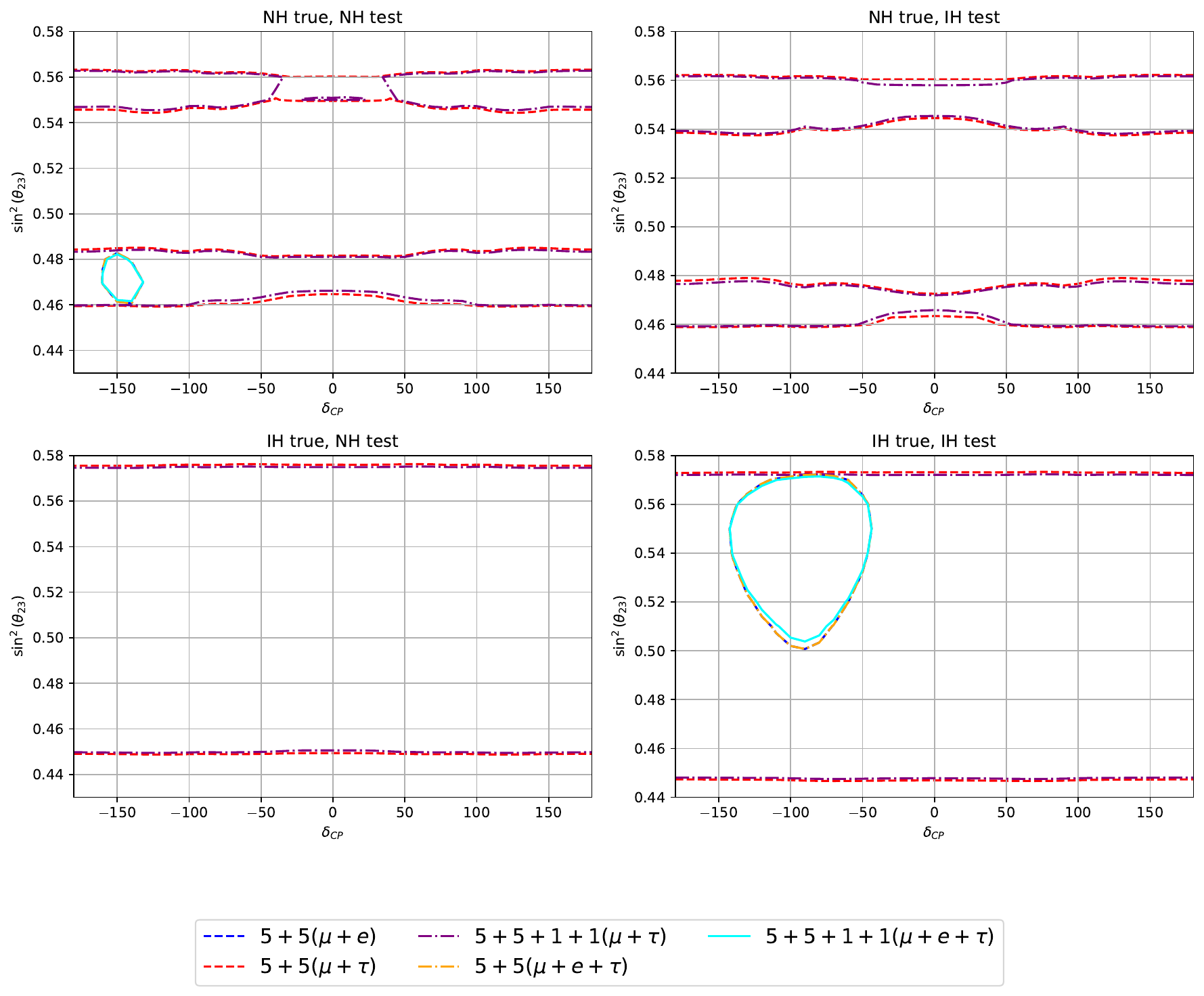}
\caption{\footnotesize{Allowed regions in the test $\dcp-sin^2\tz$ plane for standard oscillation without any NSI. The top (bottom) panels are for NH (IH) being the true hierarchy. The left (right) panels are for test hierarchy being NH (IH). }}
\label{octant-SM}
\end{figure}

In this section, we discuss the sensitivity to determine the octant of $\tz$. At first, we have considered the standard unitary $3\times3$ mixing, without the presence of any NSI. To do so, we have fixed the true values of the standard parameters at the global best-fit values taken from ref.~\cite{Esteban:2024eli}. For test parameters, we have varied $\dcp$ in the whole range of $[-180^\circ:180^\circ]$. We also varied $\sin^2\tz$ and $|\dl|$ in their $3\,\sigma$ range. Marginalisation of $\dchsq$ has been done over $|\dl|$ and the result has been presented as a contour plot on the test $\dcp-\sin^2\tz$ plane. As can be seen from fig.~\ref{octant-SM}, a $5+5(\mu+e)$ run can rule out the wrong octant for both cases of true hierarchy. However, the $5+5(\mu+\tau)$ and $5+5+1+1(\mu+\tau)$ running scenarios do not have any octant sensitivity. This is expected from our discussions in section~\ref{theory}. 

In the next step, we considered one NSI parameter at a time. We assumed the true value of the respective NSI parameter to be $0.2$. In case of non-diagonal NSI parameters, the true value of the phase associated with the NSI parameter have been considered to be $0$. The test value of the concerned NSI parameter has been varied in the range $[0:3]$. For the non-diagonal NSI parameters, the corresponding phases have been varied in the range $[-180^\circ:180^\circ]$. The marginalisation of $\dchsq$ has been done over test values of $|\dl|$ and the NSI parameter and its corresponding phase when applicable. As expected from the bi-event plots, $5+5(\mu+\tau)$ and $5+5+1+1(\mu+\tau)$ runs do not have any octant sensitivity in presence of NSI. However, there is also no octant sensitivity for the $5+5(\mu+e)$ run, in presence of NSI due to $\epsilon_{e\mu}$ and $\epsilon_{e\tau}$, and for IH being the true hierarchy. The results for $\epsilon_{e\mu}$ and $\epsilon_{e\tau}$  are presented in fig.~\ref{octant-NSI}. The results for other NSI parameters have been shown in the appendix~\ref{octant-appendix}
\begin{figure}[htbp]
\centering
\includegraphics[width=0.7\textwidth] {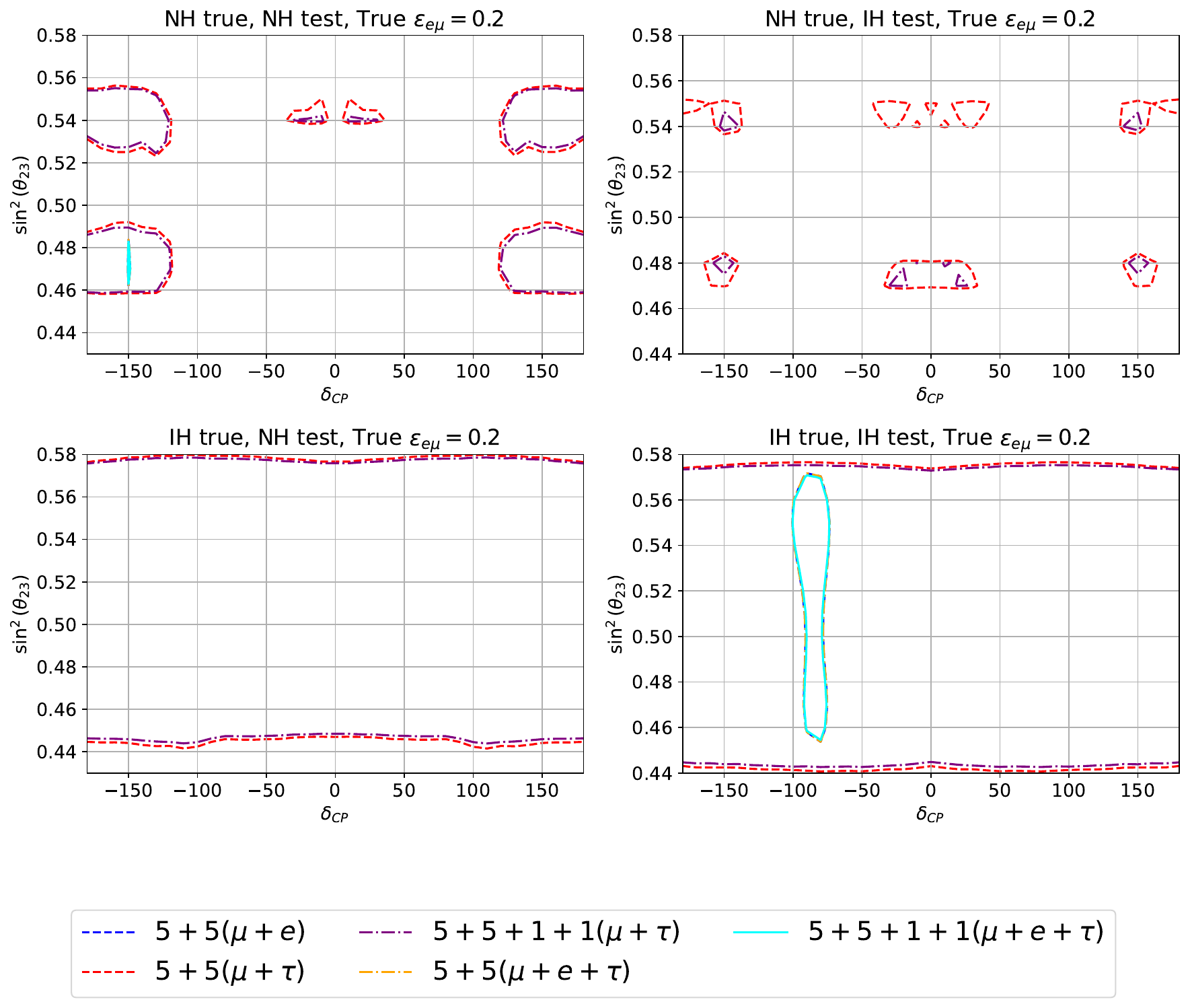}
\includegraphics[width=0.7\textwidth] {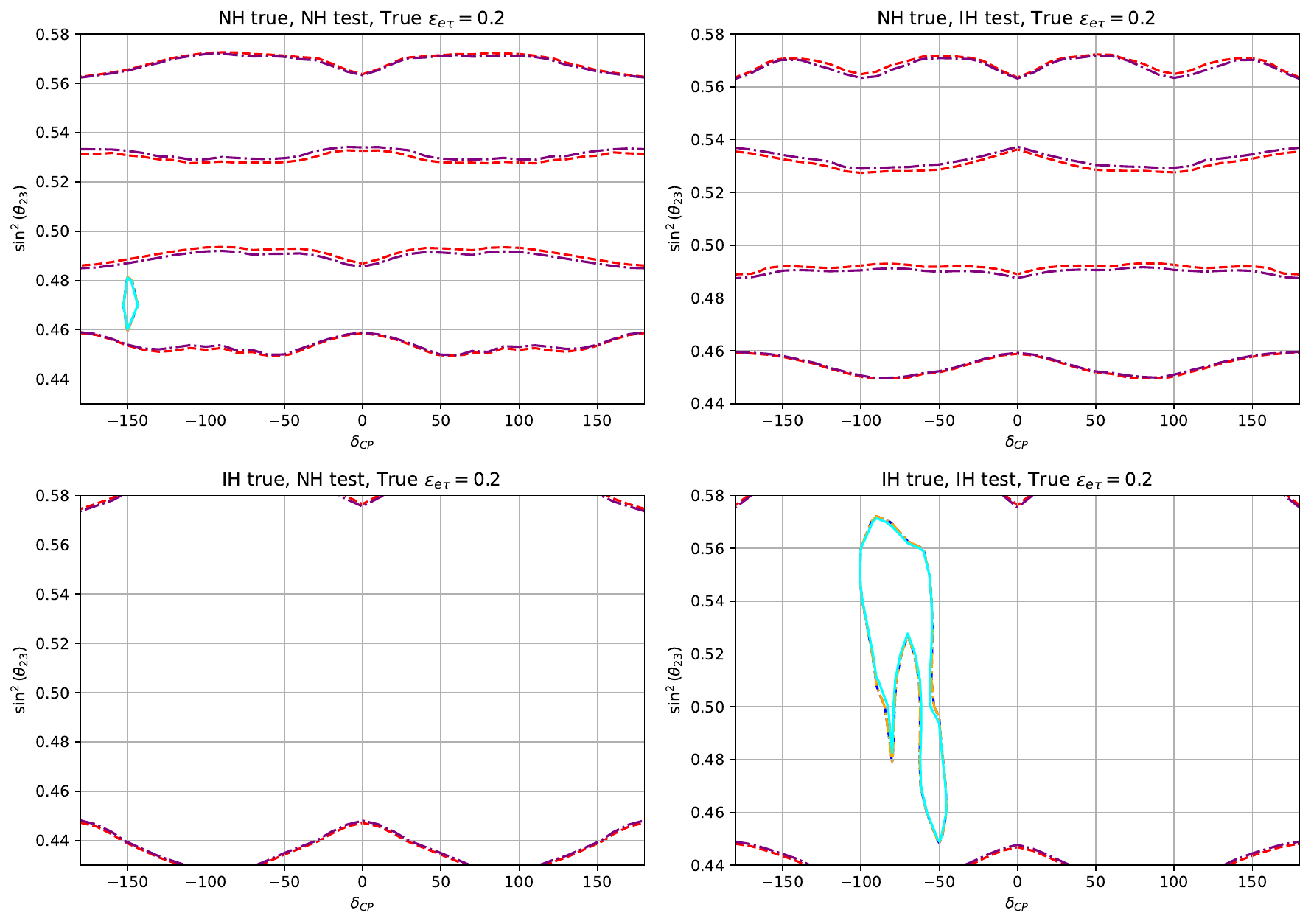}
\caption{\footnotesize{Allowed regions in the test $\dcp-sin^2\tz$ plane for oscillation with NSI due to $\epsilon_{e\mu}$ and $\epsilon_{e\tau}$. The top (bottom) panels are for NH (IH) being the true hierarchy. The left (right) panels are for test hierarchy being NH (IH). }}
\label{octant-NSI}
\end{figure}
\subsection{Determining NSI phases}
\label{phase}
\begin{figure}[htbp]
\centering
\includegraphics[width=0.7\textwidth] {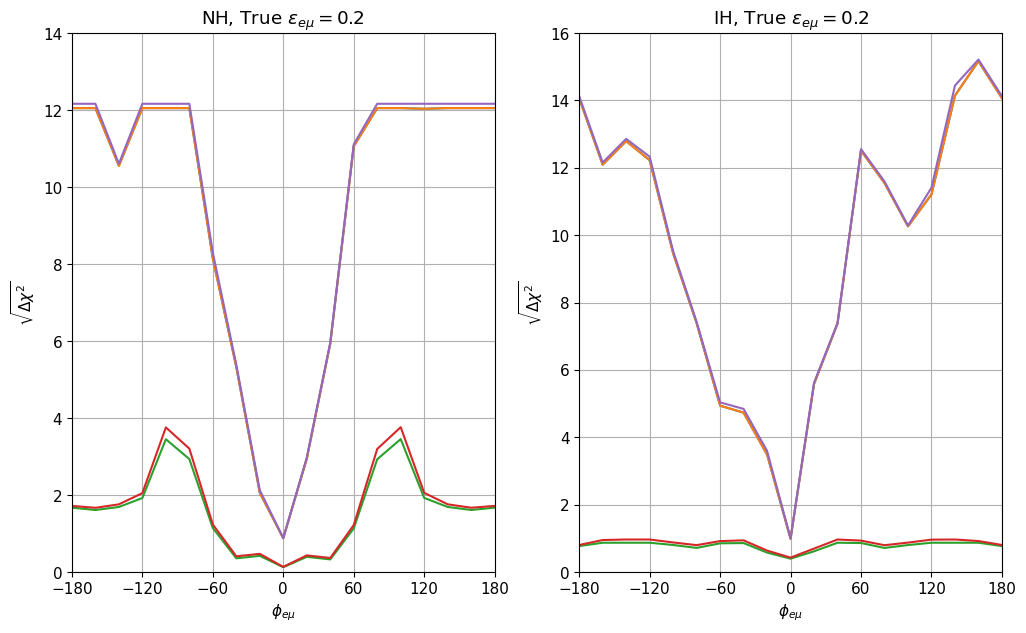}
\includegraphics[width=0.7\textwidth] {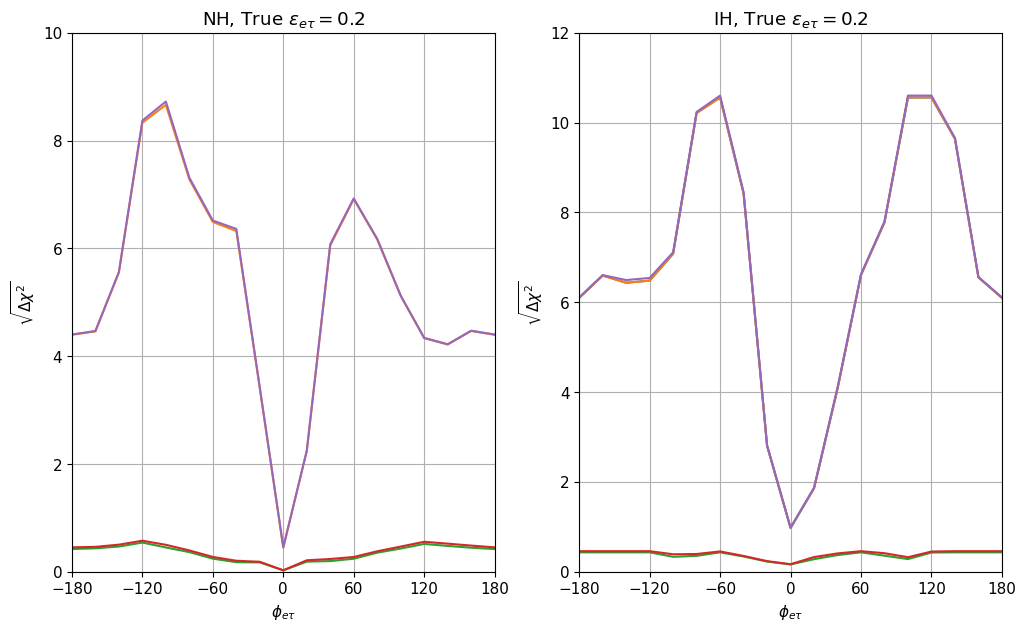}
\includegraphics[width=0.7\textwidth] {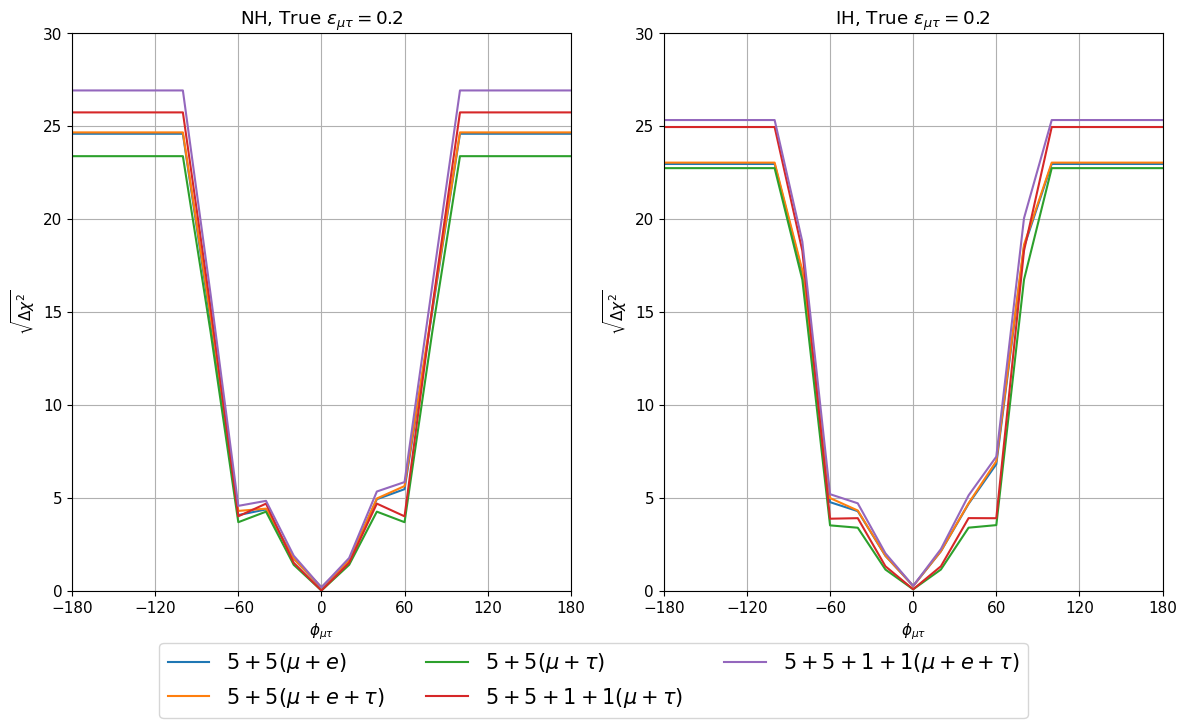}

\caption{\footnotesize{Determination sensitivity to different NSI phases with different runtimes of DUNE.}}
\label{sens-phase}
\end{figure}

In section~\ref{hierarchy}, we have seen that the hierarchy sensitivity of the $\nu_\tau$ appearance channels at DUNE in the presence of NSI due to $\epsilon_{\mu\tau}$ is reduced due to the $\phi_{\mu\tau}$-hierarchy degeneracy arising from the $\cos \phi_{\mu\tau}$ term present in the hierarchy sensitive term of $\pmt$. However, due to the presence of the same term, it is possible to measure $\phi_{\mu\tau}$ with $\nu_\tau$ appearance at DUNE in presence of NSI due to $\epsilon_{\mu\tau}$, when the hierarchy is measured with precision from the electron appearance channels of DUNE. The validity of this statement can be verified from fig.~\ref{prob-mu-tau-nh-ih-phi-0-180} as well. In this section, we investigate the potential of DUNE to measure NSI phases in case NSI is present in nature. To do that, we have fixed the true values of the standard oscillation parameters at their best-fit values.  The true values of $|\epsilon_{e\mu}|$, $\epsilon_{e\tau}$ and $\epsilon_{\mu\tau}$ (only one at a time) have been fixed to $0.2$, and the corresponding phase has been fixed to $0$. The test values of $\dcp$ have been varied in the range $[-180^\circ:180^\circ]$. The test values of $|\dl|$, and $\sin\tz$ have been varied in their $3\,\sigma$ allowed values. The test values of the magnitude of the corresponding NSI parameter have been varied in the range $[0:3]$. Test values of the corresponding phases have been varied in the range $[-180^\circ:180^\circ]$. The true and test hierarchies are the same. We have marginalised the $\dchsq$ between true and test event numbers over all the test parameters except the relevant NSI phase and the results are presented in fig.~\ref{sens-phase}. The goal of the investigation is to determine, if NSI is present and the NSI phase value chosen by nature is $0$, how well do the different running schemes of DUNE rule out non-zero values of the phase provided the mass hierarchy is already known with precision. In case of NSI due to $\epsilon_{e\mu}$, $40^\circ<\phi_{e\mu}<-40^\circ$ ($10^\circ<\phi_{e\mu}<-60^\circ$) can be ruled out at $5\,\sigma$ with only the $5+5(\mu+e)$ run when NH (IH) is the true hierarchy. The addition of $\nu_\tau$ and $\bar{\nu}_\tau$ appearance channels does not improve the sensitivity. The $5+5(\mu+\tau)$ and $5+5+1+1(\mu+\tau)$ runs rule out $\phi_{e\mu}=\pm90^\circ$ at $2\,\sigma$ confidence level for NH being the true hierarchy. This sensitivity arises from the muon disappearance channels. In case of NSI due to $\epsilon_{e\tau}$, with $5+5(\mu+e)$ run, $30^\circ<\phi_{e\tau}<-30^\circ$ can be ruled out at more than $5\,\sigma$ for both the hierarchies. The addition of $\nu_\tau$ and $\bar{\nu}_\tau$ appearance data does not improve the sensitivity. In case of NSI due to $\epsilon_{\mu\tau}$, we can see that the $5+5(\mu+e)$, $5+5(\mu+\tau)$ and $5+5+1+1(\mu+\tau)$ runs can rule out $60^\circ<\phi_{\mu\tau}<-60^\circ$ at $5\,\sigma$ C.L. However, for ruling out $90^\circ<\phi_{\mu\tau}<-90^\circ$, $5+5+1+1(\mu+\tau)$ and $5+5+1+1(\mu+e+\tau)$ runs can improve the sensitivity by $1\,\sigma$ and $2.5\,\sigma$ respectively compared to the sensitivity for the $5+5(\mu+e)$ run, when NH is the true mass hierarchy. In case of IH being the true mass hierarchy, this improvement is $2.5\,\sigma$ for both the $5+5+1+1(\mu+\tau)$ and $5+5+1+1(\mu+e+\tau)$ runs. Hence, the determination of $\nu_\tau$ and $\bar{\nu_\tau}$ appearance can improve the precision of $\phi_{\mu\tau}$ in case of NSI due to $\epsilon_{\mu\tau}$, given that $|\epsilon_{\mu\tau}|\sim 10^{-1}$ and the mass hierarchy is precisely known.

\subsection{Sensitivity to NSI parameters}
\label{sensitivity}
In this section, we investigate the sensitivities of different NSI parameters at DUNE and the effect of $\tau$ neutrino detection on the sensitivity. The true event rates were generated assuming standard 3-flavour neutrino oscillation without the presence of any NSI effect. To do so, the standard oscillation parameter values have been fixed at their global best-fit point taken from ref.~\cite{Esteban:2024eli, nufit}. For the test event rates, NSI has been considered with one parameter at a time. Test values of $\dcp$ and the NSI phase $\phi_{ij}$ for off-diagonal parameters $\epsilon_{ij}$, with $i=e,\mu$; $j=\mu,\tau$ and $i\neq j$, have been varied in the range $[-180^\circ:180^\circ]$. $\sin^2\tz$ and $|\dl|$ have been varied in the $3\,\sigma$ range taken from ref.~\cite{Esteban:2024eli, nufit}. The off-diagonal NSI parameters $|\epsilon_{e\mu}|$, $|\epsilon_{\mu\tau}|$, and $|\epsilon_{e\tau}|$ have been varied in the range $[0:3]$, where as the diagonal parameters $\epsilon_{\mu\mu}$ and $\epsilon_{\tau\tau}$ have been varied in the range $[-3:3]$. The marginalisation has been done over test values of $\dcp$, $\sin^2\tz$ and $|\dl|$, along with the phases $\phi_{ij}$ in case of non-diagonal NSI parameters. The final results have been expressed as $\sqrt{\dchsq}$ as a function of test NSI parameters. It can be observed from the figure that in case of $\epsilon_{\mu\tau}$, all the 5 different runs can rule out $|\epsilon_{\mu\tau}|>0.6$ at more than $20\,\sigma$ C.L.
For the two diagonal terms, $\epsilon_{\mu\mu}$ and $\epsilon_{\tau\tau}$, all the different run schemes have almost equivalent sensitivity. For $\epsilon_{e\mu}$, and $\epsilon_{e\tau}$, however, the $5+5(\mu+e)$ run have much better sensitivity than the $5+5(\mu+\tau)$ or $5+5+1+1(\mu+\tau)$ runs. In table~\ref{limit}, we have given the expected $90\%$ and $3\,\sigma$ constraints on different NSI parameters for different running schemes. It can be concluded that in case of $|\epsilon_{\mu\tau}|$, the $5+5+1+1(\mu+e+\tau)$ run has the best sensitivity among all the different running schemes. Also for $5+5+1+1(\mu+e+\tau)$ run, the $90\%$ constraint of $|\epsilon_{\mu\tau}|<7\times10^{-2}$ ($6\times10^{-2}$) from DUNE will be close to the IceCube constraint of $|\epsilon_{\mu\tau}|<2\times 10^{-2}$ at $90\%$ confidence level \cite{IceCubeCollaboration:2021euf}, if NH (IH) is the true hierarchy. $5+5(\mu+\tau)$ and $5+5+1+1(\mu+\tau)$ run has slightly weaker sensitivity than $5+5(\mu+e)$ run, but it is possible to provide complimentary sensitivity to $|\epsilon_{\mu\tau}|$ from the $\nu_\tau$ and $\bar{\nu}_\tau$ appearance channels. In case of NSI due to $\epsilon_{\mu\mu}$, and $\epsilon_{\tau \tau}$ the $5+5(\mu+\tau)$ and $5+5+1+1(\mu+\tau)$ run schemes have sensitivities to NSI parameters of the similar order of magnitude as in other run schemes. For $|\epsilon_{\mu\tau}|$, the sensitivity is coming from $\nu_\tau$ appearance and $\nu_\mu$ disappearance channels, and the contribution from $\nu_e$ appearance channels are negligible. This is in agreement with the theoretical discussion given in section~\ref{theory}. From eq.~ B6 and B8 of ref.~\cite{Kikuchi:2008vq}, it is clear that the sensitivity to $\epsilon_{\mu\mu}$ and $\epsilon_{\tau\tau}$ mostly comes from the muon disappearance channels due to their large statistics. For this reason, the sensitivity to $\epsilon_{\mu\mu}$ and $\epsilon_{\tau\tau}$ for the $\mu+\tau$ run schemes are comparable to the $\mu+e$ and $\mu+e+\tau$ run schemes. However, for $\epsilon_{e\mu}$ and $\epsilon_{e\tau}$, the $5+5(\mu+e)$ run has much better sensitivity compared to the run schemes without $\nu_e$ detection. We have presented the results for sensitivity to $\epsilon_{\mu\tau}$, $\epsilon_{\mu\mu}$ and $\epsilon_{\tau\tau}$ in fig.~\ref{nsi-sens}. The results for sensitivity to other two NSI parameters have been presented in the appendix~ \ref{sensitivity-appendix}.

\begin{table}
\hskip -1.3cm
\resizebox{1.18 \textwidth}{!}{
    
\begin{tabular}{|c|c|c|c|c|c|}

  \hline
  Parameters & $5+5$ &$5+5$&$5+5$&$5+5+1+1$&$5+5+1+1$\\
  & $(\mu+e)$ &$(\mu+\tau)$&$(\mu+e+\tau)$&$(\mu+\tau)$&$(\mu+e+\tau)$\\
  \hline
  $|\epsilon_{e \mu}|$ NH & $<0.04$ ($0.10$) & $<0.31$ ($0.43$) & $<0.04$ ($0.10$) & $<0.31$ ($0.43$) & $<0.04$ ($0.10$)\\
  \hspace{0.7 cm} IH & $<0.04$ ($0.09$) & $<0.25$ ($0.38$) & $<0.0.04$ ($0.09$) & $<0.25$ ($0.38$) & $<0.04$ ($0.09$)\\
  \hline
   $|\epsilon_{e\tau}|$ NH & $<0.07$ ($0.20$) & $<0.83$ ($1.02$) & $<0.07$ ($0.20$) & $<0.83$ ($1.02$) & $<0.07$ ($0.20$)\\
  \hspace{0.7 cm} IH & $<0.09$ ($0.13$) & $<0.78$ ($0.98$) & $<0.0.09$ ($0.13$) & $<0.78$ ($0.98$) & $<0.09$ ($0.13$)\\
  \hline
  $|\epsilon_{\mu\tau}|$ NH & $<0.08$ ($0.14$) & $<0.09$ ($0.17$) & $<0.08$ ($0.14$) & $<0.09$ ($0.15$) & $<0.07$ ($0.13$)\\
  \hspace{0.7 cm} IH & $<0.07$ ($0.12$) & $<0.10$ ($0.16$) & $<0.07$ ($0.12$) & $<0.09$ ($0.14$) & $<0.06$ ($0.11$)\\
  \hline
  $\epsilon_{\mu\mu}$ NH & $>-0.40$ ($-0.50$)  & $>-0.43$ ($-0.50$) & $>-0.40$ ($-0.50$) & $>-0.43$ ($-0.50$) & $>-0.40$ ($-0.50$)\\
  & $<0.11$ ($0.48$) & $<0.50$ ($0.57$) & $<0.11$ ($0.48$) & $<0.50$ ($0.57$) & $<0.11$ ($0.48$)\\
  \hspace{0.7 cm} IH & $>-0.17$ ($-0.34$) & $>-0.43$ ($-0.50$) & $>-0.17$ ($-0.34$) & $>-0.43$ ($-0.50$) & $>-0.17$ ($-0.34$)\\
  & $<0.14$ ($0.56$) & $<0.49$ ($0.54$) & $<0.14$ ($0.56$) & $<0.49$ ($0.54$)& $<0.14$ ($0.56$)\\
  \hline
  $\epsilon_{\tau\tau}$ NH & $>-0.12$ ($-0.26$)  & $>-0.54$ ($-0.63$) & $>-0.12$ ($-0.26$) & $>-0.54$ ($-0.63$) & $>-0.12$ ($-0.26$)\\
  & $<0.14$ ($0.35$) & $<0.45$ ($0.51$) & $<0.14$ ($0.35$) & $<0.45$ ($0.51$) & $<0.14$ ($0.35$)\\
  \hspace{0.7 cm} IH & $>-0.11$ ($-0.20$) & $>-0.51$ ($-0.60$) & $>-0.11$ ($-0.20$) & $>-0.51$ ($-0.60$) & $>-0.11$ ($-0.20$)\\
  & $<0.09$ ($0.25$) & $<0.42$ ($0.51$) & $<0.09$ ($0.25$) & $<0.42$ ($0.51$)& $<0.09$ ($0.25$)\\
  \hline
\end{tabular}
}
 \caption{Expected $90\%$ and $3\,\sigma$ constraints for 1 degree of freedom for different run times and for both the hierarchies. The $3\,\sigma$ constraints have been shown in the parenthesis. }
  \label{limit}
\end{table}

\begin{figure}[htbp]
\centering
\includegraphics[width=0.7\textwidth] {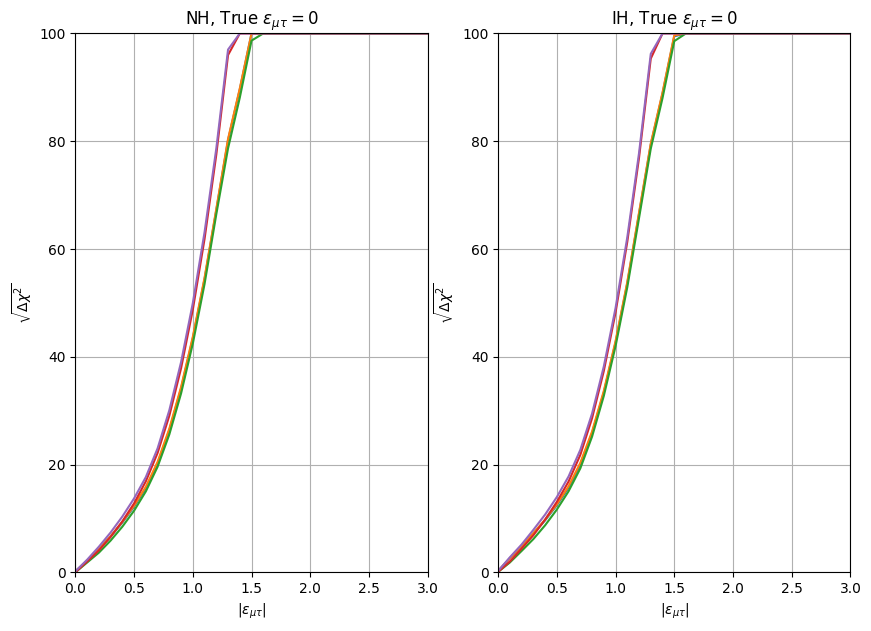}
\includegraphics[width=0.7\textwidth] {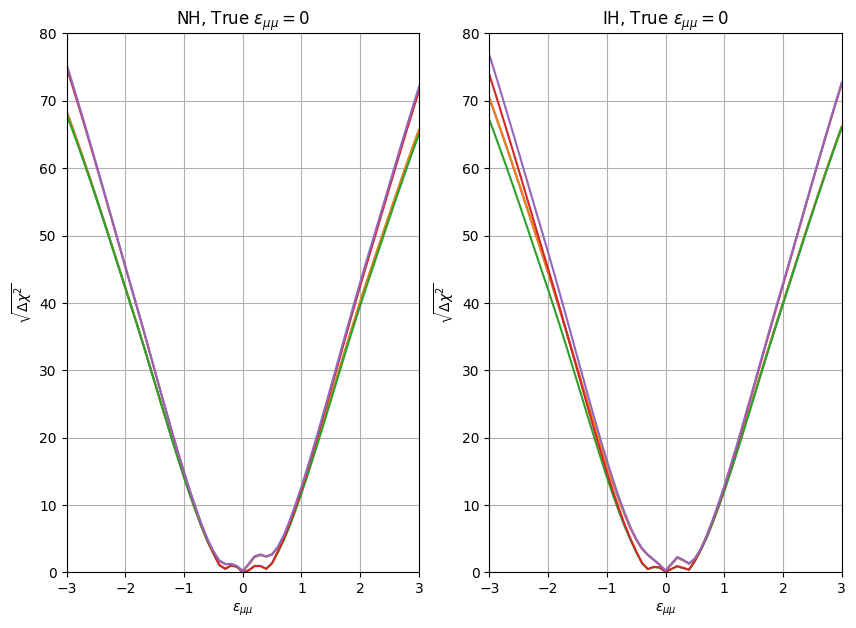}
\includegraphics[width=0.7\textwidth] {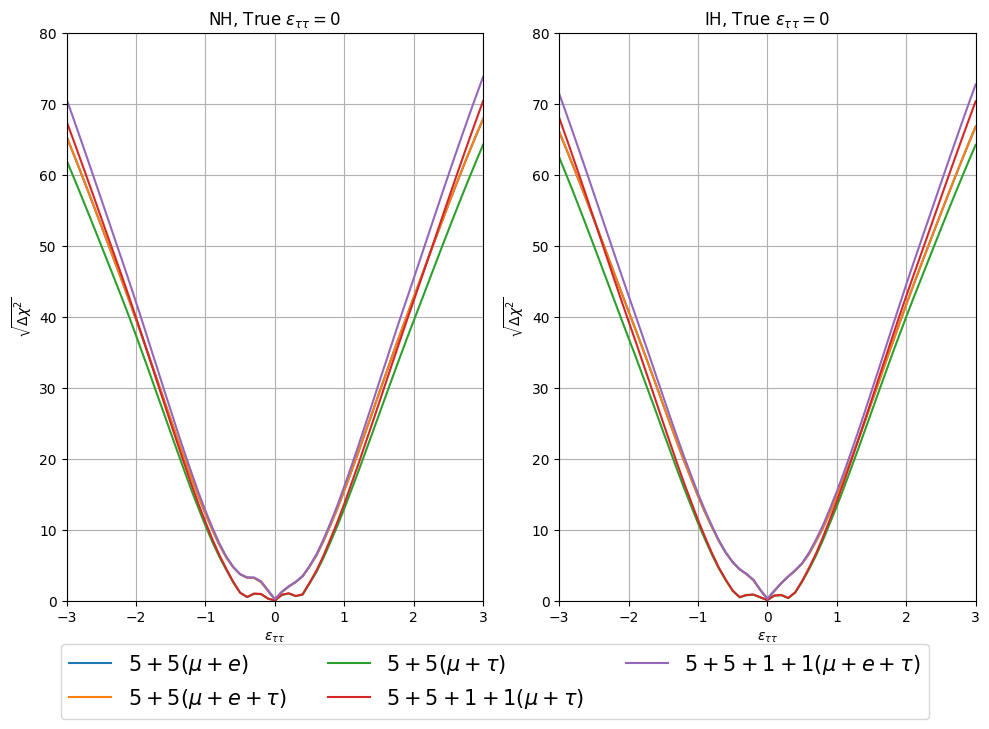}

\caption{\footnotesize{Sensitivity to  NSI parameters $\epsilon_{\mu\tau}$, $\epsilon_{\mu\mu}$ and $\epsilon_{\tau\tau}$ with different runtimes of DUNE.}}
\label{nsi-sens}
\end{figure}
\section{A short discussion on the unitarity of PMNS matrix and role of $\tau$ neutrino in determining it}
\label{unitary}
One of the main physics motivation for the determination of $\tau$ neutrino is constraining the unitary property of the third row of the PMNS matrix. The unitary property of the PMNS matrix is broken when more than three neutrino generations exist in the nature. A minimal model of extra 'sterile' neutrino consists of $3+1$ neutrino mixings dominated by the standard 3 flavours $\nu_e$, $\nu_\mu$ and $\nu_{\tau}$ with a very small perturbative contribution from a new sterile flavour $\nu_s$. $\nu_s$ consists, mainly, of a heavy mass eigenstate $\nu_4$ with mass $m_4$. If  $m_1,\, m_2,\, m_3\ll m_4$ and $\Delta_{41} = m_{4}^{2}-m_{1}^{2}=[0.1-10]\, {\rm eV}^2$, then a direct sterile neutrino search in the neutrino oscillation experiments is possible. Recent results from the IceCube experiment constrain the sterile neutrino mass and mixing using atmospheric neutrino fluxes \cite{TheIceCube:2016oqi}. Constraints on the existence of sterile neutrino have been discussed in ref.~\cite{Bryman:2019bjg, Boser:2019rta, Miranda:2018buo}, while ref.~\cite{Gupta:2018qsv, Chatla:2018sos, Choubey:2017ppj, Choubey:2017cba, Berryman:2015nua} discuss the effects of a light sterile neutrino on present and future long baseline experiments.

\begin{figure}[htbp]
\centering
\includegraphics[width=0.6\textwidth] {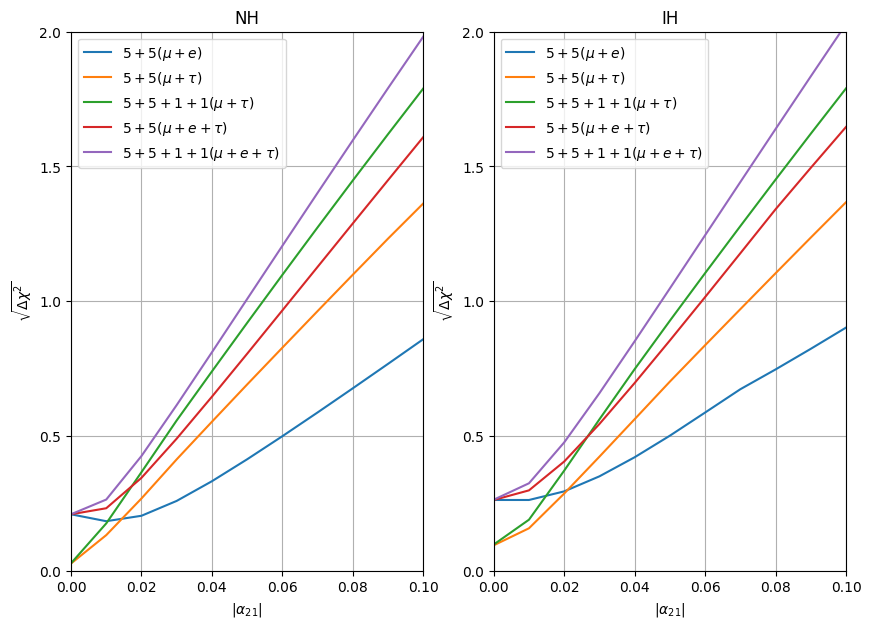}
\includegraphics[width=0.6\textwidth] {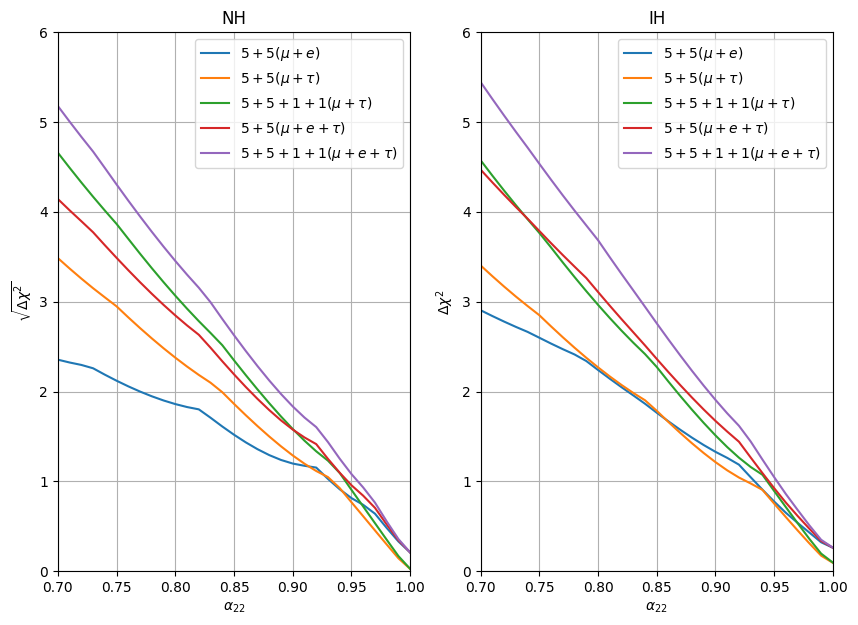}

\caption{\footnotesize{Sensitivity to non-unitary mixing parameters with different runtimes of DUNE.}}
\label{nonuni-sens}
\end{figure}

However, if the extra generation exists as iso-singlet neutral heavy leptons (NHL), then the sterile neutrinos will not take part in neutrino oscillations. In this case, their ad-mixture in charged current weak interactions will violate the unitary property of the $3\times3$ mixing matrix. The 3-flavour neutrino oscillation will then be described by a non-unitary $3\times3$ mixing matrix. The non-unitary mixing matrix, in a generalised model independent way, can be written as
\begin{equation}
    N=N_{NP}U_{3\times 3}= \left[ {\begin{array}{ccc}
   \alpha_{00} & 0 & 0 \\
   \alpha_{10} & \alpha_{11} & 0 \\
   \alpha_{20} & \alpha_{21} & \alpha_{22}
  \end{array} } \right] U \,
  \label{non-uni}
\end{equation}
where $U$ is the PMNS matrix from eq.~\ref{PMNS}.The diagonal elements $\alpha_{ii}$ of $N_{NP}$ are real, and the off-diagonal elements $\alpha_{ij}=|\alpha_{ij}|e^{i\phi_{ij}}$ are complex, with $i,j=0,1,2$ and $i>j$. The constraints on the unitary properties of the first two rows of the mixing matrix can be given by detecting electron and $\muon$ neutrinos at the neutrino oscillation experiments. However constraints on the third row mostly comes from charged lepton flavor violation (CLFV) experiments. The present $3\,\sigma$ limits for 1 degree of freedom on the unitary violation in the third row of the mixing matrix from a joined fit of neutrino oscillation and CLFV (only neutrino oscillation) experiments are \cite{Escrihuela:2016ube}:
\begin{equation}
|\alpha_{20}| < 4.4\times10^{-3} (9.8\times10^{-2}) \,;\, |\alpha_{21}| < 2.0\times10^{-3} (1.7\times10^{-2}) \,;\, \alpha_{22} > 0.9976 (0.76)
\label{eq:nubounds}
\end{equation}
The opportunity to detect $\tau$ neutrinos and anti-neutrinos in DUNE gives a unique opportunity to constrain the violation of the unitary property of the third row of $3\times3$ mixing matrix through neutrino oscillation experiments. In ref.~\cite{Denton:2021mso}, it has been shown that tau neutrino appearance from the beam data at DUNE far detector, combined with the atmospheric data of $\nu_\mu$ disappearance and $\nu_\tau$ appearance can be useful to constrain the elements of the third row of the PMNS matrix. In fig.~\ref{nonuni-sens} we have shown the expected sensitivity to $|\alpha_{21}|$ and $\alpha_{22}$ from DUNE with different running schemes. We have not shown the sensitivity to $\alpha_{20}$ because the sensitivity is very low and that is because the $e-\tau$ oscillation is negligible in experiments with $\nu_\mu$ and $\bar{\nu}_\mu$ beams. To calculate this sensitivity, we first obtained the true event rates considering standard $3\times3$ unitary mixing with the standard oscillation parameters fixed at the global best-fit values. For the test event rates, we varied $\sin^2\tz$ and $|\dl|$ in their $3\,\sigma$ ranges. $\dcp$ has been varied in the complete range $[-180^\circ:180^\circ]$. The test mass hierarchy is same as the true one. The test values of the non-unitary parameter $|\alpha_{21}|$ ($\alpha_{22}$) have been varied in the range $[0:0.1]$ ($[0.7:1]$). The phase $\phi_{21}$ associated with $\alpha_{21}$ has been varied in the range $[-180^\circ:180^\circ]$. The $\dchsq$ between the true and test event rates have been calculated and marginalized over all the test parameters except $|\alpha_{21}|$ or $\alpha_{22}$. It can be observed that for these two parameters, $\tau$ neutrino detection enhances the sensitivity of detecting these PMNS parameters. It can also be seen that $5+5(\mu+\tau)$ and $5+5+1+1(\mu+\tau)$ runs have better sensitivity than $5+5(\mu+e)$ run. The complete data set of the $5+5+1+1(\mu+e+\tau)$ run gives a $2\,\sigma$ limit of $|\alpha_{21}|<0.1$ for both the hierarchies. This limit is weaker than the global limit from only neutrino oscillation given in eq.~\ref{eq:nubounds}. For $\alpha_{22}$, $5+5+1+1(\mu+e+\tau)$ run gives a $3\,\sigma$ limit of $\alpha_{22}>0.83$ for both the hierarchies. This bound is stronger than the global limit from only neutrino oscillation data given in eq.~\ref{eq:nubounds}.

\section{Conclusions}
\label{conclusion}
In this work, we have investigated the impact of detecting $\nu_\tau$ and $\bar{\nu}_\tau$ at DUNE on constraining Non-Standard Interactions (NSI) in neutrino propagation. We demonstrated that the most significant NSI effects on $\nu_\tau$ appearance arise from $\epsilon_{\mu\tau}$, and that incorporating $\nu_\tau$ detection alongside traditional $\nu_e$ and $\nu_\mu$ channels improves sensitivity to this parameter. Our results indicate that DUNE’s ability to probe $\epsilon_{\mu\tau}$ approaches the current constraint from IceCube, making $\nu_\tau$ detection a valuable addition to NSI studies.

The three main physics goals of DUNE- namely, mass hierarchy determination, CP violation determination, and octant determination- are not affected by tau neutrino detection regardless of whether NSI effects are present or not. Although, we see that at the oscillation probability level, there are distinctions between two hierarchies for $|\epsilon_{\mu\tau}|$, the hierarchy sensitivity of $\nu_\tau$ and $\bar{\nu}_\tau$ appearance channels is lost due to the $\phi_{\mu\tau}$-hierarchy degeneracy arising due to the dependency of the hierarchy sensitive term in the probability expression on $\cos\phi_{\mu\tau}$. For any true hierarchy and true value of $\phi_{\mu\tau}$, $\pmt$ and $\pmtbar$ can be mimicked by the wrong hierarchy and $\phi_{\mu\tau}({\rm test})=180^\circ-\phi_{\mu\tau}({\rm true})$. 

For the same $\cos\phi$ dependent term, we found that the measurement of the NSI phase $\phi_{\mu\tau}$ can be enhanced through $\nu_\tau$ detection, providing a potential additional source of CP violation. This is especially relevant in scenarios where the mass hierarchy is well-determined.  The detection of $\tau$ neutrino appearances, along with $\mu$ neutrino disappearances, provide sensitivities to NSI parameters $\epsilon_{\mu\mu}$ and $\epsilon_{\tau\tau}$ which are comparable to the sensitivities from electron neutrino appearances and $\mu$ neutrino disappearances. The sensitivities to $\epsilon_{\mu\mu}$ and $\epsilon_{\tau\tau}$ mostly come from $\nu_\mu$ and $\bar{\nu}_\mu$ disappearances. In case of $|\epsilon_{\mu\tau}|$, the $5+5+1+1(\mu+e+\tau)$ run provides a slightly stronger constraint than the $5+5(\mu+e)$ run. The constraint on $|\epsilon_{\mu\tau}|$ from the $5+5+1+1(\mu+\tau)$ run is close to the IceCube constraint \cite{IceCubeCollaboration:2021euf}. This sensitivity to $|\epsilon_{\mu\tau}|$ comes from both muon disappearance and tau appearance channels. For the other two NSI parameter, considered in this paper, the $5+5(\mu+e)$ case has much better sensitivity than any other running scheme without electron neutrino and anti-neutrino appearances. Therefore, in case of $\epsilon_{\mu\tau}$, $\epsilon_{\mu\mu}$ and $\epsilon_{\tau\tau}$, it is possible to obtain an independent complimentary sensitivity by analysing the tau (anti-)neutrino appearance and muon (anti-)neutrino disappearance data alone.

Furthermore, we explored the implications of $\nu_\tau$ detection for testing the unitarity of the PMNS matrix. The ability to constrain the non-unitary parameter $\alpha_{22}$ stronger than the present global fit limit using $\nu_\tau$ appearance data presents an additional motivation for incorporating $\nu_\tau$ measurements in DUNE experiments.

Looking ahead, our results suggest that future long-baseline and atmospheric neutrino experiments, such as DUNE, IceCube, and KM3NeT, could provide complementary constraints on NSI from $\nu_\tau$ and $\bar{\nu}_\tau$ appearance channels. A combined analysis incorporating $\nu_\tau$ data from these experiments may significantly enhance the sensitivity to $\epsilon_{\mu\tau}$ and other NSI parameters. Additionally, improvements in $\nu_\tau$ identification at DUNE, such as optimized flux configurations and enhanced event reconstruction techniques, could further strengthen the experimental reach.

In conclusion, our study highlights the importance of $\nu_\tau$ detection as a complementary probe of new physics in neutrino oscillations. Future experimental efforts aimed at increasing $\nu_\tau$ detection efficiency will be crucial for refining NSI constraints and probing possible beyond Standard Model effects in the neutrino sector.

\section*{Acknowledgement}
We thank Pedro Machado and Pedro Pasquini for the valuable discussions and comments over email communications. This work reflects the views of the authors and not those of the DUNE collaboration.

\bibliographystyle{jhep}
\bibliography{referenceslist}
\appendix
\section{More discussions on fluxes and detector simulations}
\label{flux+sim}
\begin{figure}[htbp]
\centering
\includegraphics[width=0.4\textwidth] {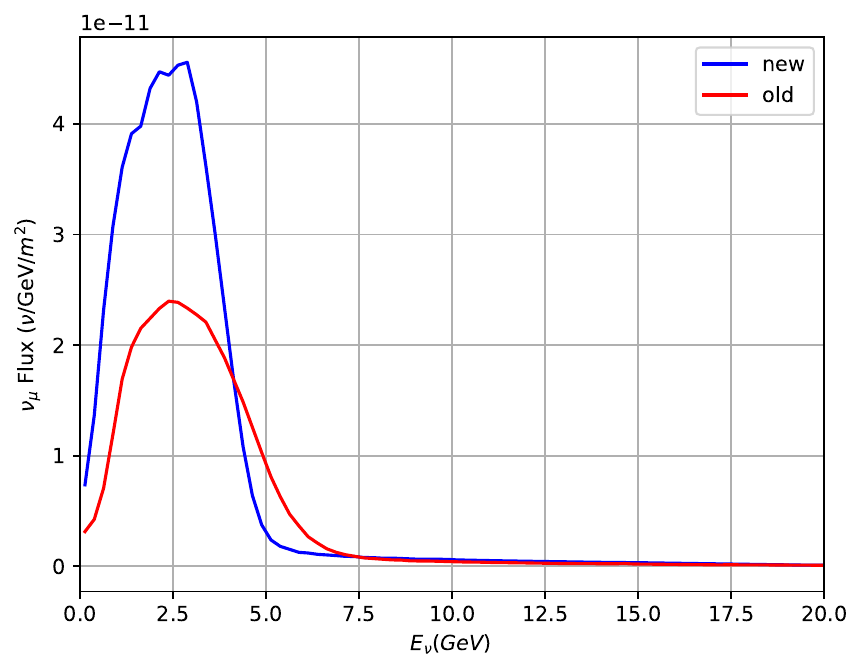}
\includegraphics[width=0.4\textwidth] {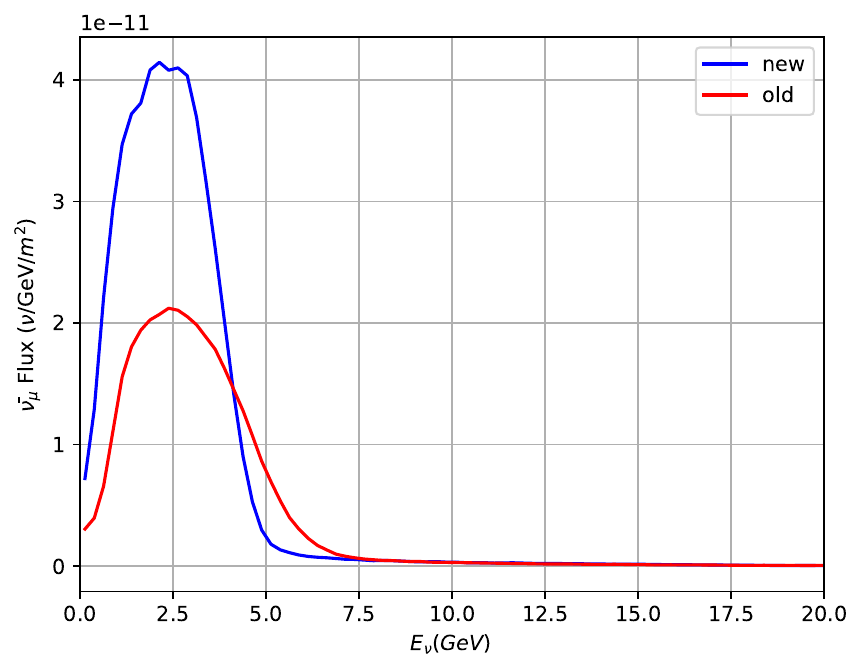}
\includegraphics[width=0.4\textwidth] {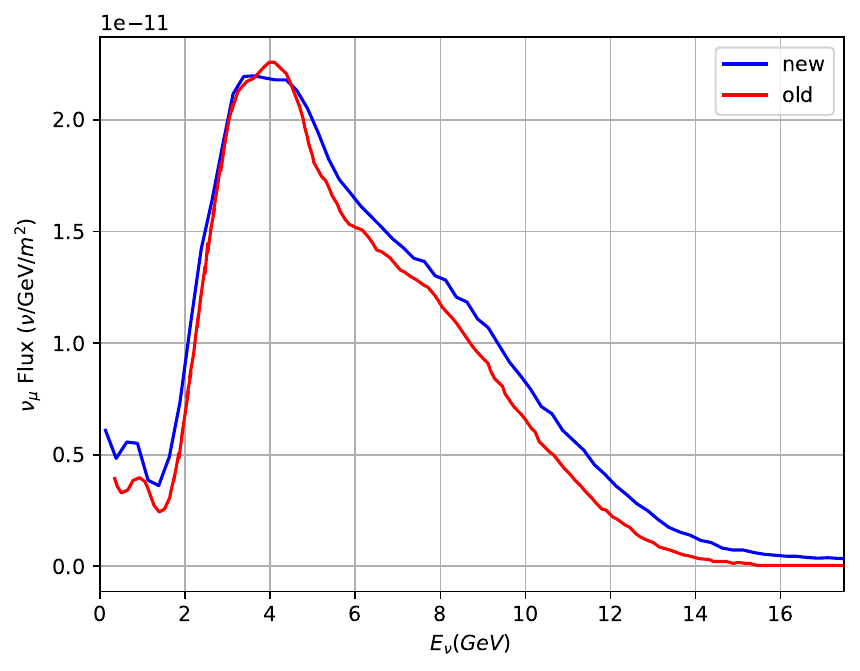}
\includegraphics[width=0.4\textwidth] {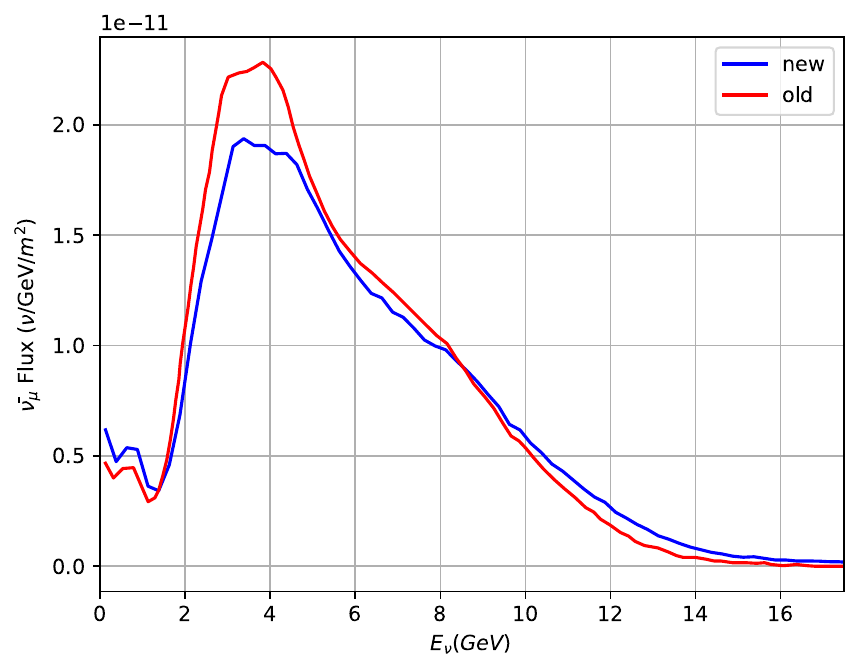}
\caption{\footnotesize{Comparison between fluxes used in our paper with those used in previous works. The top (bottom) panels are for the regular (tau optimized) fluxes, and the left (right) panels are for $\nu_\mu$ ($\bar{\nu}_\mu$) fluxes in the neutrino (anti-neutrino) beam.}}
\label{flux}
\end{figure}

In section~\ref{intro}, we mentioned that although ref.~\cite{DeGouvea:2019kea, Masud:2017bcf, Ghoshal:2019pab} discussed about the role of tau neutrino detection in the NSI analysis at DUNE, all of them had considered outdated fluxes. In fig.~\ref{flux}, we have shown the comparison between the fluxes used in our paper (referred to as "new"), and those used in the previous works (referred to as "old"). We can see that for the regular DUNE beam, the fluxes used in our paper are significantly higher than those used in previous works, and hence the former lead to higher statistics than the latter. The tau optimized fluxes in both the cases are almost equivalent. For $\nu_\mu$ disappearances, and $\nu_e$ appearances, ref.~\cite{DeGouvea:2019kea, Masud:2017bcf, Ghoshal:2019pab} have used the detector simulations provided in ref.~\cite{DUNE:2016ymp}, whereas we have used the detector simulations provided in ref.~\cite{DUNE:2021cuw}. For the tau optimized fluxes, we have followed the simulations provided in ref.~\cite{DeGouvea:2019kea}. However, they did not use tau optimized anti-neutrino beam. In section~\ref{analysis}, we have discussed the details of the simulations for tau optimized anti-neutrino beam. Ref.~\cite{Ghoshal:2019pab} used a different simulation for their tau optimized beam. For instance, they considered a constant efficiency independent of energy for all the energy bins, whereas we have considered energy dependent efficiencies for each energy bin. Details of the energy resolution functions for the tau neutrino and anti-neutrino appearance channels have not been provided in ref.~\cite{Ghoshal:2019pab}. 

\section{Octant sensitivity for NSI due to $\epsilon_{\mu\tau}$, $\epsilon_{\mu\mu}$ and $\epsilon_{\tau \tau}$}
\label{octant-appendix}
\begin{figure}[htbp]
\centering
\includegraphics[width=0.5\textwidth] {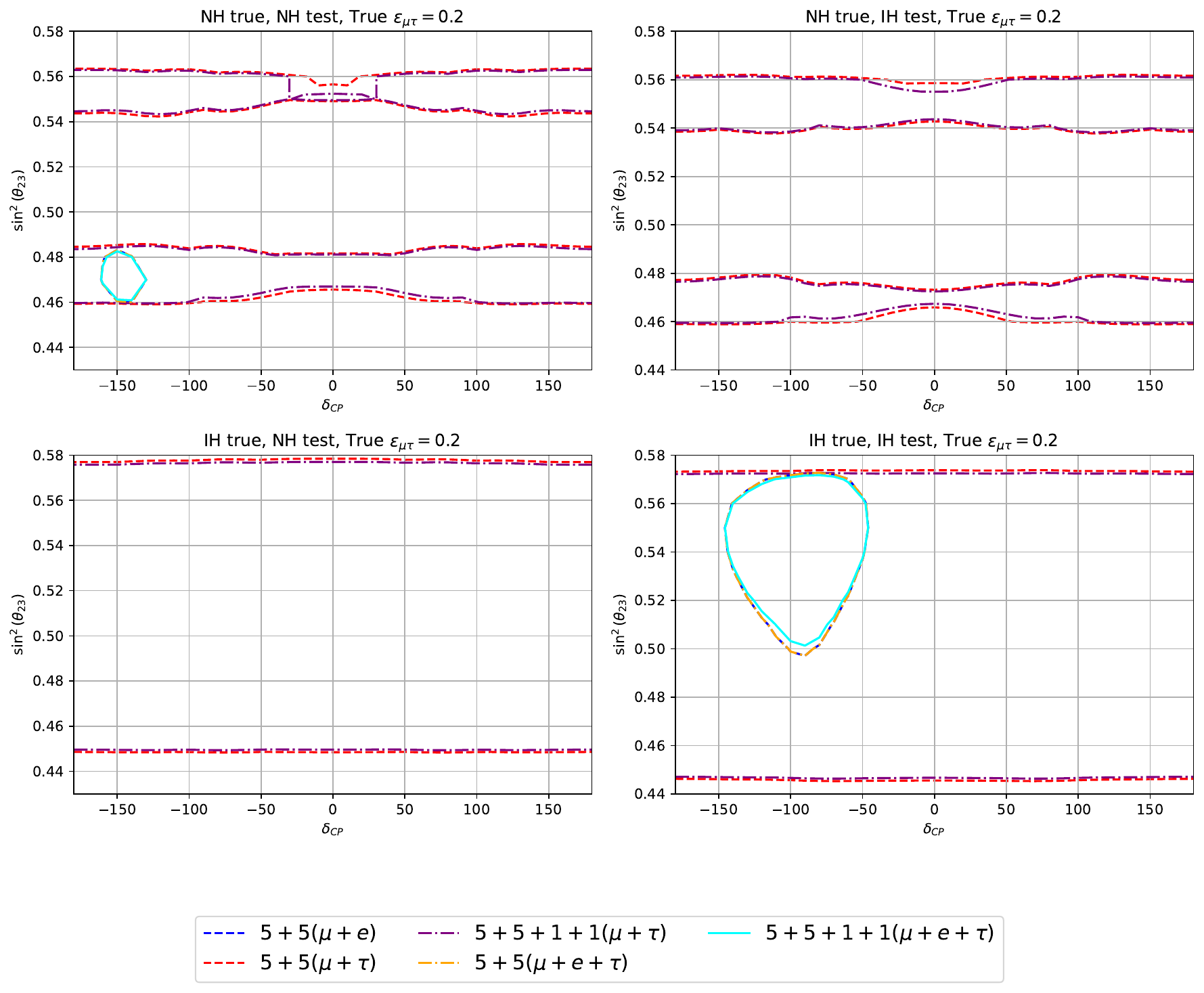}
\includegraphics[width=0.5\textwidth] {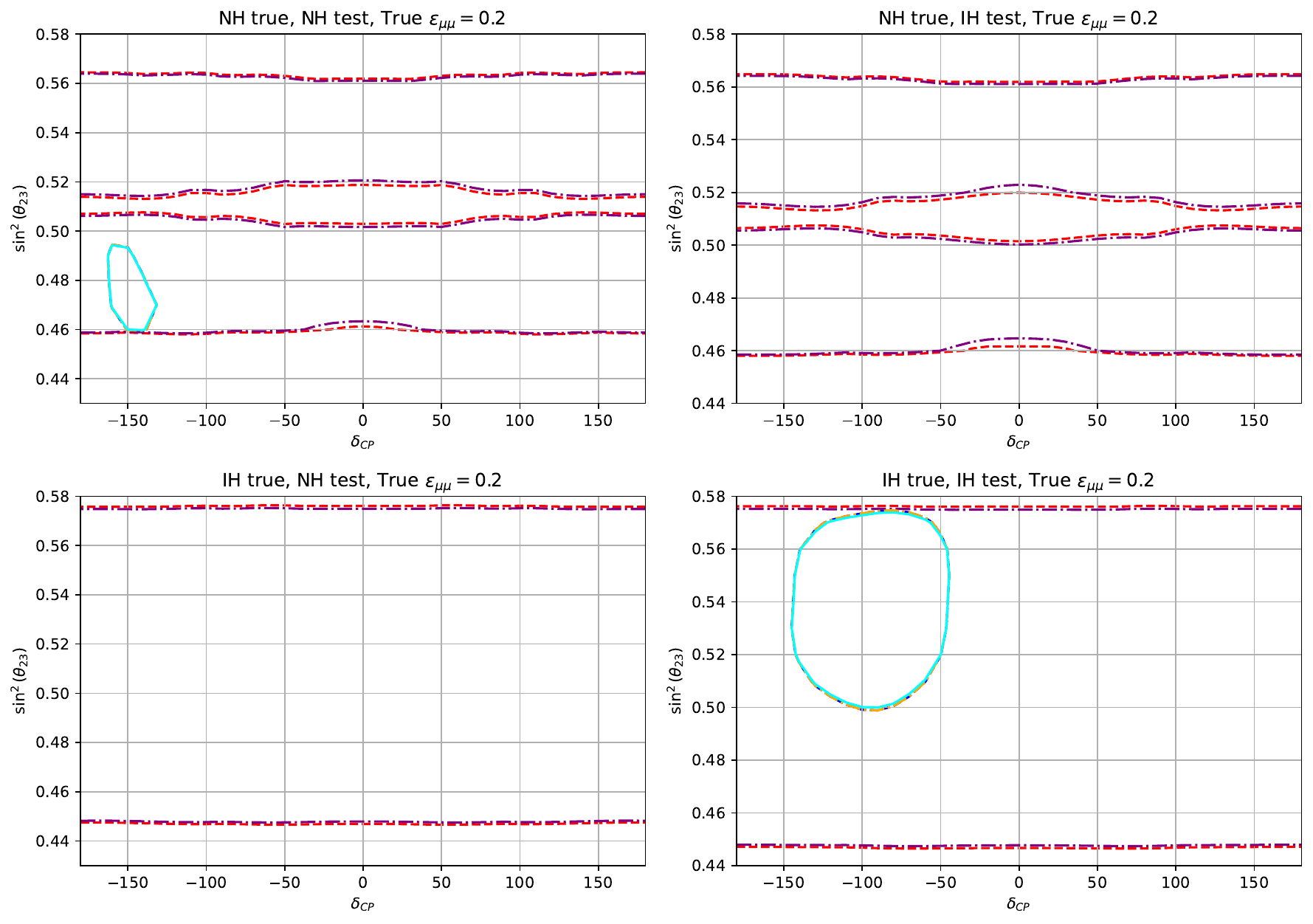}
\includegraphics[width=0.5\textwidth] {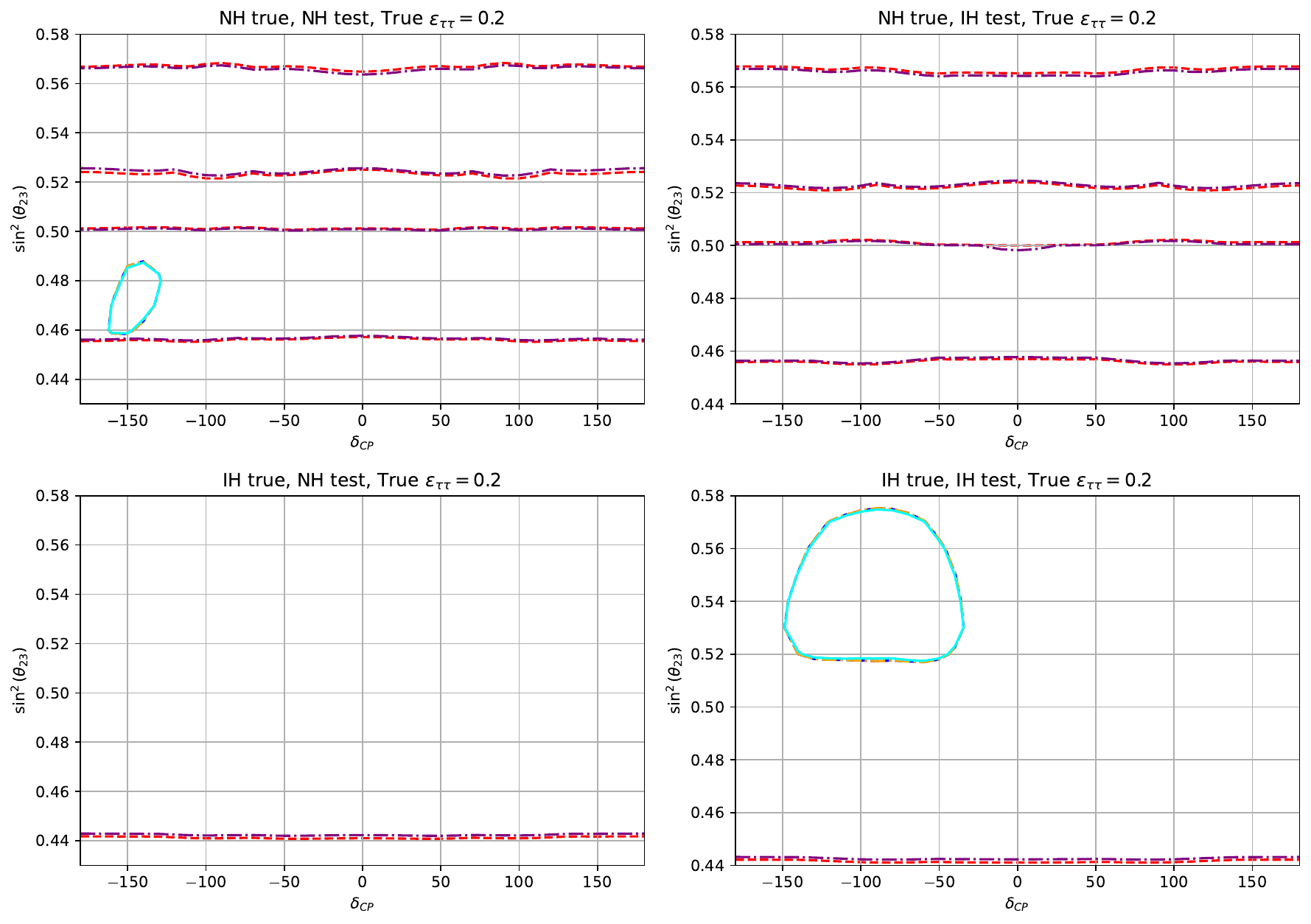}
\caption{\footnotesize{Allowed regions in the test $\dcp-sin^2\tz$ plane for oscillation with NSI due to $\epsilon_{\mu\tau}$, $\epsilon_{\mu\mu}$ and $\epsilon_{\tau \tau}$. The top (bottom) panels are for NH (IH) being the true hierarchy. The left (right) panels are for test hierarchy being NH (IH). }}
\label{octant-NSI-appendix}
\end{figure}

We have discussed about the octant sensitivity from $\tau$ neutrino and anti-neutrino appearance channels in presence of NSI in section~\ref{octant}. In fig.~\ref{octant-NSI}, we presented the results for NSI due to $\epsilon_{e\mu}$, and $\epsilon_{e\tau}$ only. In this section, we will present the octant sensitivity results for other three NSI parameters, namely $\epsilon_{\mu\tau}$, $\epsilon_{\mu\mu}$ and $\epsilon_{\tau \tau}$ in fig.~\ref{octant-NSI-appendix}. $5+5(\mu+\tau)$, $5+5+1+1(\mu+\tau)$ running schemes do not have any octant sensitivity for these three NSI parameters. $5+5(\mu+e)$ run has good octant sensitivity but it does not improve with additional data from $\nu_{\tau}$ and $\bar{\nu}_\tau$ appearance channels.

\section{Sensitivity to $\epsilon_{e\mu}$ and $\epsilon_{e\tau}$}
\label{sensitivity-appendix}
\begin{figure}[htbp]
\centering
\includegraphics[width=0.8\textwidth] {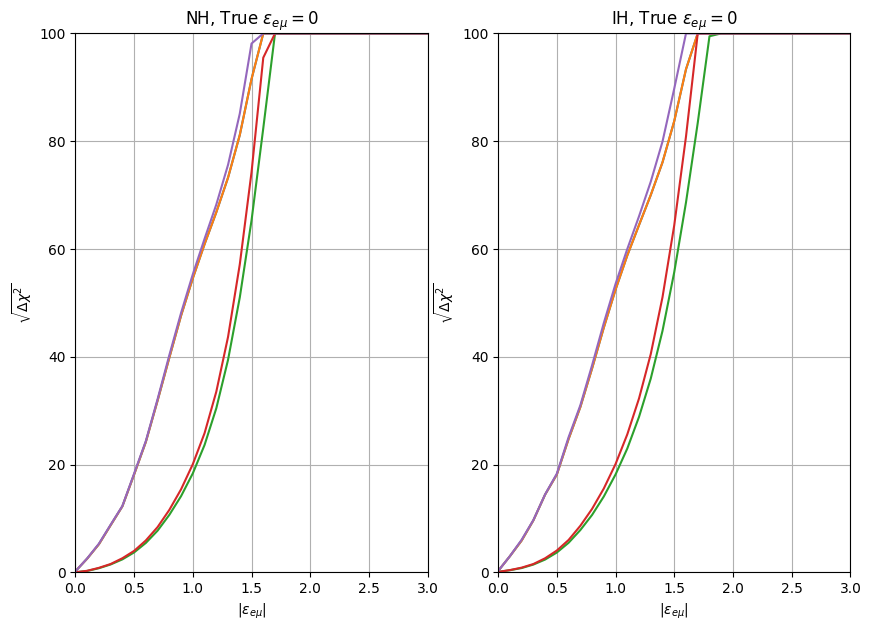}
\includegraphics[width=0.8\textwidth] {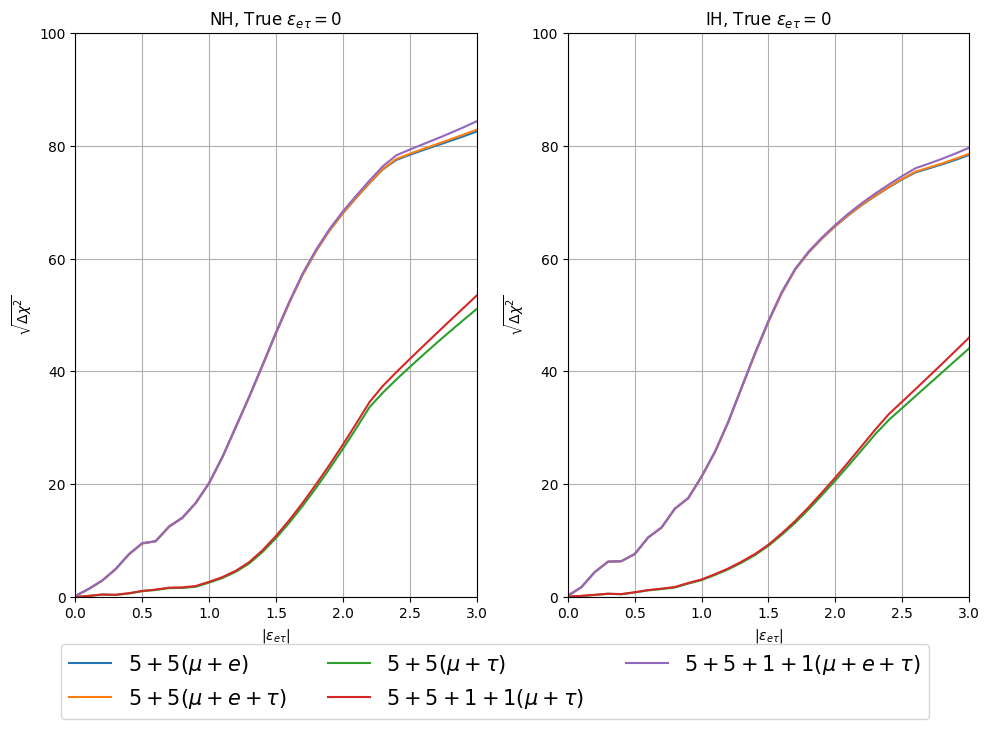}
\caption{\footnotesize{Sensitivity to different NSI parameters $\epsilon_{e\mu}$ and $\epsilon_{e\tau}$ with different runtimes of DUNE.}}
\label{nsi-sens-appendix}
\end{figure}

In section \ref{sensitivity}, we have discussed the sensitivity to different NSI parameters. In fig.~\ref{nsi-sens}, we have presented the results for sensitivity to NSI parameters for $\epsilon_{\mu\tau}$, $\epsilon_{\mu\mu}$ and $\epsilon_{\tau\tau}$.The results for sensitivity to $\epsilon_{e\mu}$ and $\epsilon_{e\tau}$ have been represented in fig.~\ref{nsi-sens-appendix}. 

\end{document}